\documentclass[a4paper,11pt]{article}
\usepackage{jcappub} % for details on the use of the package, please see the JINST-author-manual
\usepackage{lineno}
\usepackage[dvipsnames]{xcolor}
\usepackage{xfrac}
\usepackage[T1]{fontenc} % if needed
\usepackage{aas_macros}
\usepackage{natbib}
\usepackage{tabularx}
\usepackage{bm}
\usepackage{hyperref}
\usepackage{subcaption}
\usepackage{mathrsfs}

\newcommand{\bfc}{BFC}

\usepackage{xcolor}
\usepackage{soul}

\arxivnumber{1234.56789} % Only if you have one
\title{Baryonification: An alternative to hydrodynamical simulations for cosmological studies}
\author[a]{Aurel Schneider,}
\author[b]{Michael Kova\v{c},}
\author[c]{Jozef Bucko,}
\author[d, b]{Andrina Nicola,}
\author[b]{Robert Reischke,}
\author[e,f]{Sambit K. Giri,}
\author[g]{Romain Teyssier,}
\author[c]{Tilman Tr\"oster,}
\author[c]{Alexandre Refregier,}
\author[h,i]{Matthieu Schaller,}
\author[h]{Joop Schaye}

\affiliation[a]{Department of Astrophysics, University of Zurich, Winterthurerstrasse 190, 8057 Zurich, Switzerland}
\affiliation[b]{Argelander-Institut für Astronomie, Universität Bonn, Auf dem Hügel 71, D-53121 Bonn, Germany}
\affiliation[c]{Institute for Particle Physics and Astrophysics, ETH Zurich, Wolfgang Pauli Strasse 27, 8093 Zurich, Switzerland}
\affiliation[d]{Jodrell Bank Centre for Astrophysics, Department of Physics and Astronomy, The University of Manchester, Manchester M13 9PL, UK}
\affiliation[e]{Van Swinderen Institute for Particle Physics and Gravity, University of Groningen, Nijenborgh 4, 9747 AG Groningen, The Netherlands}
\affiliation[f]{Department of Astronomy and Oskar Klein Centre, AlbaNova, Stockholm University, SE-10691 Stockholm, Sweden}
\affiliation[g]{Department of Astrophysical Sciences, Princeton University, Princeton, NJ 08540, USA}
\affiliation[h]{Leiden Observatory, Leiden University, PO Box 9513, 2300 RA Leiden, the Netherlands}
\affiliation[i]{Lorentz Institute for Theoretical Physics, Leiden University, PO box 9506, 2300 RA Leiden, the Netherlands}

\emailAdd{aurel.schneider@uzh.ch}

\abstract{We present an improved baryonification (\bfc{}) model that modifies dark-matter-only $N$-body simulations to generate particle-level outputs for gas, dark matter, and stars. Unlike previous implementations, our approach first splits each simulation particle into separate dark matter and baryonic components, which are then displaced individually using the \bfc{} technique. By applying the hydrostatic and ideal gas equations, we assign pressure and temperature values to individual gas particles. The model is validated against hydrodynamical simulations from the FLAMINGO and TNG suites (which feature varied feedback prescriptions) showing good agreement at the level of density and pressure profiles across a wide range of halo masses.

As a further step, we calibrate the \bfc{} model parameters to gas and stellar mass ratio profiles from the hydrodynamical simulations. Based on these calibrations, we baryonify $N$-body simulations and compare the resulting total matter power spectrum suppressions to the ones from the same hydrodynamical simulation. Carrying out this test of the \bfc{} method at each redshift individually, we obtain a 2 percent agreement up to $k=5\,h$/Mpc across all tested feedback scenarios. We also define a reduced, 2+1 parameter \bfc{} model that simultaneously accounts for feedback variations (2 parameters) and redshift evolution (1 parameter). The 2+1 parameter model agrees with the hydrodynamical simulations to better than 2.5 percent over the scales and redshifts relevant for cosmological surveys.

Finally, we present a map-level comparison between a baryonified $N$-body simulation and a full hydrodynamical run from the TNG simulation suite. Visual inspection of dark matter, gas, and stellar density fields, along with the integrated pressure map, shows promising agreement. Further work is needed to quantify the accuracy at the level of observables. Overall, the new component-wise baryonification model offers a flexible and efficient framework for multi-probe cosmological studies.
}

\begin{document}
\maketitle
\flushbottom

\section{Introduction}
Modern cosmological galaxy surveys such as KiDS \cite{Wright:2025xka}, DES \cite{DES:2020aks}, HSC \cite{Dalal:2023olq}, Euclid \cite{Euclid:2024yrr}, LSST \cite{LSST:2008ijt}, or Roman \cite{ROMAN:2015aaa} are measuring the positions and shapes of millions to billions of galaxies, enabling precise cosmological studies of the weak lensing and galaxy clustering signal over a large range of scales and deep into the nonlinear regime. At the same time, measurements of the Sunyaev-Zel'dovich effect from cosmic microwave background (CMB) surveys, such as Planck \cite{Planck:2015koh,Planck:2015vgm}, ACT \cite{ACT:2020lcv,AtacamaCosmologyTelescope:2020wtv}, and SPT \cite{SPT:2014wbo,SPT-SZ:2021gsa}, as well as X-ray observations from missions like eROSITA \cite{Bulbul:2024mfj} provide valuable information on the distribution of intergalactic gas at cosmological scales. Together, these surveys are advancing our understanding of the gas, stellar, and dark matter clustering signal at scales relevant for both cosmology and extragalactic astrophysics. However, the increased quality of observational data requires better theoretical models at both large and small cosmological scales.

Theoretical predictions on small cosmological scales are particularly challenging. While gravity-only $N$-body simulations yield precise (and converged) results for the nonlinear clustering process, they fail to capture non-gravitational effects caused by hydrodynamical processes. A key factor is the feedback energy from active galactic nuclei (AGN) that drives gas out of galaxy groups and clusters, significantly impacting the cosmological signal \citep[see e.g. Ref.][]{Rudd:2008, vanDaalen:2011xb,Chisari:2019tus,Schaller:2024jiq}.

Cosmological hydrodynamical simulations attempt to account for key physical phenomena from gravitational clustering to gas cooling, star formation, stellar and AGN feedback processes. However, these simulations require their empirical sub-grid star-formation and feedback models to be calibrated against observations \cite{McCarthy:2016mry,Kugel:2023wte}. The significant computational costs, combined with the need for calibration, make running hydrodynamical simulations both a challenging and resource-intensive task.

The baryonification method \citep{Schneider:2015wta} offers a complementary approach for cosmological studies. It relies on small modifications of $N$-body simulations based on empirical functions for the gas, stellar, and dark matter profiles around galaxy groups and clusters. The method can be calibrated to observations or hydrodynamical simulations, providing a computationally efficient way to model cosmological probes. Several models and implementations, either based on baryonification or related methods, have been developed over the past few years. They are based on small displacements of particles  \citep{Schneider:2018pfw, Arico:2019ykw,  Kacprzak:2022pww, Anbajagane:2024nzx}, the painting of new haloes onto an existing $N$-body simulation \citep{Osato:2022znr, Williams:2022zma}, or analytically motivated halo model approaches \cite{Semboloni2011aaa,Semboloni2013aaa, Fedeli:2014gja, Debackere:2019cec, Mead:2020qgo, Troster:2021gsz}.

The baryonification method has been shown to accurately capture the baryonic feedback effects on the power spectrum \citep{giri2021emulation}, the bi-spectrum \citep{Arico:2020yyf,Burger:2025meh} and other summary statistics such as weak lensing peaks \citep{Weiss:2019jfx, Lee:2022jyg}, as well as gas and stellar fractions from X-ray data \citep{Schneider:2019xpf,Grandis:2023qwx}, dispersion measures of Fast Radio Bursts \citep{2024arXiv241117682R}, or profiles from the kinetic Sunyaev-Zel'dovich effect \citep{Schneider:2021wds}. It has been applied to cosmological parameter studies using data from KiDS \citep{Schneider:2021wds,Bucko:2022kss} or DES \citep{DES:2022eua, Arico:2023ocu, DES:2024iny}, and it can be used for cosmological studies combining probes \cite[e.g.][]{Zennaro:2024dyy, ACT:2025llb}. However, baryonification provides three-dimensional density fields and is therefore not limited to summary statistics. Indeed, the method has been used for the first simulation-based cosmological inference analysis with neural networks \citep{Fluri:2022rvb,Fluri:2019qtp}. Note that such approaches remain exploratory and their accuracy needs to be further tested, especially considering the improved precision of future survey data.

In this paper, we present an updated version of the baryonification model from \cite{Schneider:2015wta,Schneider:2018pfw,giri2021emulation}. Instead of displacing the original $N$-body simulation particles, we first split each $N$-body particle into a dark matter and a baryonic particle before displacing them at different rates. After the displacement, the baryonic particles are either assigned to become a star or a gas particle using a random process based on the density profiles.

The new \bfc{} method presented in this paper allows us to produce individual density fields for dark matter, gas, and stars. For each gas particle, we can furthermore assign pressure and temperature values based on the hydrostatic equation and the ideal gas law. This extension is a first step towards multi-survey cosmological analyses at the map level combining, for example, weak lensing data with galaxy clustering, X-ray, the Sunyaev-Zel'dovich effect, and other observables.

Besides introducing a new method for the baryonification  process, we also present a detailed comparison study using measured profiles from the FLAMINGO \cite{Schaye:2023jqv,Kugel:2023wte} and TNG \cite{Springel:2017tpz,Pillepich:2017fcc} hydrodynamical simulation suites. We test and refine the \bfc{} parametrisation for the gas, stellar, dark matter, and pressure profiles and, in particular, we refine the model for the dark matter back-reaction process.

After updating the parametrisation, we test the \bfc{} model by comparing the power spectra to those from various hydrodynamical simulations. Instead of directly fitting to the power spectra, we fit the \bfc{} model to the simulated gas and stellar mass fractions as a function of radius. We then compare the power spectrum of the baryonified box with the one from the corresponding hydrodynamical simulation. This is a much harder test for the \bfc{} method that is based on the connection between halo profiles and clustering statistics.

Finally, we visually compare the \bfc{} model with the TNG-100 simulation at the particle level. We plot the TNG gas, stellar, and dark matter densities, as well as the integrated electron pressure map next to the results from the baryonification model. This consists of a first step towards a validation of the \bfc{} method for field-level analysis. Further investigations of cosmological observables at the field-level will be carried out in a future publication.

The paper is structured as follows: In Sec.~\ref{sec:model} we summarise the baryonification approach, and we present the parametrisation of the profiles and the dark matter back-reaction model. Sec.~\ref{sec:comparison} consists of a comparison with the FLAMINGO and TNG-300 hydrodynamical simulations at the profile-level. In Sec.~\ref{sec:profile2PS} we test the validity of baryonification by pre-fitting the \bfc{} model to simulated gas and stellar fractions before comparing the power spectrum with the one from the same simulation. Sec.~\ref{sec:PressTemp} is dedicated to the pressure and temperature treatment and Sec.~\ref{sec:FieldLevel} provides a first comparison at the level of density and pressure maps. We conclude our work in Sec.~\ref{sec:Conclusion} and provide further information in the appendices.

\section{Baryonification model}\label{sec:model}
The baryonification (\bfc{}) model is a post-processing tool to modify gravity-only $N$-body simulations, thereby emulating the effects of gas, stars, and baryonic feedback. The method is based on empirical halo profiles informing an algorithm to displace particles around halo centres. Although the displacements are spherically symmetric, the complex structure of the cosmological density field with nodes, filaments, and voids is retained by the method.

We present an updated version of the  original model published in Refs. \cite{Schneider:2015wta, Schneider:2018pfw}. Next to several modifications regarding the parametrisation of profiles and the dark matter back-reaction model, the major change introduced here is that dark matter, gas, and stellar components are now treated as independent fields. Furthermore, we present a model for the gas temperature and pressure.

Throughout the paper, we define the virial mass and radius using a density threshold of  200 times the critical density ($\rho_\mathrm{c}$). Note that, in contrast to the background density ($\rho_\mathrm{b})$, the critical density is redshift dependent even in co-moving coordinates.

\subsection{Displacement}
As a first step, each particle of the original gravity-only $N$-body simulation output is duplicated, creating a dark matter and a baryonic particle. The particles from both species start with the same positions but are displaced differently. The particle masses are renormalised to obtain the correct global baryon and dark matter fractions.

The displacement for the dark matter (dm) and the baryonic (bar) particles is performed radially around each host halo centre according to the function
\begin{equation}\label{dispalcement}
d_{\rm A}(r_{\rm i}|M_{200},c_{200},\theta_{\rm \bfc{}}) = r_{\rm A,f}(M_A)-r_{\rm A, i}(M_{\rm A}),\hspace{1cm}{\rm A}=\lbrace{\rm dm,\, bar}\rbrace ,
\end{equation}
which depends on the halo mass ($M_{200}$), the halo concentration ($c_{200}$), and the \bfc{} model parameters (${\theta_{\rm bfc}}$). The functions $r_{\rm A, i}(M)$ and $r_{\rm A, f}(M)$ correspond to the inverted cumulative mass profiles of the initial, gravity-only ($M_{\rm i}$) and the final, baryonified ($M_{\rm f}$) state. The cumulative mass profiles are obtained by integrating the density profiles, i.e.,
\begin{equation}
M_{\rm A,i}(r) = \int_0^r\mathrm{d}s\;s^2\rho_{\rm A,i}(s),\hspace{1cm} M_{\rm A,f}(r) = \int_0^r\mathrm{d}s\;s^2\rho_{\rm A,f}(s).
\end{equation}
For the dark matter component, the initial and final profiles are given by
\begin{eqnarray}\label{DMinitialfinal}
\rho_{\rm dm, i}(r) = f_{\rm dm}\left[\rho_{\rm nfw}(r)+ \rho_{\rm 2h}(r)\right],\hspace{0.5cm} \rho_{\rm dm,f}(r) = \rho_{\rm dm}(r) + f_{\rm dm} \rho_{\rm 2h}(r),
\end{eqnarray}
where $f_{\rm dm}=\Omega_{\rm dm}/\Omega_{\rm m}$ is the fraction of dark matter. The truncated Navarro-Frenk-White (NFW) profile ($\rho_{\rm nfw}$), the adiabatically modified dark matter profile ($\rho_{\rm dm}$) and the two-halo term ($\rho_{\rm 2h}$) are introduced in Sec.~\ref{sec:DMOprofile},~\ref{sec:2hprofile}, and~\ref{sec:ACMprofile}. For the baryonic component, the initial and final density profiles are given by
\begin{eqnarray}\label{BARinitialfinal}
\rho_{\rm bar, i}(r) = f_{\rm bar}\left[\rho_{\rm nfw}(r)+\rho_{\rm 2h}(r)\right],\hspace{0.5cm}\rho_{\rm bar, f}(r) = \rho_{\rm gas}(r)+\rho_{\rm star}(r) + f_{\rm bar}\rho_{\rm 2h}(r),
\end{eqnarray}
where $f_{\rm bar}=\Omega_{\rm b}/\Omega_{\rm m}$ is the cosmic baryon fraction. The gas profile is subdivided into a dominant hot gas (hga) and an inner, cold gas (iga) component, while the stellar profile consists of the central galaxy (cga) and a satellite (sga) component:
\begin{equation}
\rho_{\rm gas}(r)=\rho_{\rm hga}(r) + \rho_{\rm iga}(r),\hspace{1cm}\rho_{\rm star}(r)=\rho_{\rm cga}(r)+\rho_{\rm sga}(r).
\end{equation}
All gas and stellar density profiles are defined in Sec.~\ref{sec:HGAprofile}, \ref{sec:IGAprofile}, \ref{sec:CGAprofile}, and \ref{sec:SGAprofile}.

Because the final density profile of the baryons differs significantly from the initial one, unlike the dark matter profile, which remains relatively unchanged, baryonic particles experience much larger average displacements than dark matter particles. As a result, baryonic particles from neighbouring haloes, including sub-haloes, are often displaced much farther out than their dark matter counterparts. This may lead to a spatial offset between the baryonic and dark matter components of haloes close to larger structures. To prevent any misalignment, we treat baryonic particles in neighbouring haloes differently: if a baryonic particle belongs to a neighbouring halo, we displace it using the dark matter prescription. This ensures that baryons and dark matter within a neighbouring halo are moved coherently and remain part of the same bound structure.

After the displacement process, the baryonic particles are assigned to become either a star or a gas particle. The assignment is done randomly according to radius-dependent probabilities ($P_{\rm gas}$, $P_{\rm star}$) that are proportional to the ratio of the profiles
\begin{equation}
P_\mathrm{i}(r) = \frac{\rho_\mathrm{i}(r)}{\rho_{\rm gas}(r)+\rho_{\rm star}(r)}, \hspace{1.5cm} i=\lbrace {\rm gas,\ star}\rbrace.
\end{equation}
At small radii, $P_{\rm star}$ is close to one and most particles become star particles. Further out, the gas profile dominates, and most particles are assigned the status of a gas particle. We assume that all stars reside within the virial radius. As a consequence, all baryonic particles that do not belong to any halo automatically become gas particles. 

Regarding satellite galaxies inside the virial radius, we provide two modelling options. The first option is to ignore all sub-haloes in the halo catalogue and to distribute all satellite stars radially around the host (according to the satellite galaxy profile $\rho_{\rm sga}$). Although this is a very simplistic approach, it is good enough for many applications in cosmology. The second option is to assign part (or all) of the satellite stars to the sub-haloes from the halo catalogue. We populate each sub-halo with the number of stars expected from the parametrised stellar fraction (see $f_{\rm cga}$ in Eq.~\ref{fstar}). Any remaining satellite stars are distributed radially around the halo centre. The total mass adds up to the stellar mass of the particular halo. In the rare case that there are not enough satellite stars to populate all sub-haloes, we reduce the number by an equal rate for all sub-haloes.

\begin{figure*} 
\centering\includegraphics[width=0.995\textwidth,trim=0.05cm 0.0cm 0.1cm 0.0cm, clip]{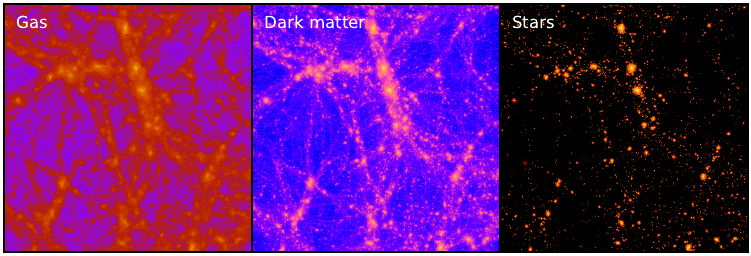}
\caption{\label{fig:gas_dm_stars_sat} Baryonified density fields of the gas (left), dark matter (centre), and stellar components (right). Here we show a randomly selected sub-box of $L=24$ Mpc/$h$ from a periodic simulation box of $L=128$ Mpc/$h$ with $N=512^3$ particles.}
\label{fig:densityslices}
\end{figure*}

In Fig.~\ref{fig:densityslices} we show a sub-box of $L=24$ Mpc/$h$ (in height, width and depth) of a baryonified $N$-body simulation output with periodic box of $L=128$ Mpc/$h$ and $N=512^3$ particles. The gas, dark matter, and stellar densities (with populated sub-haloes) are plotted from left to right. The density field of the gas is puffed up and smoother compared to the dark matter density. This is a direct consequence of the stronger particle displacements for baryonic particles, a fact that can be attributed to additional pressure forces and energy injection from baryonic feedback. The stars, on the other hand, are strongly concentrated around the highest DM density peaks and sub-peaks.

\subsection{NFW profile}\label{sec:DMOprofile}
The initial density profiles defined in Eq.~(\ref{DMinitialfinal}) and (\ref{BARinitialfinal}) consist of a truncated NFW profile and a two-halo term. Here we define the former, while the latter is introduced in Sec.~\ref{sec:2hprofile}. Following Ref.~\cite{Baltz:2007vq} and \cite{Oguri:2011vj}, the truncated NFW profile can be written as
\begin{equation}
\rho_{\rm nfw}(r)=\frac{\rho_{\rm nfw,0}}{x_\mathrm{s}(1+x_\mathrm{s})^2}\frac{1}{(1+x_\mathrm{t}^2)^2},
\label{eq:nfw}
\end{equation}
where $x_\mathrm{s}=r/r_\mathrm{s}$ and $x_\mathrm{t}=r/r_\mathrm{t}$. While the scale radius ($r_\mathrm{s}$) is defined via the halo concentration $c_{200} = r_{200}/r_\mathrm{s}$, the outer truncation radius is given by $r_\mathrm{t}=\varepsilon r_{200}$. In \cite{Schneider:2015wta} and \cite{ Schneider:2018pfw} $\varepsilon$ was assumed to be 4. Here we use a modified prescription
\begin{equation}
\varepsilon(\nu) = \varepsilon_0 + \varepsilon_1\nu,
\end{equation}
where $\nu=\delta_\mathrm{c}/\sigma$ is the peak height ($\delta_\mathrm{c}=1.686$ denoting the critical overdensity of spherical collapse and $\sigma$ the standard deviation of the linear overdensity field). We set $\varepsilon_0=4$ and $\varepsilon_1=0.5$ which provides a slightly better fit to DMO simulations \citep[see Ref.][]{Diemer:2014xya}.

The normalisation $\rho_{\rm nfw,0}$ can be expressed in terms of the virial mass ($M_{200}$) and radius ($r_{200}$), i.e.,
\begin{equation}\label{rhonfw0}
\rho_{\rm nfw,0} = M_{200}\left[4\pi\int_0^{r_{200}}\mathrm{d}s\, s^2 y_{\rm nfw}(s)\right]^{-1},
\end{equation}
where $\rho_{\rm nfw}\equiv\rho_{\rm nfw,0}y_{\rm nfw}$. The integral in Eq.~(\ref{rhonfw0}) can be solved analytically, and the expression is given in the Appendix of Ref.~\cite{Schneider:2015wta}.

%The normalisation parameter $\rho_{\rm nfw, 0}$ is defined so that the total 

\subsection{Two-halo profile}\label{sec:2hprofile}
For the two halo profile, we assume
\begin{equation}\label{rho2h}
\rho_{\rm 2h}(r) =f_{\rm excl}(r)\left[1+b(\nu)\xi_{\rm lin}(r)\right]\Omega_\mathrm{b}\rho_{\rm crit},
\end{equation}
where $b(\nu)$ is the halo bias, $\xi_{\rm lin}(r)$ is the linear correlation function, and $\rho_{\rm crit}$ the critical density of the Universe. The halo bias is calculated analytically using the peak-background split method \citep{Sheth:1999mn}. Compared to the original model in \cite{Schneider:2018pfw}, we introduce an empirical function
\begin{equation}
f_{\rm excl}(r) = 1-\exp\left[\alpha_{\rm excl}\, r/r_{200})\right]n ,
\end{equation}
that accounts for the halo exclusion effect. A best fit to the DMO profiles shown in Fig.~\ref{fig:DMOprofiles} is obtained with $\alpha_{\rm excl}=0.4$. 

Note that Eq.~(\ref{rho2h}) corresponds to a simplified version of the two-halo term from the halo-model formalism \citep{Hayashi:2007uk}. It does neither correctly reproduce the small scale behavior nor account for effects caused by different baryonic feedback. However, this is not a concern for the baryonification process, where the two-halo profile is only used to  assure that the displacement function (Eq.~\ref{dispalcement}) converges to zero at very large scales. Without two-halo term, the mass profile $M(r)$ becomes flat towards large scales, which means its inverse function $r(M)$ is not well-defined. We refer the reader to the Refs. \citep{Schneider:2015wta,Schneider:2018pfw} for more information. A comparison between the full simplified and full two-halo formula can be found in the Appendix of Ref.~\cite{Kovac:2025aaa}.

\subsection{Hot gas profile}\label{sec:HGAprofile}
We assume that the gas profile is made of a hot and cold gas component. The hot gas component (hga), which dominates in the regime of galaxy groups and clusters, is described by the profile
\begin{equation}\label{hgaprofile}
\rho_{\rm hga}(r) =\rho_{\rm hga,0}\left[1+\left( \frac{r}{r_\mathrm{c}}\right)^{\alpha}\right]^{-\beta(M_{200})/\alpha}\left[1+\left(\frac{r}{r_\mathrm{t}}\right)^{\gamma}\right]^{-\frac{\delta}{\gamma}},
\end{equation}
where $r_\mathrm{c}=\theta_\mathrm{c} r_{200}$ is the core radius and $r_{\rm t}=\varepsilon r_{200}$ the truncation radius. This functional form has been shown in Ref.~\cite{Schneider:2018pfw} to provide a good match to cluster profiles from X-ray observations.

Throughout the paper, the exponents of the core and truncation radii in Eq.~(\ref{hgaprofile}) are fixed to $\alpha=1$ and $\gamma=3/2$. The truncation radius is assumed to be the same as for the truncated NFW profile (see Eq.~\ref{eq:nfw}). The slope $\beta$ of the hga profile depends on the halo mass. Taking into account the effects of AGN feedback, less massive haloes generally exhibit shallower gas profiles. This effect is parametrised by the function \citep{giri2021emulation}
\begin{equation}\label{betafct}
\beta(M_{200})= \frac{3(M_{200}/M_\mathrm{c})^{\mu}}{1+(M_{200}/M_\mathrm{c})^{\mu}},
\end{equation}
that is equal to 3 at the largest mass scales, gradually decreasing to zero towards lower halo masses ($M_{200}$). The free model parameters $M_\mathrm{c}$ and $\mu$ describe the typical mass scale and the smoothness of this transition.

Note that compared to the previous model from Ref.~\cite{Schneider:2018pfw}, we have replaced the gas ejection radius with the truncation radius of the dark matter only (DMO) profile. This allows us to reduce the number of free parameters while maintaining the versatility of the profile. The maximum reach of the ejected gas can now be regulated using the $\delta$ parameter.

\subsection{Inner gas profile}\label{sec:IGAprofile}
In contrast to Ref.~\cite{Schneider:2015wta} and \cite{Schneider:2018pfw}, we also assume the presence of an inner gas component describing the cooling gas that is involved in the process of star formation. The cold, inner gas has a negligible effect on cosmology and has so far been implicitly modelled as part of the stellar component. However, it is visible in the inner part of gas profiles from hydrodynamical simulations, especially at higher redshifts. We add it here for the sake of completeness.

The inner gas component is described by the profile
\begin{equation}\label{igaprofile}
\rho_{\rm iga}(r) = \frac{\rho_{\rm iga,0}}{r^3}\exp\left(-r/r_{200}\right),
\end{equation}
which consists of a simple power law shape followed by an exponential truncation at the virial radius. We have checked that this functional form agrees well with the FLAMINGO and TNG-300 gas profiles at very low radii.

\subsection{Central galaxy profile}\label{sec:CGAprofile}
For the profile of the central galaxy, we assume
\begin{equation}\label{cgaprofile}
\rho_{\rm cga}(r) = \frac{\rho_{\rm cga,0}}{r^2}\exp(-r/R_{\rm cga}),
\end{equation}
where $R_{\rm cga}=0.03r_{200}$ is the galactic scale-radius. Compared to previous implementations, we have increased $R_{\rm cga}$ by a factor of two, and we have changed the scale dependence of the outer suppression from $\exp(-r^2)$ to $\exp(-r)$. These modifications lead to a better match with the FLAMINGO and TNG hydrodynamical simulations (see Sec.~\ref{sec:comparison}).

\subsection{Dark matter profile}\label{sec:ACMprofile}
The final dark matter profile deviates from the NFW shape due to gravitational interactions with stars and gas \cite[e.g.][]{Blumenthal:1985qy,Gnedin:2004cx, Abadi:2010aaa,Teyssier:2011aaa,Schaller:2014uwa, Velmani:2022una}. In particular, the central galaxy leads to a steepening of the inner profiles, an effect usually referred to as adiabatic contraction. A more subtle relaxation effect can be observed further out, caused by the ejection of gas as a result of feedback processes.

This back-reaction effect on the dark matter caused by the stellar and gas components has to be accounted for in the \bfc{} model. We provide a detailed discussion of different modelling choices in Appendix~\ref{app:back-reaction}. Here we present our default model, which will be used in the main analysis of the paper. It is based on an empirical function of the form
\begin{eqnarray}\label{ACM}
%\frac{r_\mathrm{i}}{r_\mathrm{f}}-1=\frac{Q_0}{1+(r_\mathrm{i}/r_{\rm step})^{n_{\rm step}}}
\xi-1=\frac{Q_0}{1+(r_\mathrm{i}/r_{\rm step})^{3/2}}
+Q_1 f_{\rm cga}\left[\frac{M_{\rm cga}(r_\mathrm{i})}{M_{\rm nfw}(r_\mathrm{i})}-1\right]+\nonumber\\
Q_1f_{\rm iga}\left[\frac{M_{\rm iga}(r_\mathrm{i})}{M_{\rm nfw}(r_\mathrm{i})}-1\right]+Q_2 f_{\rm hga}\left[\frac{M_{\rm hga}(r_\mathrm{i})}{M_{\rm nfw}(r_\mathrm{i})}-1\right],
\end{eqnarray}
describing the ratio $\xi=r_\mathrm{i}/r_\mathrm{f}$ between the initial radius $r_\mathrm{i}$ and the final radius $r_\mathrm{f}$ of a DM mass shell before and after the correction process. The first term (proportional to $Q_0$) corresponds to a smooth step function that describes the contraction of dark matter due to the inflow and cooling of gas before the onset of feedback. The parameter $r_{\rm step}$ is fixed to $r_{\rm step} = (\epsilon/\epsilon_{0})r_{200}$. The second and third terms (proportional to $Q_1$) describe the contraction due to stars and cold gas in the centre of haloes. The fourth term (proportional to $Q_2$) is responsible for the expansion caused by the ejection of gas by feedback.

Following the parameter inference analysis presented in Appendix~\ref{app:back-reaction}, we assume $Q_0=0.075$, and $Q_1=0.25$ for all simulations. The remaining parameter $Q_2$ depends on the resolution of the simulation and is fixed to $Q_2=0.5$ for the three FLAMINGO m9, $Q_2=0.7$ for the FLAMINGO m8, and $Q_2=0.8$ for the TNG-300 simulation (see Fig.~\ref{fig:AC_params} and \ref{fig:AC_params_z} in Appendix~\ref{app:back-reaction} as well as Table \ref{tab:sims} for an overview over the simulations). Note that we have also investigated potential dependencies of the back-reaction parameters on redshift and baryonic feedback.

The resulting dark matter mass and density profiles are obtained by solving
\begin{equation}
M_{\rm dm}(r)=f_{\rm dm}M_{\rm nfw}(\xi r),\hspace{0.5cm}\rho_{\rm dm}(r)=\frac{f_{\rm dm}}{4\pi r^2}\frac{\mathrm{d}}{\mathrm{d}r}M_{\rm nfw}(\xi r),
\end{equation}
where $M_{\rm nfw}$ is the truncated NFW profile introduced in Sec.~\ref{sec:DMOprofile}.

\subsection{Satellite galaxy profile}\label{sec:SGAprofile}
The satellite galaxies (sga) and halo stars are assumed to follow the dark matter distribution. Including the back-reaction effect, we obtain
\begin{equation}
\rho_{\rm sga}(r)=\frac{f_{\rm sga}}{4\pi r^2}\frac{\mathrm{d}}{\mathrm{d}r}M_{\rm nfw}(\xi r),
\end{equation}
where $f_{\rm sga}$ corresponds to the fraction of satellite galaxies and $\xi$ is obtained via Eq.~(\ref{ACM}).

\subsection{Fractions and normalisations}
The gas and central galaxy profiles are defined up to a normalisation parameter $\rho_{\chi,0}$, where $\chi={\rm \lbrace hga,\, iga,\, cga\rbrace}$. This normalisation parameters can be calculated as follows
\begin{equation}
\rho_{\chi,0} = f_{\chi}M_{\rm tot}\left[4\pi \int_0^{\infty} \mathrm{d}s y_{\chi}(s)s^2\right]^{-1},\hspace{0.5cm}\rho_{\chi}(r)\equiv\rho_{\chi,0}y_{\chi}(r),
\end{equation}
where $M_{\rm tot}$ is the total halo mass (not to be confused with the virial mass). It is obtained by integrating the truncated NFW profile to infinity, i.e.,
\begin{equation}
M_{\rm tot}=4\pi\int_0^{\infty} {\rm d}s s^2\rho_{\rm nfw}(s).
\end{equation}
Let us now define the fractions ($f_{\chi}$) of all components. The total stellar fraction and the fraction of stars in the central galaxy are parametrised as 
\begin{equation}\label{fstar}
f_{\rm star}(M_{200})=\frac{N_{\rm star}}{\left(\frac{M_{200}}{M_{\rm star}}\right)^{-\zeta}+ \left(\frac{M_{200}}{M_{\rm star}}\right)^{\eta}},
\hspace{0.5cm}
f_{\rm cga}(M_{200})=\frac{N_{\rm star}}{\left(\frac{M_{200}}{M_{\rm star}}\right)^{-\zeta} + \left(\frac{M_{200}}{M_{\rm star}}\right)^{\eta+d\eta}},
\end{equation}
with $\zeta=1.376$ and $M_{\rm star}=2.5\times 10^{11}$ M$_{\odot}$/h. The remaining parameters $N_{\rm star}$, $\eta$, and $d\eta$ are free \bfc{} model parameters. For the satellite fraction, we assume 
\begin{equation}
f_{\rm sga}(M_{200})=f_{\rm star}(M_{200})-f_{\rm cga}(M_{200}).
\end{equation}
Eq.~(\ref{fstar}) is motivated by the Moster abundance matching relation \cite{Moster:2012aaa}. In Ref.~\cite{Schneider:2018pfw} it is shown that the relation provides a good fit to abundance matching results, as well as direct estimates of both the central galaxy fraction and the total stellar to halo mass relation.

The fraction of cold, inner gas is tightly connected to the central galaxy itself. We therefore define $f_{\rm iga}$ to be proportional to $f_{\rm cga}$, i.e.,
\begin{equation}\label{figa}
f_{\rm iga}(M_{200}) = c_{\rm iga}f_{\rm cga}(M_{200}),
\end{equation}
with the proportionality factor $c_{\rm iga}$ that is a free model parameter. 

With a fixed stellar and inner gas fraction, we obtain the following hot gas fraction:
\begin{equation}
f_{\rm hga}(M_{200})=f_{\rm bar}-f_{\rm star}(M_{200})-f_{\rm iga}(M_{200}),
\end{equation}
where the cosmic baryon fraction is simply given by $f_{\rm bar}=\Omega_{b}/\Omega_{m}$. No additional free parameters are introduced here. In addition, note that all stellar and gas fractions are defined so that 
\begin{equation}
f_{\chi}=M_{\chi}(\infty)/M_{\rm tot}(\infty)
\end{equation}
where $M_{\chi}$ are the mass profiles of the individual components ($\chi$) without the two-halo-term. This means that the fractions, although functions of $M_{200}$, are equal to the mass fractions at infinity and not at the virial radius.

\subsection{Summarising the model}
The baryonification model consists of empirical profiles with functional shapes that are motivated by simulations and observations. There are a number of free parameters that can be set to different values, resulting in different strength of the baryonic effects. The model parameters are summarised in Table \ref{table:params}. 

Two basic setups are discussed throughout this paper. The first one consists of a general case, where all eight parameters are kept free. No redshift evolution of parameters are considered. Instead, we treat each redshift individually, refitting all parameters. In the second case, we only keep $M_\mathrm{c}$ and $\delta$ as free parameters. $M_\mathrm{c}$ is furthermore allowed to linearly evolve with redshift, with a free slope. This leaves us with a three-parameter model covering all scales and redshifts.

{\renewcommand{\arraystretch}{1.5}
\begin{table}[h]
    \centering
    \small
    \begin{tabularx}{\textwidth}{ccccl}
    \hline
        Name & \bfc{}& red-\bfc{} & Eq. & Description\\
        \hline
        $\theta_\mathrm{c}$ & free & fixed & (\ref{hgaprofile}) & Core size of the gas density profile ($\rho_{\rm hga}$). \\
        $M_\mathrm{c}$ & free & free & (\ref{betafct}) & Halo mass scale at which the slope of $\rho_{\rm hga}$ is exactly -3/2. \\
        $\mu$ & free & fixed &(\ref{betafct}) & Controls the rate the slope of $\rho_{\rm hga}$ varies with halo mass.\\
        $\delta$ & free & free  & (\ref{hgaprofile}) & Slope of the outer truncation of $\rho_{\rm hga}$.\\
        $c_{\rm iga}$ & free & fixed & (\ref{figa}) & Normalisation of the inner (cold) gas fraction.\\
        $\eta$ & free & fixed & (\ref{fstar}) & High-mass slope of the total stellar-to-halo fraction ($f_{\rm star}$).\\
        $d\eta$ & free & fixed & (\ref{fstar}) & High-mass slope of the central stellar-to-halo fraction ($f_{\rm cga}$).\\
        $N_{\rm star}$ & free & fixed & (\ref{fstar}) &  Normalisation parameter of $f_{\rm star}$ and $f_{\rm cga}$. \\
        \hline
        $Q_0$ & fixed & fixed & (\ref{ACM}) & Regulates pre-feedback DM contraction.\\
        $Q_1$ & fixed & fixed & (\ref{ACM}) & Regulates inner contraction caused by the central galaxy.\\
        $Q_2$ & fixed & fixed & (\ref{ACM}) & Regulates outer expansion due to feedback.\\
        \hline
    \end{tabularx}
    \caption{\label{table:params}Parameters of the baryonification model. For the full model (\bfc{}) all 8 parameters are kept free, while the reduced model (red-\bfc{}) only has two free parameters. A third one is added later on to describe the redshift evolution of the signal. The dark matter back-reaction parameters shown in the lower part of the table always remain fixed.}
    \label{tab:my_label}
\end{table}}

There are a number of potential improvements of the baryonification method that we have not addressed so far. First of all, we have not modified the velocities of simulation particles. It has been shown recently that the velocity of gas is affected by non-gravitational interactions until well beyond the virial radius \cite{Ondaro-Mallea:2024lhp}. To what extent this affects probes such as the kinematic Sunyaev-Zeldovich effect requires further investigation. Second, it is well known that the presence of gas and stars modify the shape of haloes, making them more spherical \cite[e.g.][]{Bryan:2012mw,Chua:2021oqe}. This effect becomes apparent at rather small scales below the virial radius and has so far been ignored in the baryonification model. Finally, the currently implemented galaxy model remains fairly crude: we assume simple stellar profiles without distinguishing between galaxy shapes and colors nor accounting for galaxy assembly biases. A more detailed galaxy model will be needed for realistic galaxy clustering studies.

\section{Comparison to simulations}\label{sec:comparison}
In this section, we compare the baryonification (\bfc{}) density profiles to measured profiles from simulations. We start with the gravity-only case before going through the gas, dark matter, and stellar profiles from hydrodynamical simulations. For all comparisons with hydrodynamical simulations, we jointly fit the gas and stellar density profiles, simultaneously varying all gas and stellar parameters of the model. We restrict ourselves to redshift zero. However, note that similar outcomes are obtained at higher redshifts. 
%For all comparisons to hydrodynamical simulations, we jointly fit the gas and stellar density profiles using a Markov-Chain Monte Carlo (MCMC) inference model with a Gaussian likelihood.

\subsection{Simulations}
Our comparison is based on the Illustris TNG \citep{Springel:2017tpz,Pillepich:2017fcc, Nelson:2017cxy,Naiman:2017aaa,Marinacci:2017wew} and the FLAMINGO simulation suites \citep{Schaye:2023jqv,Kugel:2023wte, Schaller:2024jiq}. For most cases, we restrict ourselves to the highest-resolution runs (TNG-300 and FLAMINGO m8) with their fiducial baryonic feedback implementations. For some cases we also compare to the mid-resolution FLAMINGO simulations (FLAMINGO m9, FLAMINGO m9 Jet, and FLAMINGO m9 fgas-8$\sigma$), which feature different feedback strength and prescriptions.

Some of the specifics of the simulations used in this paper are summarised in Table~\ref{tab:sims}. Note that the simulations have different box sizes and resolutions. While the FLAMINGO m8 simulation has a box size that is more than thirty times larger in volume compared to the one from TNG-300, its resolution is more than a factor of ten lower. The resolution of the FLAMINGO m9 simulations is eight times lower than that of FLAMINGO m8. The resolution of the simulations is important because it affects the convergence of the profiles around haloes. In Appendix~\ref{app:FLAMINGOResolution} we show that the FLAMINGO m9 simulation diverges from the FLAMINGO m8 simulation at smaller radii. The effect is small for the baryonic mass profiles, but more substantial for the dark-matter profiles. This is an important caveat, especially in the context of the dark-matter back-reaction effect.

{\renewcommand{\arraystretch}{1.5}
\begin{table}[h]
    \centering
    \small
    \begin{tabularx}{\textwidth}{lcccccr}
        \hline
        Simulation Suite & Name & Reference & $L_{\rm box}$ [Mpc/$h$] & $m_{\rm gas}$ [M$_{\odot}$/$h$] & $m_{\rm dm}$ [M$_{\odot}$/$h$] & $N_{\rm dm}$\\
        \hline
        TNG & 300-1 & \cite{Springel:2017tpz, Pillepich:2017fcc} & 205 & $7.6\times10^6$ & $4.0\times10^7$ & $2500^3$\\
        TNG & 300-1-Dark & \cite{Springel:2017tpz, Pillepich:2017fcc} & 206 & - & $4.8\times10^7$ & $2500^3$\\
        TNG & 100-2 & \cite{Springel:2017tpz, Pillepich:2017fcc} & 75 & $7.6\times10^6$ & $4.0\times10^7$ & $910^3$\\
        TNG & 100-2-Dark & \cite{Springel:2017tpz, Pillepich:2017fcc} & 75 & - & $4.8\times10^7$ & $910^3$\\
        FLAMINGO & m8 & \cite{Schaye:2023jqv,Kugel:2023wte} & 681 & $9.1\times10^7$ & $4.8\times10^8$ & $3600^3$\\
        FLAMINGO & m8\_DMO & \cite{Schaye:2023jqv,Kugel:2023wte} & 681 & - & $5.7\times10^8$ & $3600^3$\\
        FLAMINGO & m9 & \cite{Schaye:2023jqv,Kugel:2023wte} & 681 &  $7.3\times10^8$ & $3.8\times10^9$ & $1800^3$\\
        FLAMINGO & m9 Jet & \cite{Schaye:2023jqv,Kugel:2023wte} & 681 & $7.3\times10^8$ & $3.8\times10^9$& $1800^3$\\
        FLAMINGO & m9 fgas-8$\sigma$ & \cite{Schaye:2023jqv,Kugel:2023wte} & 681 & $7.3\times10^8$ & $3.8\times10^9$& $1800^3$\\
        FLAMINGO & m9\_DMO & \cite{Schaye:2023jqv,Kugel:2023wte} & 681 &  - & $4.6\times10^9$ & $1800^3$ \\
        \hline
    \end{tabularx}
    \caption{Hydrodynamical and gravity-only (DMO) simulations used in this paper. We provide the name of the simulation suite and the run, the relevant references, as well as the box size ($L_{\rm box}$),  the resolution of the gas ($m_{\rm gas}$) and the dark matter particles ($m_{\rm dm}$), and the total dark matter particle number ($N_{\rm dm}$). Note that all hydrodynamical simulations have the same amount of baryonic than dark matter particles.}
    \label{tab:sims}
\end{table}}

In the following sections, we use density and mass profiles from the TNG-300 and FLAMINGO simulations\footnote{For the TNG simulations we use halo positions from the {\tt Subfind} halo finder \citep{Springel:2000qu,Dolag:2008ar}, for FLAMINGO we rely on the halo positions from the SOAP halo catalogues based on {\tt VELOCIraptor} \citep{Elahi:2019wap}.}. These profiles are obtained by defining the halo centres based on the most-bound particle, drawing concentric circles (of equal distance in logarithmic space) around each halo centre, and measuring the mean density and mass of each shell. We randomly down-sample the halo-catalogue by a factor of ten to simplify the analysis. The largest circle was placed at 10 Mpc/$h$ from the centre. This is why the median profile drops to zero at $r/r_{200}\sim 10$ in the largest halo-mass bin of Figs.~\ref{fig:DMOprofiles}-\ref{fig:DMprofiles} (best visible in the lower right panels).

\begin{figure*} 
\centering
\includegraphics[width=1.0\textwidth,trim=2.6cm 2.36cm 3.6cm 0.5cm, clip]{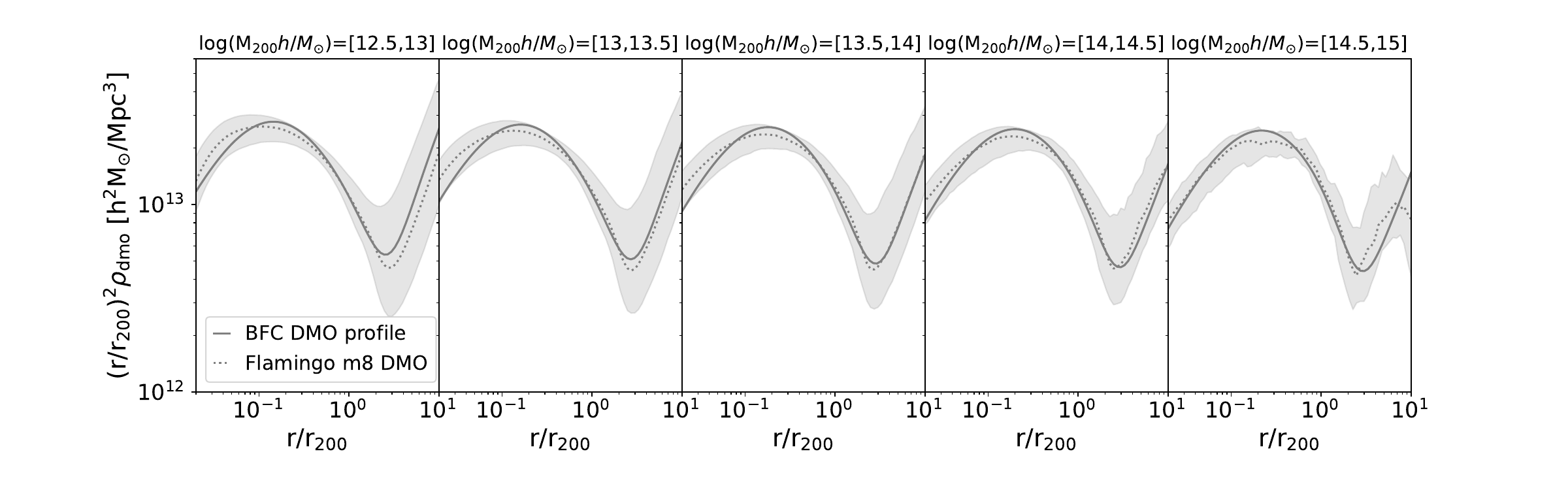}\\
\includegraphics[width=1.0\textwidth,trim=2.6cm 0.5cm 3.6cm 1.4cm, clip]{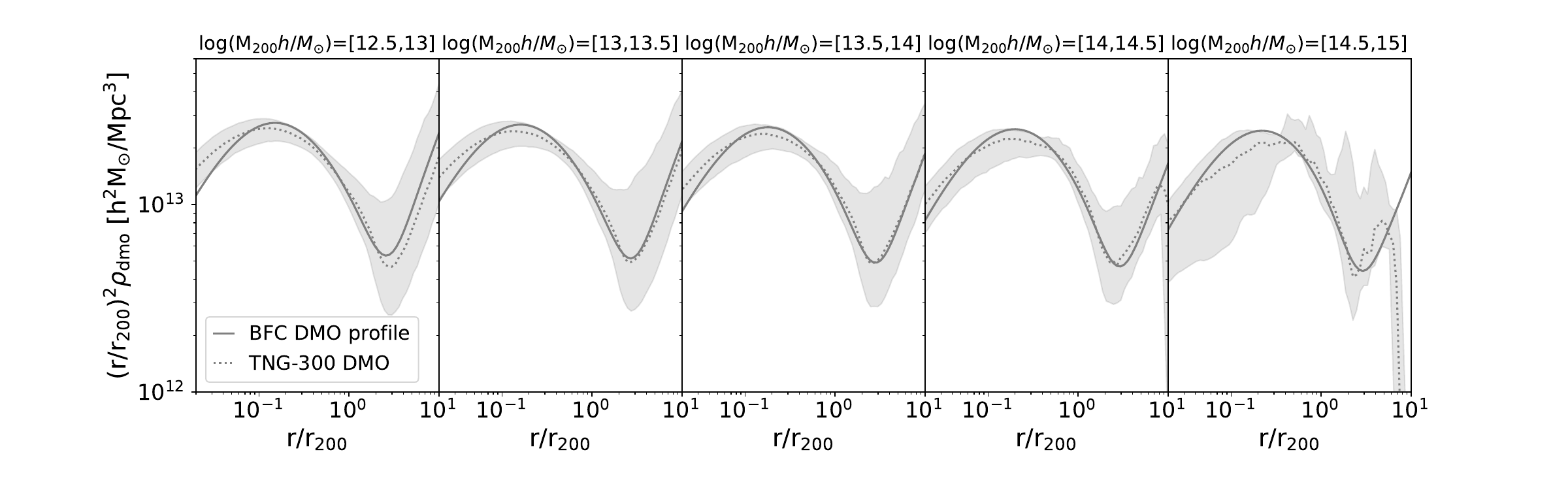}
\caption{\label{fig:DMOprofiles}Truncated NFW profile with two-halo term from the \bfc{} model (solid line) compared to the mean profiles (dotted line) and 68 \% scatter (grey band) from the DMO only counterparts of the FLAMINGO m8 (top panels) and TNG-300 (bottom panels) simulations.}
\label{fig:DMOprofiles}
\end{figure*}

\subsection{Initial dark matter profile}
The initial profile used for the dark-matter and baryonic displacement (see Eq.~\ref{DMinitialfinal} and \ref{BARinitialfinal}) consists of the truncated NFW profile ($\rho_{\rm nfw}$) plus the two-halo term ($\rho_{\rm 2h}$). The resulting total profile $\rho_{\rm dmo}=\rho_{\rm nfw}+\rho_{\rm 2h}$ has been shown to be in good agreement with halo profiles from gravity-only simulations \citep{Oguri:2011vj}.

In Fig.~\ref{fig:DMOprofiles} we show the density profiles from the gravity-only runs of the FLAMINGO m8 and the TNG-300 simulations separated into five equal mass bins in log space in the range of $3\times 10^{12}-10^{15}$ M$_{\odot}/h$. The dotted grey lines and shaded bands illustrate the median value and the 68 percentile scatter of all haloes in a given mass bin. The solid grey lines show the \bfc{} profile for the median halo mass of each bin. They provide a good (but not a perfect) match to the simulations. The maximum of the curves (corresponding to the position where $\rho_{\rm dmo}$ has a slope of -2) agrees very well for high-mass halos but is slightly displaced for lower masses. This may be a consequence of a slight mismatch in concentrations. As measurements of concentrations are not currently available for the TNG and Flamingo runs, we relied on the concentration–mass relation of Ref.~\cite{Dutton:2014xda} instead.

\begin{figure*} 
\centering
\includegraphics[width=1.0\textwidth,trim=2.6cm 2.37cm 3.6cm 0.5cm, clip]{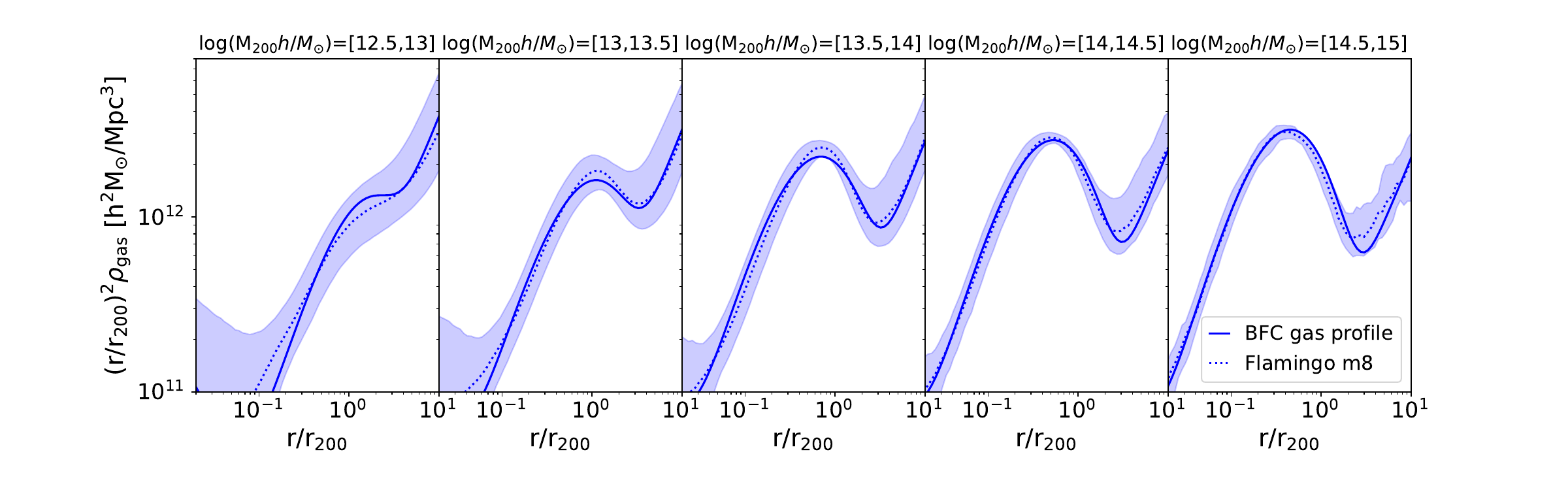}\\
\includegraphics[width=1.0\textwidth,trim=2.6cm 0.5cm 3.6cm 1.4cm, clip]{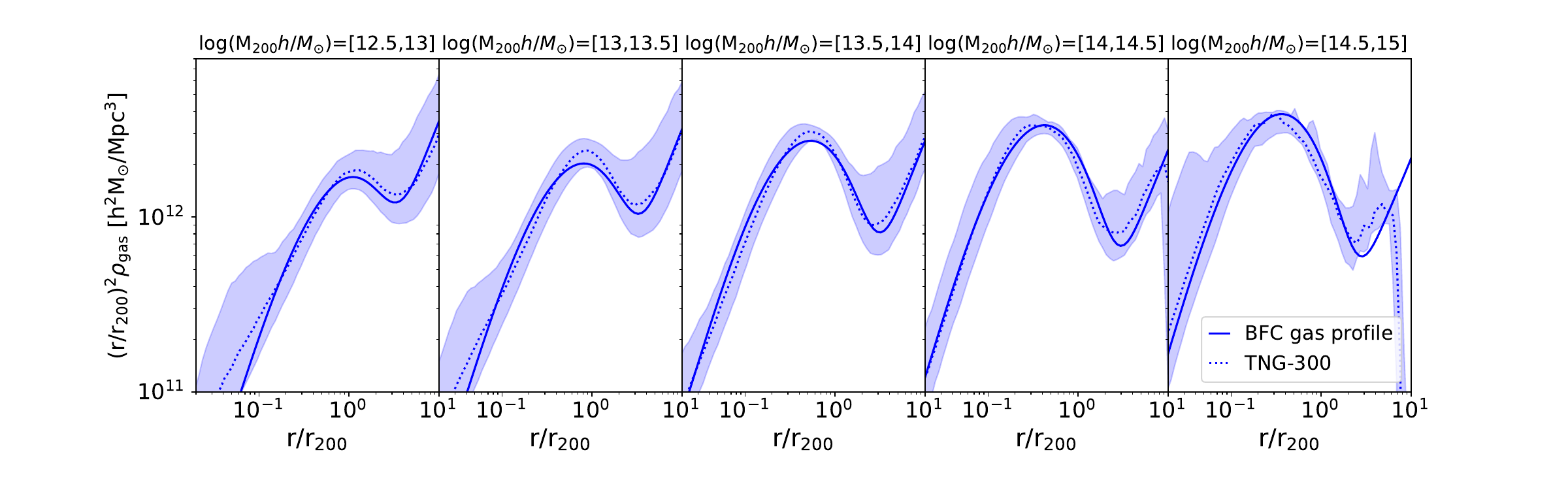}
\caption{\label{fig:GASprofiles}Gas profiles from the \bfc{} model (solid line) compared to measurements from hydrodynamical simulations (dashed line with 68 \% scatter band). The top panels show results from FLAMINGO m8, while the bottom panels are from the TNG-300 simulations. The \bfc{} model parameters are obtained via a simultaneous fit to the gas and the stellar profiles for each simulation.}
\end{figure*}

\subsection{Gas profile}
In this section, we compare the gas profile of the \bfc{} model to the profiles from FLAMINGO m8 and TNG-300. The gas component is composed of the hot and the inner (cold) gas profiles, as well as the corresponding two-halo term. The latter is multiplied by the cosmic gas fraction, i.e. the baryonic minus the stellar fraction\footnote{We use the cosmic gas fractions from the simulations. Note, however, that it can also be obtained analytically with the help of the halo model.}.

In Fig.~\ref{fig:GASprofiles} we show the gas profiles from the FLAMINGO m8 and TNG-300 simulations (top and bottom). The dotted blue lines and surrounding shaded areas correspond to the median and the 68 percent scatter of the gas profiles in a given halo mass bin. The best-fit profile from the \bfc{} model (Eq.~\ref{hgaprofile}) is shown in solid blue.

From Fig.~\ref{fig:GASprofiles} we can draw some general conclusions. The \bfc{} model provides a good general fit to the simulations at all mass and radial scales investigated. The profile is strongly dominated by the hot gas component. The subdominant inner gas is only visible as an upturn at very small radii in the lowest mass bins of the FLAMINGO m8 profiles. There are visible differences between the FLAMINGO m8 and the TNG-300 profiles, which are caused by differences in the feedback prescription. The \bfc{} model is capable of reproducing these different shapes. We have checked that this is also the case for the FLAMINGO m9, m9 Jets, and m9 fgas-8$\sigma$ simulations, which have stronger differences in their feedback prescriptions.

\begin{figure*} 
\centering
\includegraphics[width=1.0\textwidth,trim=2.6cm 2.36cm 3.6cm 0.5cm, clip]{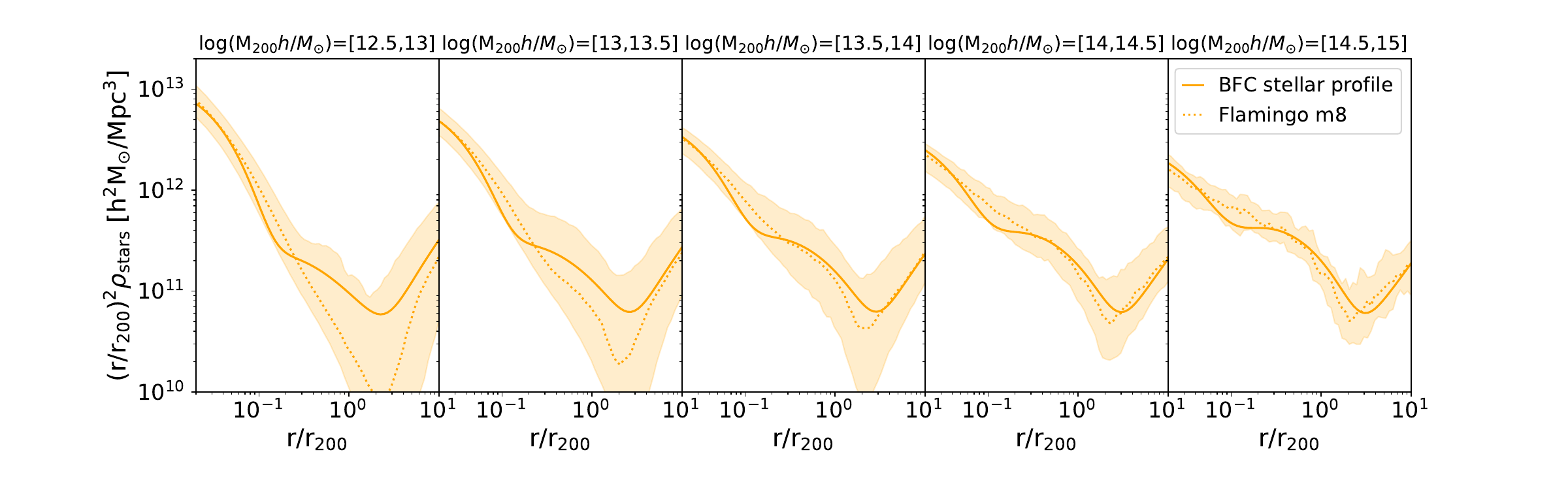}\\
\includegraphics[width=1.0\textwidth,trim=2.6cm 0.5cm 3.6cm 1.4cm, clip]{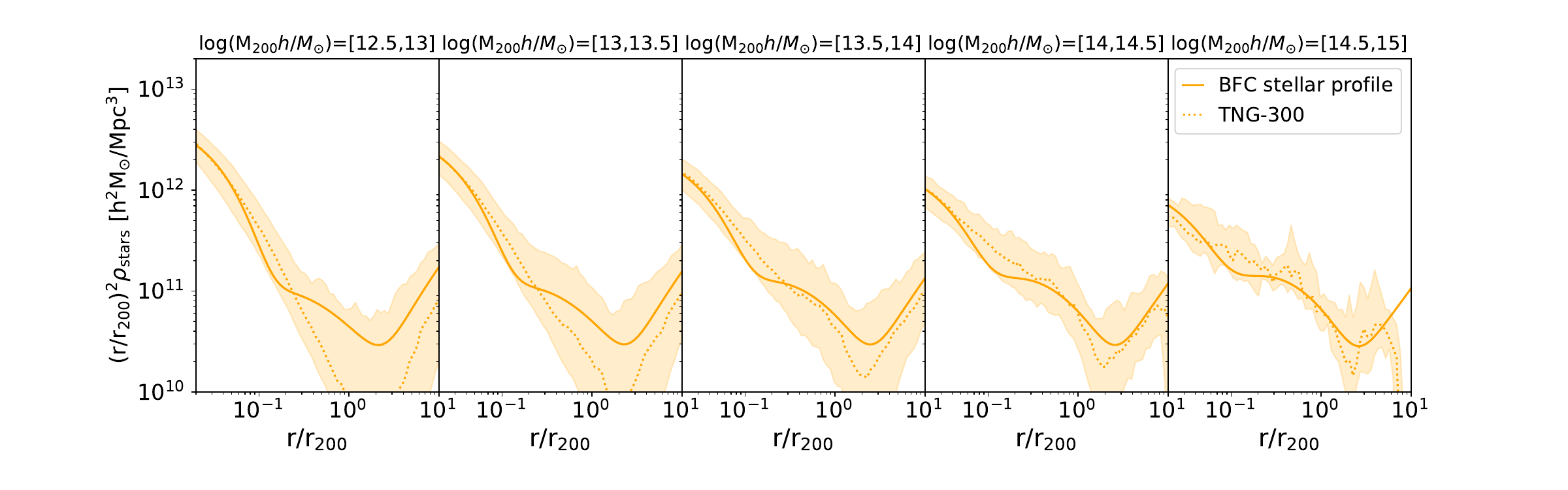}
\caption{\label{fig:STELLARprofiles}Stellar profiles from the \bfc{} model (solid line) compared to measurements from hydrodynamical simulations (dashed line with 68\% scatter band). The top panels show results from the FLAMINGO m8 run, while the bottom panels are from the Illustris TNG-300 simulation. The \bfc{} model parameters are selected via a simultaneous fit to the gas and the stellar profiles for each simulation.}
\end{figure*}

\subsection{Stellar profile}
The stellar profile ($\rho_{\rm star}$) is composed of stars from the central galaxy ($\rho_{\rm cga}$) and from satellite galaxies ($\rho_{\rm sga}$). The latter also contain the stellar halo, i.e., stars that are spherically distributed around the halo centre without belonging to any satellite galaxy.

In Fig.~\ref{fig:STELLARprofiles} we show the stellar profiles from the FLAMINGO m8 and the TNG-300 simulations (top and bottom panels). The orange dotted lines correspond to the median profile, while the shaded bands show the 68 percent scatter in the given mass bin. The solid orange lines show the \bfc{} profiles for comparison.

The inner part of the profile is fitted well by the profile of the central galaxy (Eq.~\ref{cgaprofile}) over all mass bins. Further out, at the transition where the profile becomes dominated by the contribution from satellites, the slope of the \bfc{} profile becomes shallower. In the simulations, this characteristic change in slope is only visible at cluster scales above $M_{200}\sim 10^{14}$ M$_{\odot}$/$h$. For lower halo masses, the profile decreases as a power law until the two-halo term takes over beyond the virial radius.

In summary, Fig.~\ref{fig:STELLARprofiles} shows that the \bfc{} stellar parametrisation works well for clusters, while it is more approximate at the level of individual galaxies and galaxy groups. However, for our purposes, we are only interested in two main stellar properties: the size of the central galaxy affecting the adiabatic contraction process and the total amount of stars in the halo because more stars means less gas that can be blown out by feedback. We therefore do not further improve (and complicate) the current parametrisation.

\begin{figure*} 
\centering
\includegraphics[width=1.0\textwidth,trim=2.6cm 2.38cm 3.6cm 0.5cm, clip]{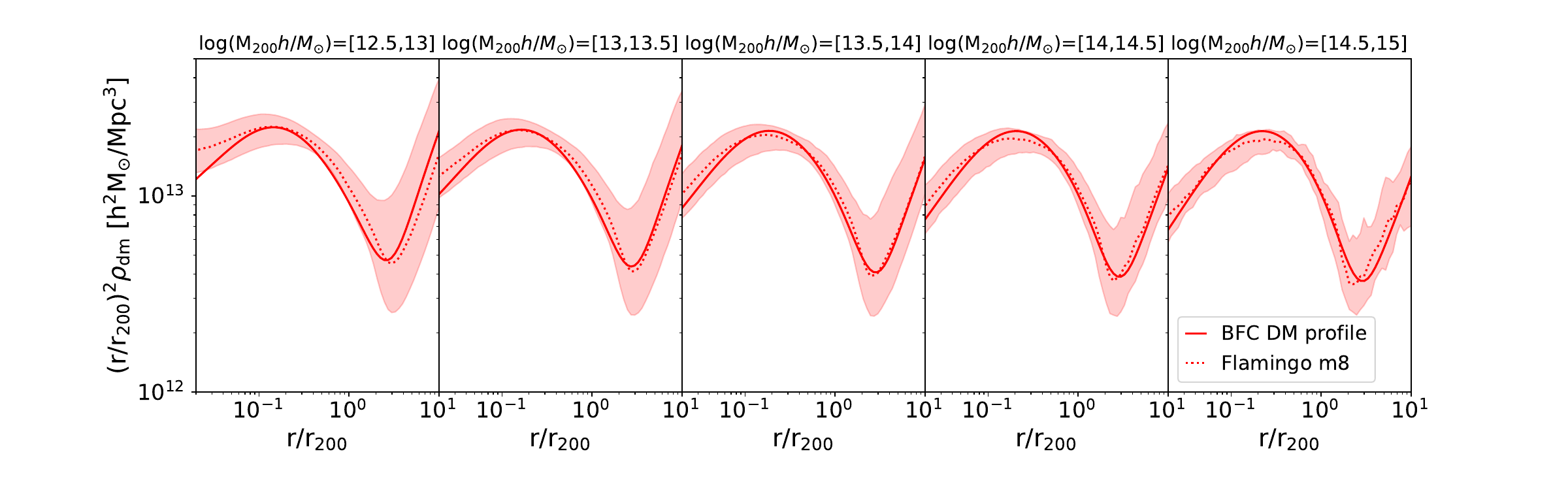}\\
\includegraphics[width=1.0\textwidth,trim=2.6cm 0.5cm 3.6cm 1.4cm, clip]{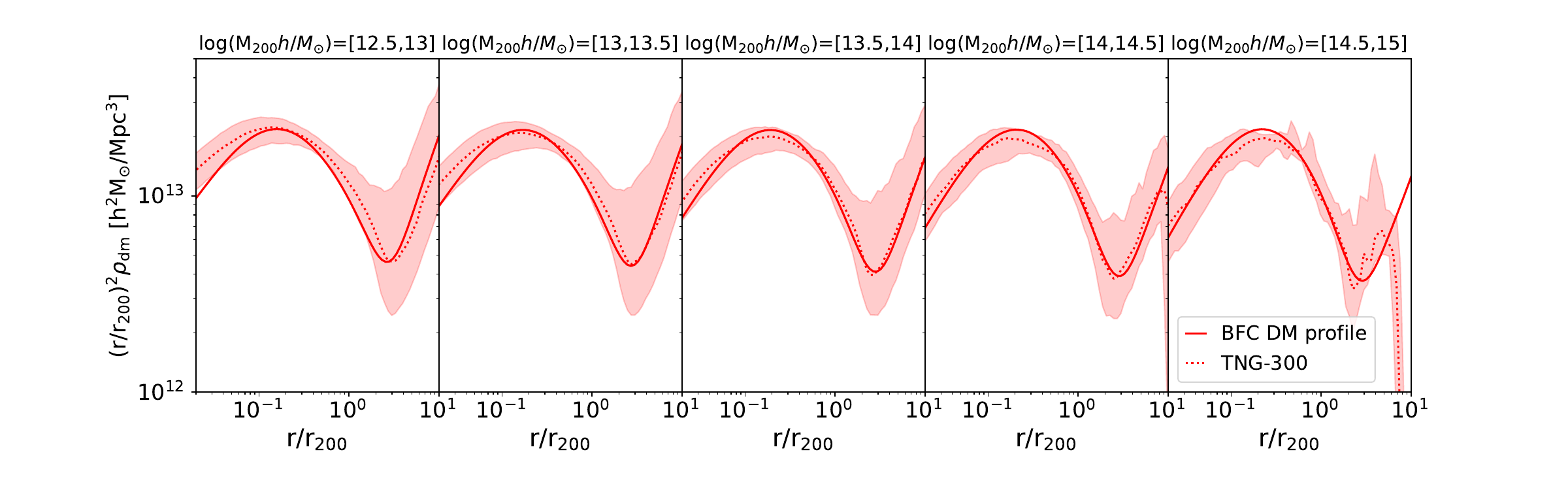}
\caption{\label{fig:DMprofiles}Dark matter (DM) profiles from the \bfc{} model (solid line) compared to measurements from hydrodynamical simulations (dashed line with 68 percent scatter band). The top panels show results from the FLAMINGO m8 run, while the bottom panels are from the Illustris TNG-300 simulation. The \bfc{} model parameters are selected via a simultaneous fit to the gas and the stellar profiles for each simulation.}
\end{figure*}

\subsection{Dark matter profile}
Given the best-fit \bfc{} parameters derived from jointly fitting the gas and stellar profiles, we can use our model to predict the resulting DM profiles ($\rho_{\rm dm}$) using the dark matter back-reaction model discussed in Sec.~\ref{sec:ACMprofile}. Together with the two-halo term ($\rho_{\rm{2h}}$) we obtain the total halo profile that is modified by the presence of stars and gas.

The best fitting parameters for the dark-matter back-reaction model are presented in Appendix~\ref{app:back-reaction}. We find parameter shifts due to the resolution of the TNG-300 and FLAMINGO simulations and, to a minor degree, due to the feedback recipe. The latter is small and therefore ignored in this paper.

The DM profiles are shown in Fig.~\ref{fig:DMprofiles}. As before, the top and bottom panels correspond to the FLAMINGO m8 and the TNG-300 simulations. The median profiles from the simulations are shown as dotted lines, with the shaded red area representing the 68 percent scatter. Not surprisingly, the two simulations exhibit very similar profiles for the DM component. A slight difference is visible at very small radii, where the FLAMINGO m8 run shows a larger amount of adiabatic contraction compared to the TNG-300 simulation.

The \bfc{} model shown as solid red line provides a good match to the hydrodynamical simulations, especially in the high halo-mass regime. The agreement is non-trivial, given that the profiles clearly differ from the dark-matter-only (DMO) case presented in Fig.~\ref{fig:DMOprofiles}. There are visible signs of contraction at very small radii (below $r\sim0.1 r_{200}$) and the profiles are slightly more relaxed further out. The result is a more gradual transition from the shallower inner to the steeper outer part of the DM profiles that is well recovered by the \bfc{} model. More details on the difference between DM profiles from hydrodynamical and gravity-only simulations can be found in Appendix~\ref{app:back-reaction}.

\section{From profiles to the power spectrum}\label{sec:profile2PS}
In the previous section we compared the density profiles of the gas, dark matter, and stars from the baryonification model with the profiles measured in hydrodynamical simulations. We now go a step further and investigate whether the \bfc{} model is able to predict the correct total matter power spectrum suppression due to feedback effects.

It is possible to directly fit \bfc{} model parameters to the power spectrum suppressions of hydrodynamical simulations, thereby obtaining high accuracies. In Ref.~\cite{giri2021emulation} it is shown that even when reducing the number of free \bfc{} model parameters from seven to three, the agreement remains better than 2 percent for $k$-values below 10 $h$/Mpc.

However, a much harder test for the \bfc{} model is to first fit all parameters to halo properties of a given hydrodynamical simulation, before predicting the power spectrum and comparing it with the one from the same simulation. This exercise was carried out in Ref.~\cite{giri2021emulation} using the gas-to-halo and stellar-to-halo mass fractions at $r_{500}$. It was found that for a sample of six different hydrodynamical simulations, the \bfc{} model was able to reproduce the suppressions of the power spectra at the level of about 5 percent if previously fitted to the mass fractions at $r_{500}$. 

\begin{figure*} 
\centering
\includegraphics[width=0.99\textwidth,trim=0.5cm 5.0cm 6.5cm 0.3cm, clip]{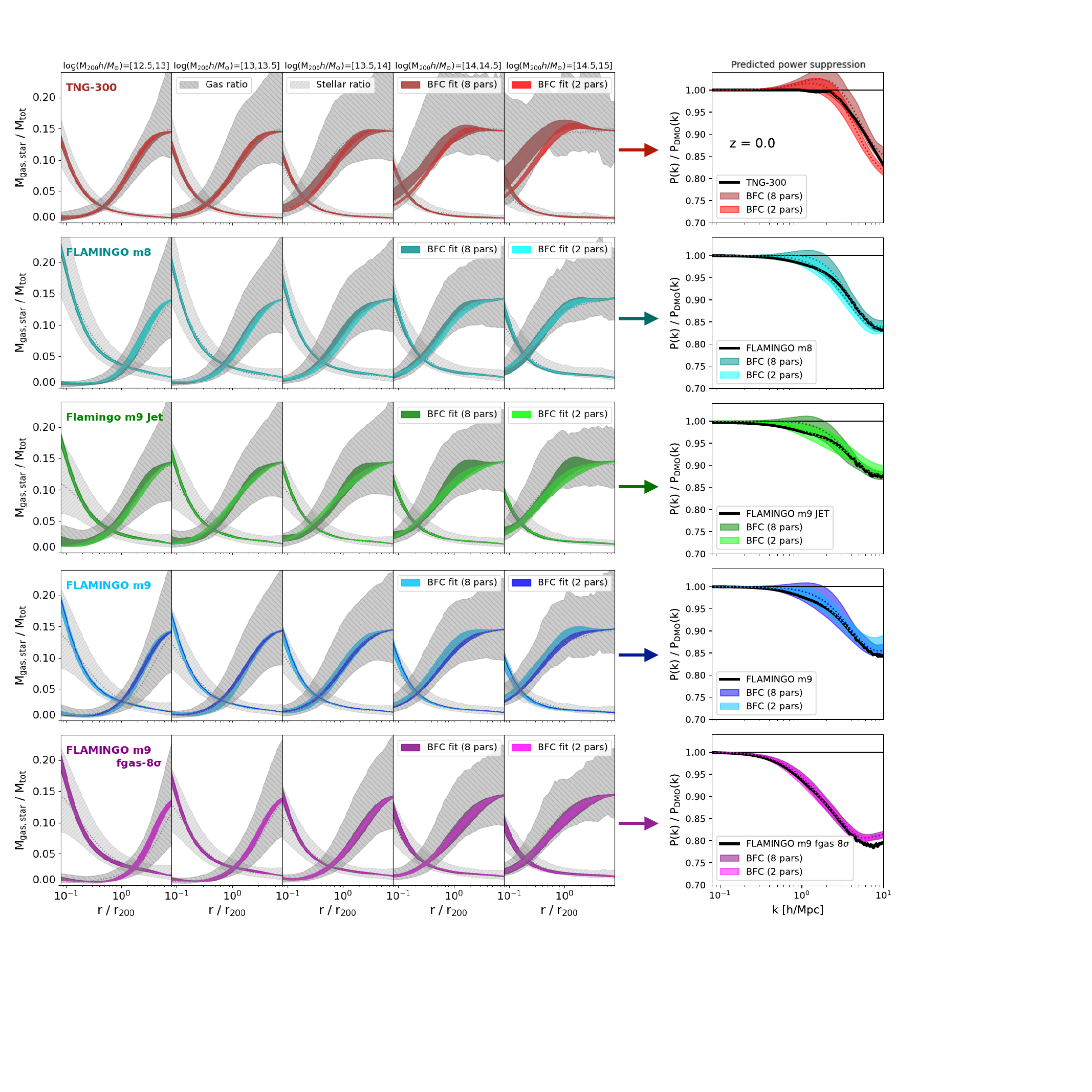}
\caption{\label{fig:ratiofit} Visualisation of the \bfc{} validation procedure at $z=0$. The \bfc{} parameters are first fitted to the simulated gas and stellar mass fractions as a function of radius (left) before the same best-fitting \bfc{} model is used to predict the matter power spectrum (right). We fit to results from the TNG-300, the FLAMINGO m8, and the three FLAMINGO m9 simulations with different feedback (top to bottom). \emph{Left:} Median and 68 percent scatter of the simulated gas and stellar mass fractions are shown as dotted lines with surrounding dark-grey and light-grey hatched areas. The dark- and bright-coloured bands correspond to the \bfc{} fits with 8 and 2 free parameters (we show posteriors at 90 percent confidence level). Each panel corresponds to a separate halo mass bin, delimited by $\log (M_{200}h/M_{\odot})=\lbrace12.5,13,13.5,14,14.5, 15\rbrace$. \emph{Right:} The predicted baryonic power suppression from the \bfc{} model (dotted lines with surrounding dark- and bright-coloured bands) are compared to the results from the hydrodynamical simulations (solid black lines).}
\end{figure*}

In this section, we go a step further and carry out the following exercise: we first fit the \bfc{} model to the median gas and stellar mass fraction profiles from a given hydrodynamical simulation; once the \bfc{} model parameters are determined via the fitting process, we baryonify an $N$-body simulation and compare its matter power spectrum to the one from the same hydrodynamical simulation. This procedure provides a test of the baryonification method in general. Furthermore, it allows us to better estimate the theoretical error attributed to studies in which the matter power spectrum suppression is predicted based on observables related to galaxy groups and clusters. For example, X-ray observations of the gas and stellar fractions \citep{giri2021emulation,  Grandis:2023qwx} or studies of the kinetic Sunyaev-Zel'dovich effect \citep{Schneider:2021wds, DES:2024iny}.

The gas and stellar mass fraction profiles are defined as follows:
\begin{equation}
R_{\rm gas}(r) = \frac{M_{\rm gas}(r)}{M_{\rm tot}(r)},\hspace{1.0cm}R_{\rm star}(r) = \frac{M_{\rm star}(r)}{M_{\rm tot}(r)},
\end{equation}
where $M_{\rm gas} = M_{\rm hga}+M_{\rm iga}$ is the gas, $M_{\rm star} = M_{\rm cga}+M_{\rm sga}$ the stellar, and $M_{\rm tot}=M_{\rm hga}+M_{\rm iga}+M_{\rm cga} + M_{\rm sga} + M_{\rm cdm}$ the total mass profile. For each simulation, we measure $R_{\rm gas}(r)$ and $R_{\rm star}(r)$ in five different halo mass bins, equally spaced in $\log M_{200}h/M_\odot$ from 12.5 to 15.

The fitting is performed twice, once with the full eight free parameters ($\log M_\mathrm{c}$, $\mu$, $\delta$, $\theta_\mathrm{c}$, $\eta$, $d\eta$, $N_{\rm star}$, $c_{\rm iga}$) and once assuming a reduced \bfc{} model with a sub-set of only two free parameters ($\log M_\mathrm{c}$,$\delta$), while the other parameters ($\mu$, $\theta_\mathrm{c}$, $c_{\rm IGA}$, $N_{\rm star}$, $\eta$, d$\eta$) remain fixed throughout the inference procedure. 
All fits (for the full and the reduced \bfc{} model) are obtained using the MCMC inference method, which not only provides the best fitting values but also the associated posterior contours. We assume 25 radial bins between 0.1 and $10\times r_{200}$ for each halo mass bin. The error on each data point corresponds to the scatter between haloes within a given mass bin. Note that these errors are likely correlated. We ignore these correlations here because we are not interested in absolute posteriors but rather in understanding how uncertainties on the mass ratio profiles translate into uncertainties on the power spectrum. However, studies including observed profile data will have to account for error correlations (as we have done in the companion paper \cite{Kovac:2025aaa}).

Regarding the reduced (2-parameter) \bfc{} model, we assume the stellar parameters ($N_{\rm star}$, $\eta$, $d\eta$) to be known. This means that for each fit, we fix the stellar parameters to their highest likelihood values from the full 8-parameter fit to the same data. The reasoning behind this (admittedly very optimistic) assumption is that, with real data from a galaxy survey, stellar parameters can be estimated via the abundance matching method: the galaxy-halo connection is obtained by rank-matching all galaxies by stellar mass to haloes of a given cosmological model.

The remaining gas parameters $\theta_\mathrm{c}$ and $c_{\rm IGA}$ are fixed to $\theta_\mathrm{c}=\theta_{\mathrm{c},0}(1+z)^{1/2}$ and $c_{\rm IGA}=c_{\rm IGA,0}(1+z)^{3/2}$ with $\theta_{\mathrm{c},0}=0.3$ and $c_{\rm IGA}=0.1$, respectively. These relations provide a reasonable estimate for the redshift evolution of $\theta_\mathrm{c}$ and $c_{\rm IGA}$ in the TNG-300 and FLAMINGO simulations. The relations are highlighted as dashed grey lines in Fig.~\ref{fig:bfc_zevol} of the following section (see the discussion there). Note that we use the same redshift evolution independently of the simulation or the feedback recipe.

\subsection{Redshift zero}
Let us start with the case of $z=0$. In the five columns on the left-hand side of Fig.~\ref{fig:ratiofit}, we show the gas and stellar mass ratio profiles at different halo-masses. The hatched grey bands show the measurement from the hydrodynamical simulations (light-grey for the stars and dark-grey for the gas), the coloured bands correspond to the \bfc{} fit with 8 (darker colour) and 2 (brighter colour) free parameters (shown at the 90 percent confidence level).

The stellar and gas mass ratio profiles of the different hydrodynamical simulations show similar trends. Deep inside the haloes, at one tenth of the virial radius, the stellar mass is much larger than the gas mass. Moving out towards the virial radius and beyond, the stellar mass fraction strongly decreases while the gas mass fraction becomes larger, approaching the cosmic gas fraction at very large scales.

Both \bfc{} models with 8 and 2 parameters provide good fits to the gas and stellar mass fraction profiles over all halo-mass bins. Only at very small scales around $r\sim 0.1 r_{200}$ some deviations are visible, at least regarding the three Flamingo m9 simulations. Note, however, that these simulations are not well converged especially at scales deep inside the virial radius (see Appendix \ref{app:FLAMINGOResolution}).

As a next step, we baryonify $N$-body simulation boxes, randomly selecting \bfc{} models out of the MCMC chains (obtained from the fits to the mass ratio profiles). Specifically, we employ \textsc{Pkdgrav3}\footnote{\url{https://bitbucket.org/dpotter/pkdgrav3}.} \cite{Potter_pkdgrav3_2017} to generate an $N$-body simulation with box size of $L = 256\,\mathrm{Mpc}/h$ and resolution of $512^3$ particles. The halo finding is performed using AHF\footnote{AHF is based on the spherical overdsnity method iteratively removing unbound particles. The halo positions are obtained based on the location of the most bound particle.}, the Amiga Halo Finder \citep[AHF,][]{Knollmann:2009}. We measure the power spectrum of each baryonified box, calculating the ratio with respect to the DMO case. This baryonic power spectrum suppression is compared to the one from the hydrodynamical simulations. The results are shown in the right-most column of Fig.~\ref{fig:ratiofit}. The dark- and bright-coloured bands correspond to the 90 percent confidence-level contours of the \bfc{} predictions, while the solid black lines show the results from the hydrodynamical simulations.

\begin{figure*} 
\centering
\includegraphics[width=0.99\textwidth,trim=0.5cm 5.0cm 6.5cm 2.0cm, clip]{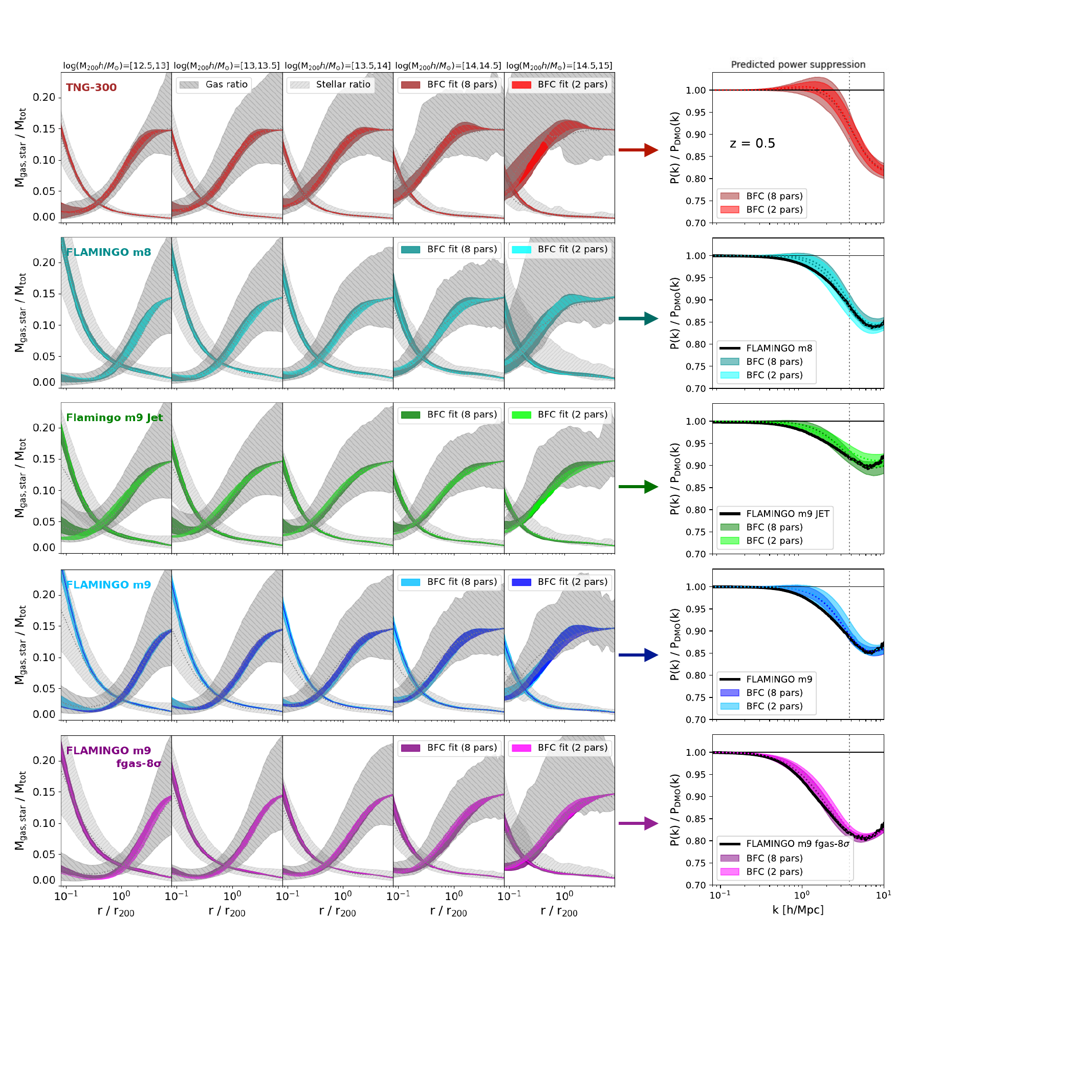}
\caption{\label{fig:ratiofit_z0.5}Visualisation of the BFC validation procedure at $z=0.5$. All denotations are the same as in Fig.~\ref{fig:ratiofit}.  The vertical dotted lines in the panels to the right indicate the maximum
$k$-value corresponding to an angular scale of $\ell = 5000$ at redshift $z=0.5$, which is the expected maximum target value of Euclid \cite{Euclid:2019clj,Schneider:2019snl}.}
\end{figure*}

We note that the power spectra from the \bfc{} model are in good agreement with the hydrodynamical simulations over all scales. The starting point and absolute depth of the suppression, as well as the small-scale upturn, are generally well recovered. Quantitatively, the agreement is at the level of 2 percent or better for all simulations and over all $k$-values between 0.1 and 10$\;h$/Mpc. This is true for both the full 8-parameter and the reduced 2-parameter \bfc{} model. As expected, the contours of the latter are somewhat narrower,  which is a direct consequence of the reduced freedom in the model parametrisation.

The conclusions above are based on the mass ratio profiles of the gas and stellar components. We note that a similar exercise could be performed with other profile data such as density profiles, absolute mass profiles etc. In Appendix~\ref{app:densityprofiles}, we investigate what happens if the BFC model is calibrated to the gas and stellar density profiles shown in Fig.~\ref{fig:GASprofiles} and \ref{fig:STELLARprofiles}. We observe a similar power suppression signal that agress at the level of about 1-2 percent for $k<5$ h/Mpc. Note that the analysis presented in Appendix~\ref{app:densityprofiles}  is limited to the TNG-300 and the Flamingo m8 simulations at $z=0$.

\begin{figure*} 
\centering
\includegraphics[width=0.99\textwidth,trim=0.5cm 5.0cm 6.5cm 2.0cm, clip]{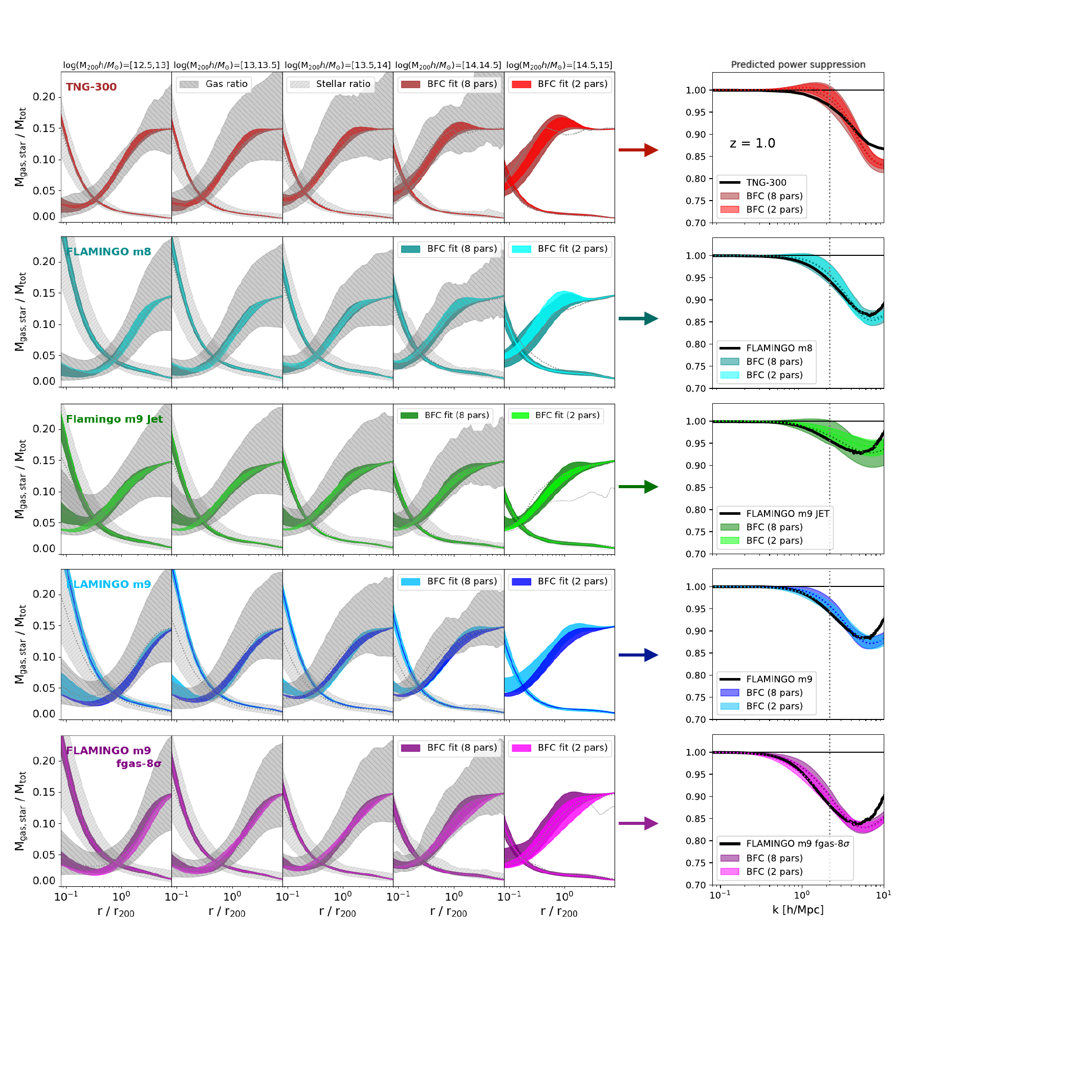}
\caption{\label{fig:ratiofit_z1}Visualisation of the BFC validation procedure at $z=1.0$. All denotations are the same as in Fig.~\ref{fig:ratiofit}.  The vertical dotted lines in the panels to the right indicate the maximum
$k$-value corresponding to an angular scale of $\ell = 5000$ at redshift $z=1.0$, which is the expected maximum target value of Euclid \cite{Euclid:2019clj,Schneider:2019snl}.}
\end{figure*}

\begin{figure*} 
\centering
\includegraphics[width=0.99\textwidth,trim=0.5cm 5.0cm 6.5cm 2.0cm, clip]{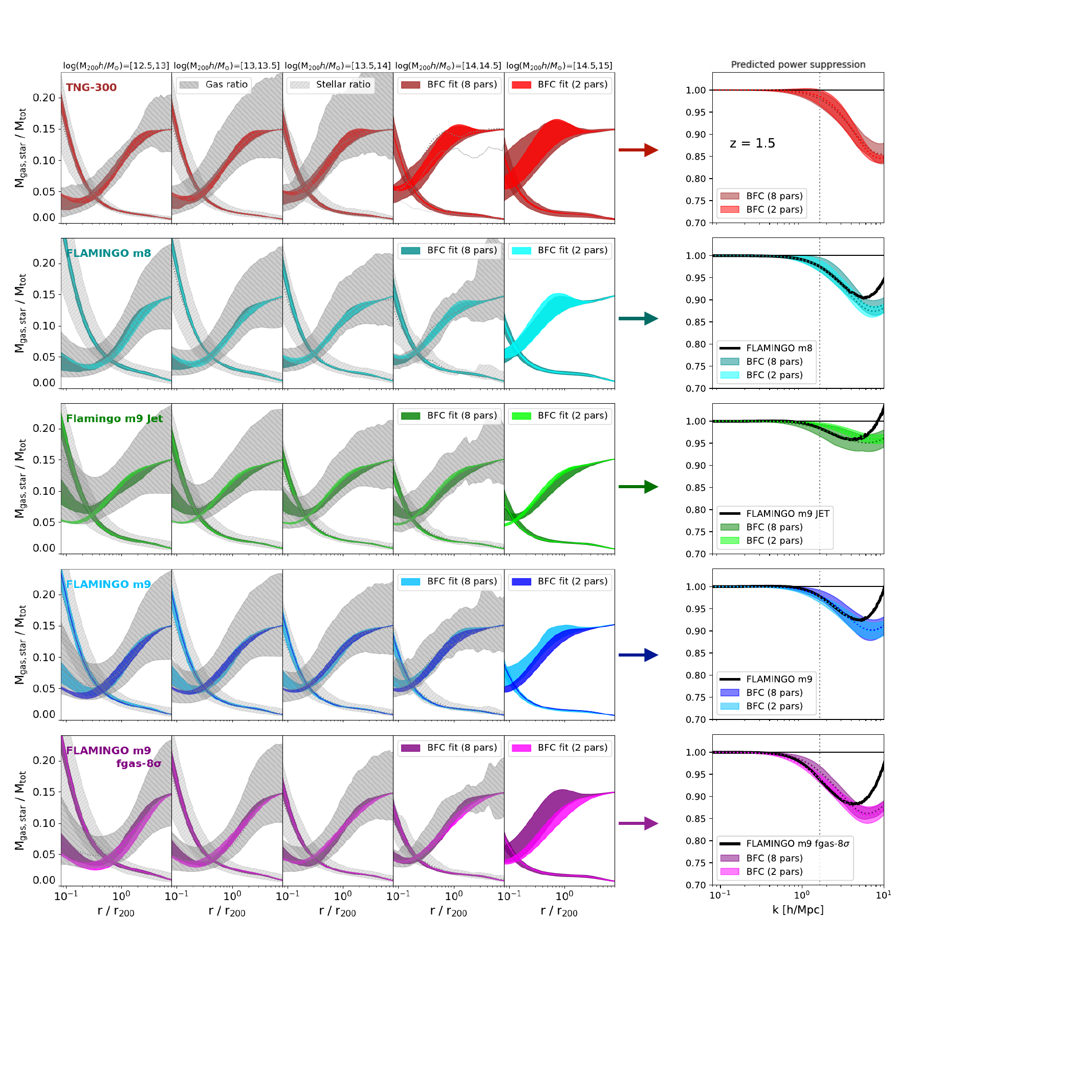}
\caption{\label{fig:ratiofit_z1p5}Visualisation of the \bfc{} validation procedure at $z=1.5$. All denotations are the same as in Fig.~\ref{fig:ratiofit}.  The vertical dotted lines in the panels to the right indicate the maximum
$k$-value corresponding to an angular scale of $\ell = 5000$ at redshift $z=1.5$, which is the expected maximum target value of Euclid \cite{Euclid:2019clj,Schneider:2019snl}.}
\end{figure*}

\subsection{Higher Redshifts}
We now perform the same analysis at $z=0.5$, $z=1$, and $z=1.5$ refitting the \bfc{} parameters at each redshift. The corresponding mass ratio profiles are illustrated in the five columns on the left-hand-side of Figs.~\ref{fig:ratiofit_z0.5},~\ref{fig:ratiofit_z1}, and~\ref{fig:ratiofit_z1p5}. The dotted grey lines with surrounding hatched areas again show the median values and scatter from the TNG-300 and the four FLAMINGO simulations, while the 8- and 2-parameter \bfc{} models are plotted as dark-coloured and light-coloured bands. Note that for the $z=1$ case, the largest halo-mass bin has not been included in the fitting procedure, as it is too sparsely populated to allow a meaningful measurement of the scatter. The same is true at $z=1.5$, except for TNG-300, where the two largest halo mass bins have been omitted.

Figs.~\ref{fig:ratiofit_z0.5}-~\ref{fig:ratiofit_z1p5} show notable differences with respect to the case of $z=0$. First of all, the stellar profile becomes more prominent when going to higher redshifts, especially at small radii. This is likely a consequence of the the explicit redshift dependence of the virial radius, defined in terms of the critical density ($\rho_{\rm crit}$). Second, the gas fraction shows an upturn towards very small radii. This upturn is barely visible at $z=0.5$ but becomes more prominent at $z=1$ and 1.5. It is caused by the inner, cold gas component that becomes more important towards higher redshifts.

%Figs.~\ref{fig:ratiofit_z0.5}-~\ref{fig:ratiofit_z1p5} show notable differences with respect to the case of $z=0$. First of all, the stellar mass becomes more important when going to higher redshifts, especially at small radii. This is a sign of a more prominent central galaxy with respect to the halo mass. Second, the gas fraction shows an upturn towards very small radii. This upturn is barely visible at $z=0.5$ but becomes more prominent at $z=1$ and 1.5. It is caused by the inner, cold gas component that becomes more important towards higher redshifts.

The power spectra at $z=0.5$, 1.0, and 1.5 are illustrated in the right-most column of Figs.~\ref{fig:ratiofit_z0.5},~\ref{fig:ratiofit_z1}, and~\ref{fig:ratiofit_z1p5}. The results from the hydrodynamical simulations correspond to the solid lines, while the predictions from the \bfc{} model are shown as dark-coloured and light-coloured bands. Note that, as before, the \bfc{} model parameters are not fitted to the power spectra, but they are obtained from the fits to the gas and stellar mass ratios profiles.

The agreement between the power spectra of the \bfc{} model and the hydrodynamical simulations is comparable to the level observed at redshift $z=0$, but gets slightly worse with increasing redshift. This is especially true at high $k$-values, where the hydrodynamical simulations show a stronger upturn compared to the \bfc{} model. This growing discrepancy is mainly caused by the modelling of the dark matter back-reaction, which is tuned to work at low redshift (see the discussion in Appendix~\ref{app:back-reaction} for more information). Another potential issue is that, at higher redshifts, halo mass bins below our plotting range become more important. In the absence of profiles, the \bfc{} model is forced to extrapolate into this regime, which is potentially biasing the result.

Note that the agreement between the \bfc{} model and the simulations is much better when we ignore the smallest scales. All modes below $k\sim 5\, h$/Mpc agree at the 2 percent level or better. This is true for all simulations and for all redshifts we investigated. In what follows, we will argue that this is enough for upcoming Stage-IV galaxy surveys.

Recent forecasts anticipate a maximum multipole moment of $\ell_{\rm max}=3000$ for LSST \cite{LSSTDarkEnergyScience:2018jkl} and $\ell_{\rm max}=5000$ for Euclid \cite{Euclid:2019clj, Schneider:2019snl} assuming an optimistic scenario. Based on these estimates, we can obtain the maximum Fourier-mode
\begin{equation}
k_{\rm max}=\ell_{\rm max}/\chi,
\end{equation}
where $\chi(z)$ is the comoving angular distance to redshift $z$. Assuming $\ell_{\rm max}=5000$, we obtain maximum $k$-modes of $k_{\rm max}=3.8$ $h/$Mpc for $z=0.5$, $k_{\rm max}=2.2$ $h/$Mpc for $z=1.0$, and $k_{\rm max}=1.6$ $h/$Mpc for $z=1.5$, respectively. These maximum values are shown as vertical dotted lines in Figs.~\ref{fig:ratiofit_z0.5}-\ref{fig:ratiofit_z1p5}. They lie well below the limit of $k=5$ $h/$Mpc, where the \bfc{} model starts to diverge from the simulation results. We conclude that the \bfc{} model is able to predict the power spectrum suppression due to feedback at the 2 percent level for all relevant scales and redshifts of upcoming (Stage-IV) cosmological surveys.

%\section{Redshift evolution of \bfc{} parameters}
\section{A \bfc{} model that accounts for redshift evolution}
In this section, we examine the explicit redshift dependence of the \bfc{} model parameters. We use this information to construct a 2+1 parameter \bfc{} model, where 2 parameters account for the strength of feedback and 1 parameter describes the explicit redshift evolution. Instead of simply assessing the model's capability to reproduce the various baryonic power suppression signals, we again follow the more ambitious approach of first fitting the \bfc{} model parameters to the gas and stellar ratio profiles, before comparing the resulting power suppression signal with the one from hydrodynamical simulations.

\begin{figure*} 
\centering
\includegraphics[width=0.99\textwidth,trim=0.1cm 0.0cm 0.0cm 0.0cm, clip]{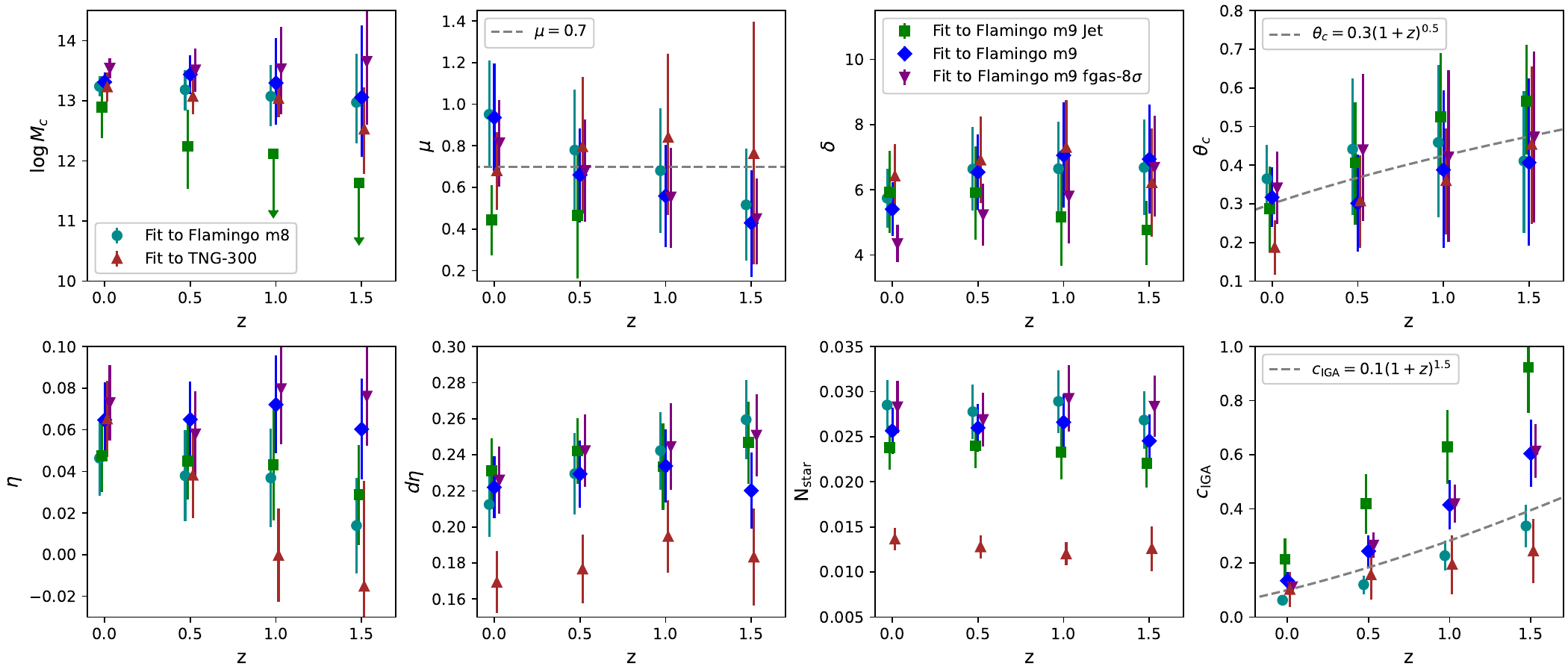}
\caption{\label{fig:bfc_zevol} Redshift evolution of the 8 \bfc{} model parameters. The model has been fitted to the gas and stellar ratios of the TNG and FLAMINGO simulations shown in Figs.~\ref{fig:ratiofit}-\ref{fig:ratiofit_z1p5}. The data points with arrows are upper limits, corresponding to cases where the model is prior limited. The dashed grey lines show the values to which $\mu$, $\theta_\mathrm{c}$, and $c_{\rm IGA}$ have been fixed in the reduced 2-parameter \bfc{} model. Note that the data points have been slightly displaced around the values $z=\left[0,0.5,1.0,1.5\right]$ to improve the readability of the plots.}
\end{figure*}

We start by investigating the redshift evolution of the full 8-parameter \bfc{} model. In Fig.~\ref{fig:bfc_zevol} we plot the median and the 68 \% posterior scatter at different redshifts. The posteriors are obtained from the chains presented in Sec.~\ref{sec:profile2PS}. Each panel corresponds to one \bfc{} model parameter, the coloured symbols showing the fits to different simulations. While there is significant scatter between the simulations, some general trends are visible: the gas parameters $\log M_\mathrm{c}$ and $\mu$ tend to decrease towards higher redshifts, while $\delta$ seems to slightly increase for most simulations. These trends indicate a general reduction of baryonic feedback effects towards higher redshifts, which is in agreement with the decreasing power spectrum suppression observed in hydrodynamical simulations \cite[e.g.][]{Springel:2017tpz,vanDaalen:2019pst,Schaller:2024jiq}. Regarding the stellar parameters, we observe that $\eta$ tends to decrease with redshift, while $d\eta$ increases and $N_{\rm star}$ stays roughly constant. This means that, while the central galaxy becomes less dominant, the total stellar component becomes more important towards high redshifts.

The grey dashed lines in Fig.~\ref{fig:bfc_zevol} denote the values to which the parameters $\mu$, $\theta_\mathrm{c}$, and $c_{\rm IGA}$ are fixed in the reduced 2-parameter \bfc{} model. The stellar parameters $\eta$, $d\eta$, and $N_{\rm star}$ are fixed individually for each simulation and redshift. The idea behind this approach is that these values can be obtained via abundance matching the observed galaxies to the haloes from the assumed cosmology.

\begin{figure*} 
\centering
\includegraphics[width=0.99\textwidth,trim=0.2cm 0.2cm 0.35cm 0.0cm, clip]{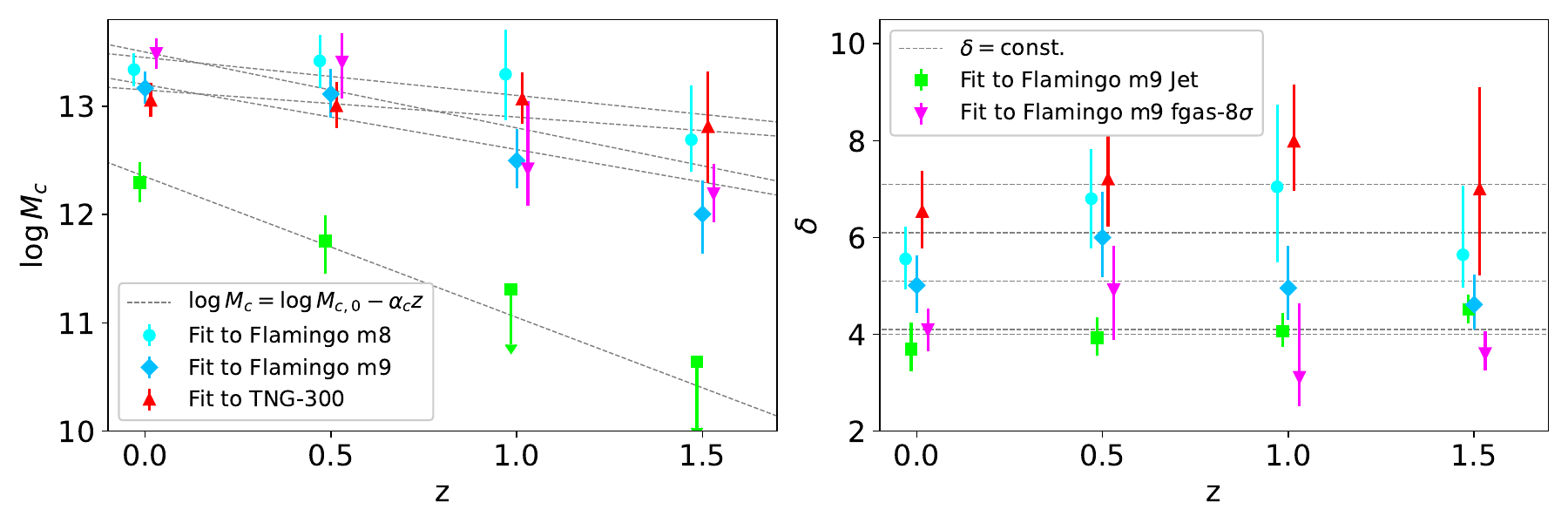}\\
\includegraphics[height=4.94cm,trim=0.0cm 0.0cm 1.2cm 0.4cm, clip]{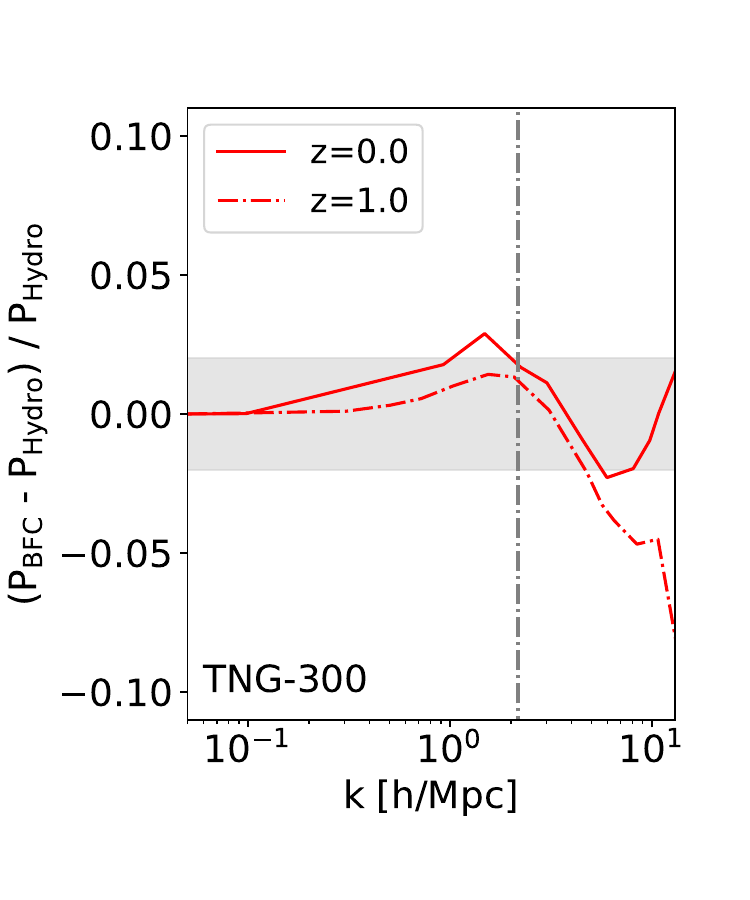}
\includegraphics[height=4.94cm,trim=3.15cm 0.0cm 1.2cm 0.4cm, clip]{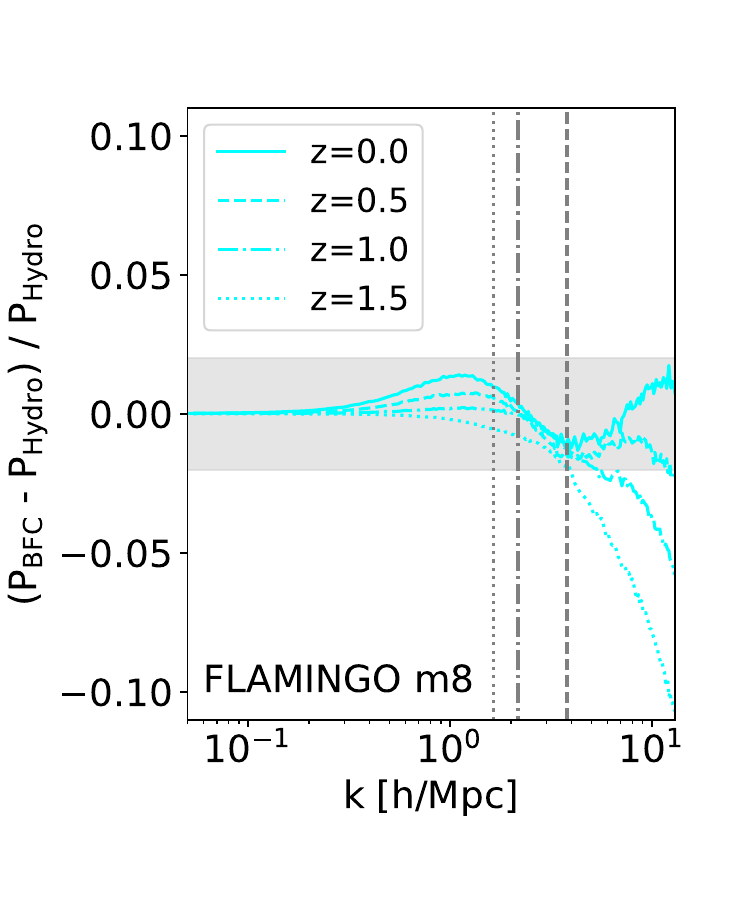}
\includegraphics[height=4.94cm,trim=3.15cm 0.0cm 1.2cm 0.4cm, clip]{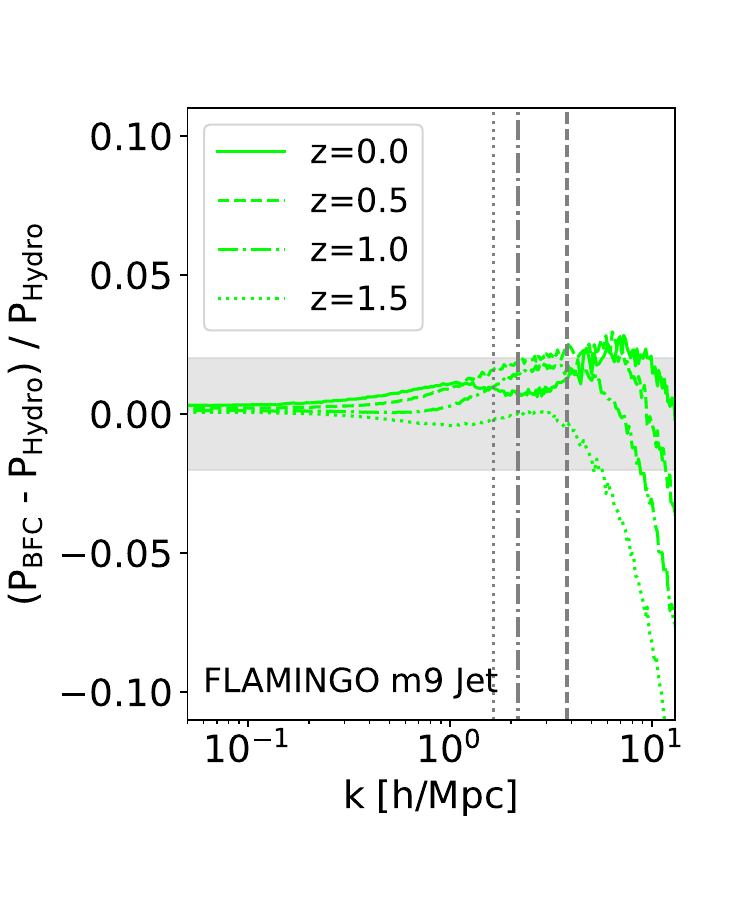}
\includegraphics[height=4.94cm,trim=3.15cm 0.0cm 1.2cm 0.4cm, clip]{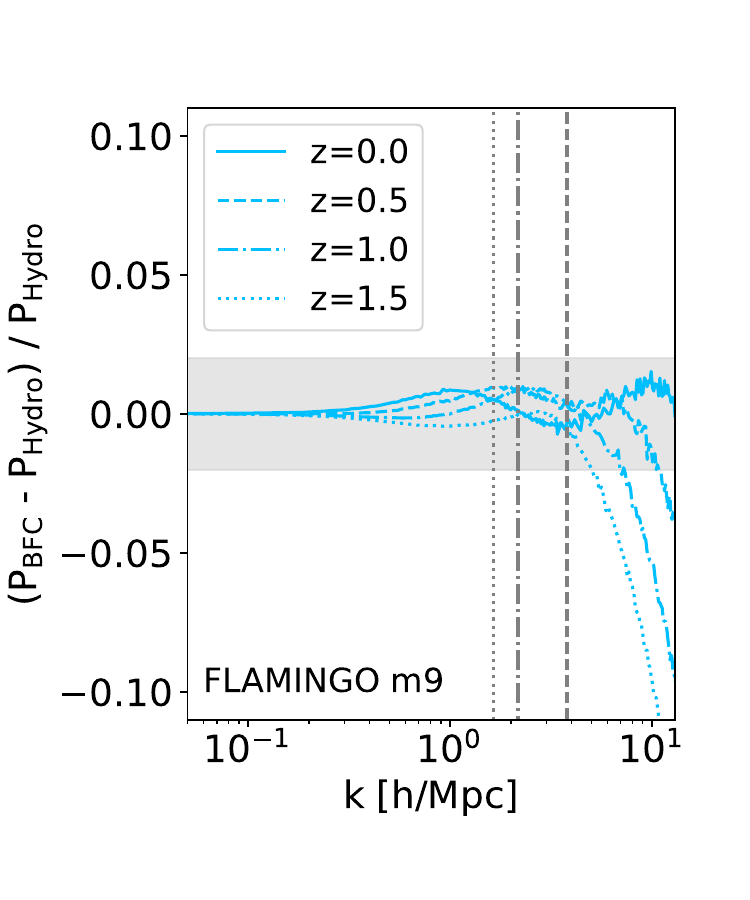}
\includegraphics[height=4.94cm,trim=3.15cm 0.0cm 1.0cm 0.4cm, clip]{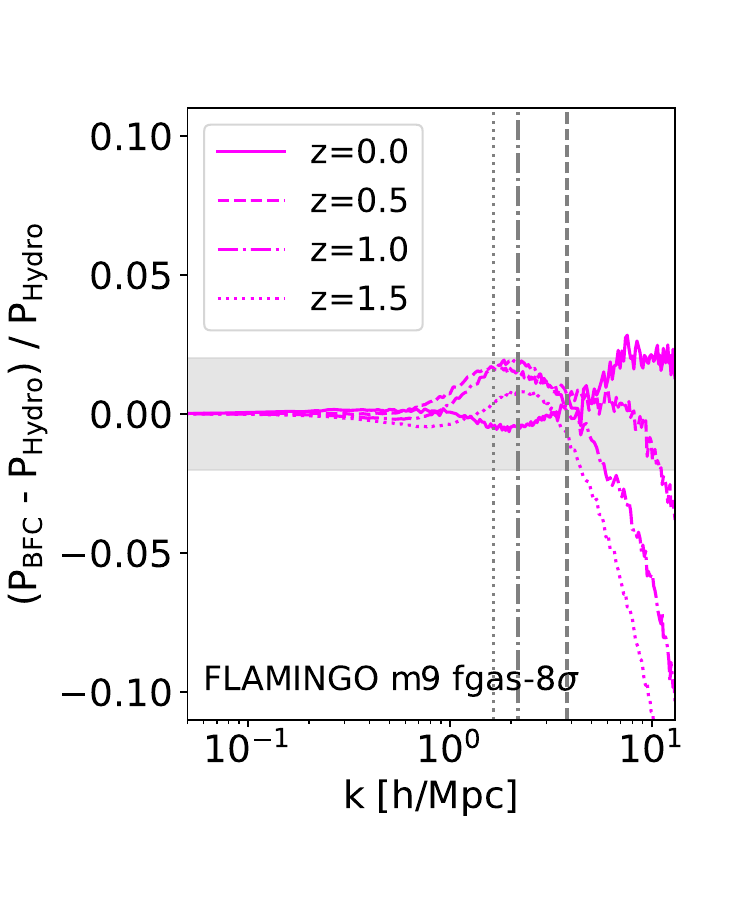}
\caption{\label{fig:bfc_reduced_params_zevol} \emph{Top:} Redshift evolution of the 2-parameter \bfc{} model constrained on the mass ratio profiles as shown in Figs.~\ref{fig:ratiofit}-\ref{fig:ratiofit_z1p5} (coloured symbols with error bars corresponding to the 68 percent posteriors). The dashed grey lines show the linear and constant lines for $\log M_\mathrm{c}$ and $\delta$ that are fitted to the data points. The slope $\alpha_\mathrm{c}$ of the linear $\log M_\mathrm{c}$-relation corresponds to the third free model parameter. \emph{Bottom:} Error of the power spectrum obtained from the 3-parameter  ($\log M_\mathrm{c}$, $\alpha_\mathrm{c}$, $\delta$) \bfc{} model compared to the hydrodynamical simulations at different redshifts. The vertical lines show the $k$-values corresponding to $l=5000$, the smallest scales potentially accessible by Stage-IV lensing surveys \cite{Euclid:2019clj}.}
\end{figure*}

As a next step, we investigate the redshift evolution of the reduced 2-parameter \bfc{} model. The upper panels of Fig.~\ref{fig:bfc_reduced_params_zevol} show the median and the 68 percent confidence limits derived from the posterior for the two free parameters ($\log M_\mathrm{c}$, $\theta_\mathrm{c}$). The constraints for the FLAMINGO m9 Jet simulation are strongly prior dominated at $z=0.5$ and higher. The highest likelihood points are found for arbitrarily small values of $\log M_\mathrm{c}$ with completely unconstrained $\mu$. We account for this by replacing the data points of $\log M_\mathrm{c}$ with upper limits.

We observe clear trends in the redshift evolution of the two \bfc{} parameters. The gas parameter $\log M_\mathrm{c}$ decreases with increasing redshift, a trend that is visible in all five simulations. The decrease can be approximated by a linear function
\begin{equation}\label{Mc_z}
\log\left[M_\mathrm{c}/M_{\mathrm{c},0}\right]=-\alpha_\mathrm{c} z,
\end{equation}
where the value of $\alpha_\mathrm{c}$ depends on the simulation. Different fits to the data points are plotted as dashed grey lines. 
%They correspond to ($\log M_{\mathrm{c},0}$, $\alpha_\mathrm{c}$) =  (13.15, 0.25), (13.45, 0.35), (12.35, 1.3), (13.20, 0.6), (13.50, 0.7) for TNG-300, FLAMINGO m8, m9 Jet, m9, and m9 fgas-8$\sigma$, respectively.
The second gas parameter $\delta$, on the other hand, does not show any clear redshift dependence. We approximate it by a constant
\begin{equation}
\delta = \delta_{0},
\end{equation}
as indicated by the horizontal grey lines in Fig.~\ref{fig:bfc_reduced_params_zevol}.

Including the simple redshift dependence of the $M_\mathrm{c}$ parameter, we are left with a \bfc{} model with 2+1 free parameters which describes different baryonic feedback scenarios and is applicable to different redshifts. The model is well suited for cosmological analyses, where the number of nuisance parameters has to be kept as low as possible.

We compare the reduced 3-parameter \bfc{} model against the FLAMINGO and TNG hydrodynamical simulations using the parameter values provided by the fits shown as grey dashed lines. The differences between the \bfc{} model and the hydrodynamical simulations are shown in the bottom panels of Fig.~\ref{fig:bfc_reduced_params_zevol}. The deviations remain below 2 percent for most scales that are relevant for cosmology. This is illustrated by the grey horizontal lines that indicate the maximum $k$-modes which contribute to the angular power spectrum, assuming a cut beyond $\ell=5000$. This is an optimistic cut even for Stage-IV lensing surveys such as Euclid (and has been used as the optimistic scenario in the official Euclid forecast paper, see \cite{Euclid:2019clj}).

In summary, the proposed 2+1 parameter \bfc{} model provides a self-consistent prescription for baryonic feedback effects over different redshifts. Fitting the three model parameters to simulated gas and stellar ratio profiles between $z=0$ and 1.5, the model is able to predict the power spectrum suppression to better than 2.5 percent across the same redshift range.

\section{Pressure and temperature profiles}\label{sec:PressTemp}
One of the main advantages of the new baryonification method introduced in this work is its ability to jointly model the individual gas, star and dark matter fields. With this at hand, it becomes possible to combine cosmological probes from different observables. The weak lensing signal, for example, traces the total matter field, while the X-ray signal depends on the density, temperature, and chemical composition of the gas. Another promising observable is the Sunyaev-Zel'dovich (SZ) effect. The thermal SZ effect depends on the thermal electron pressure of the gas, while the kinetic SZ effect depends on the gas density and velocity fields. 

In order to model SZ and X-ray observations, we require a pressure and temperature field. From the density and mass profiles, it is possible to obtain approximate pressure and temperature profiles for galaxy groups and clusters \citep[see e.g. Ref.][]{Arico:2024pvt}. We do this by assuming hydrostatic equilibrium plus a non-thermal pressure correction. The hydrostatic equilibrium equation is given by
\begin{equation}
\frac{\mathrm{d}P_{\rm tot}}{\mathrm{d}r}=-\rho_{\rm hga}(r)\frac{GM_{\rm dmb}(r)}{r^2},
\end{equation}
where $P_{\rm tot}$ denotes the total pressure of a halo, and $\rho_{\rm hga}$ and $M_{\rm dmb}$ are the hot gas density profile and the total mass profile, respectively. The total pressure profile consists of a thermal and a non-thermal component. It is obtained by integrating
\begin{equation}\label{totpressure}
P_{\rm tot}(r)=G\int_r^{\infty}\mathrm{d}s\,\rho_{\rm hga}(s)\frac{M_{\rm dmb}(s)}{s^2}.
\end{equation}
In order to obtain the thermal pressure, we need to subtract the non-thermal contribution from Eq.~(\ref{totpressure}). Using the empirical non-thermal pressure term from Ref.~\cite{Shaw2010} (which, to some extent, may also capture deviations from hydrostatic equilibrium), we can write
\begin{equation}\label{pressureprofile}
P_{\rm th}(r)=\left[1-\alpha_{\rm nth}(z)(r/r_{\rm 200})^{n_{\rm nth}}\right]P_{\rm tot},
\end{equation}
where
\begin{equation}
\alpha_{\rm nth}(z) = \alpha_{\rm nth,0}f(z),\hspace{1cm}f(z)={\rm min}\left[(1+z)^{\beta_{\rm nth}},\left(f_{\rm max}-1\right)\tanh(\beta_{\rm nth}z)+1\right],
\end{equation}
and $f_{\rm max}=4^{-n_{\rm nth}}/\alpha_{\rm nth,0}$. Following Ref.~\cite{Shaw2010}, the parameters are fixed to $\alpha_{\rm nth}=0.18$, $\beta_{\rm nth}=0.5$, and $n_{\rm nth}=0.8$. Note that we set the thermal pressure to zero whenever Eq.~(\ref{pressureprofile}) becomes negative. This generally happens in the outskirts of the halo, well beyond the virial radius.

\begin{figure*} 
\centering
\includegraphics[width=1.0\textwidth,trim=2.2cm 1.9cm 3.6cm 0.3cm, clip]{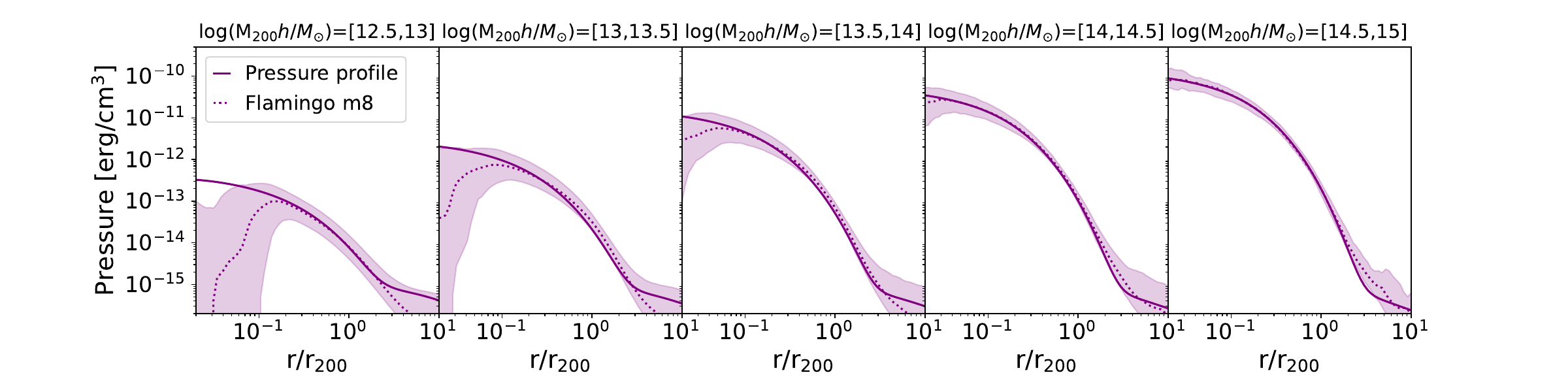}\\
\includegraphics[width=1.0\textwidth,trim=2.2cm 0.5cm 3.6cm 1.2cm, clip]{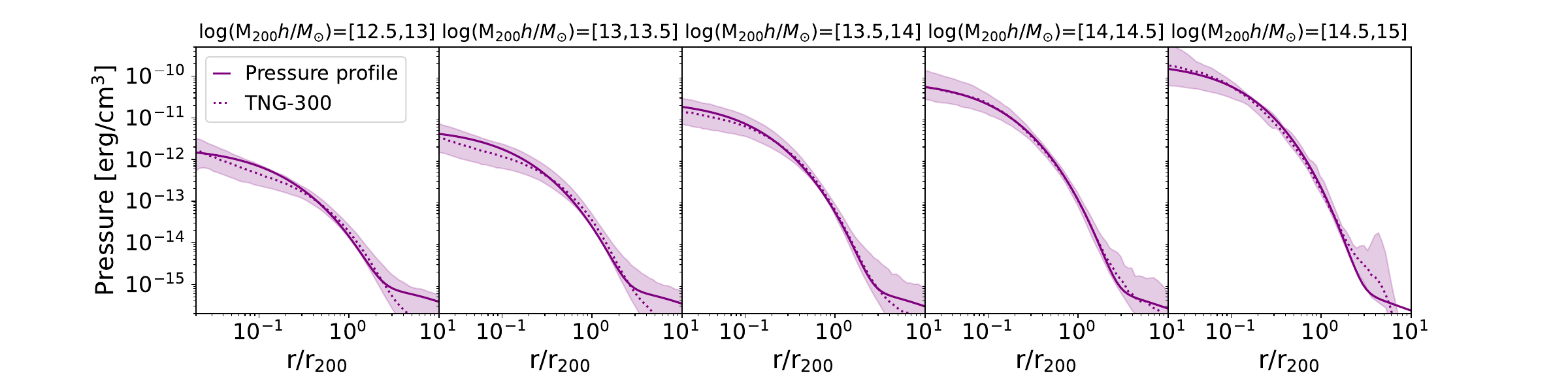}
\caption{\label{fig:PRESSUREprofiles}Pressure profiles from the baryonification model (solid lines) compared to results from the FLAMINGO m8 (top) and TNG-300 simulations (bottom). The dotted lines and shaded areas correspond to the median and 68\% scatter from  the simulations. Note that this is not a fit, the \bfc{} model parameters are taken from the fit to the gas and stellar profiles from Sec.~\ref{sec:comparison}.}
\end{figure*}

The temperature profile is finally obtained via the ideal gas equation
\begin{equation}
T(r) = \frac{m_P\mu_m P_{\rm th}(r)}{k_B\rho_{\rm hga}(r)},
\end{equation}
where $\mu_m=0.6125$ is the mean molecular weight of the ionised gas and $m_P$ is the proton mass. The profiles $P_{\rm th}(r)$ and $\rho_{\rm hga}(r)$ are obtain via Eqs.~(\ref{pressureprofile}) and (\ref{hgaprofile}).

In Fig.~\ref{fig:PRESSUREprofiles}, we show the pressure profiles from the FLAMINGO m8 and the TNG-300 simulations\footnote{The simulated pressure profiles are obtained from the total gas density and the mass-weighted temperature values using the ideal gas equation.}. The dashed line corresponds to the median value, and the coloured band to the 68 \% scatter in a given halo mass bin. The solid line consists of Eq.~(\ref{pressureprofile}) plus a two halo contribution corresponding to Eq.~(\ref{rho2h}) with the term $\Omega_{\rm b} \rho_{\rm crit}$ being replaced by the mean pressure of the simulation. Note that the \bfc{} model parameters have not been fitted to the pressure profiles here. Instead, we use the best-fitting parameters obtained from the gas and stellar density profiles of Sec.~\ref{sec:model}.

The baryonification model is capable of reproducing the pressure profiles from the simulations. This is especially true for massive galaxy groups and clusters that are fully virialised, while there are some small but visible deviations at the lower mass bins\footnote{The downturn at very small radii visible in the FLAMINGO m8 simulations is a resolution effect.}. The fact that the pressure profiles are matched with model parameters that were previously calibrated on gas and stellar density profiles further highlights the self-consistency of the \bfc{} method, not only across mass- and redshift, but also acoss different cosmological probes.

%\begin{figure*} 
%\centering
%\includegraphics[width=0.9\textwidth,trim=2.5cm 1.5cm 3.5cm 1.8cm, clip]{Figs/TEMPprofiles_FLAMINGOHR.png}\\
%\includegraphics[width=0.9\textwidth,trim=2.5cm 0.0cm 3.5cm 1.8cm, clip]{Figs/TEMPprofiles_TNG.png}
%\caption{\label{fig:PRESSUREprofiles}Temperature profiles from the baryonification model compared to results from FLAMINGO HR (top) and TNG-300 (bottom). Note that the bfc parameters used here correspond to the model fitted to the density profiles.}
%\end{figure*}

\section{Comparison at the field level}\label{sec:FieldLevel}
In contrast to other analytical and semi-numerical approaches, the baryonification method is performed at the level of simulation particles. This means that we naturally obtain the full three-dimensional fields of e.g. densities, pressure, or temperature. In principle, these fields can be used to carry out any simulation-based inference analysis, to measure statistics such as power spectra, bi-spectra and other higher-order statistics, or to directly pursue field-level inference methods.

Currently, it is unclear how well the \bfc{} model performs at the field-level. An earlier \bfc{} version has been applied within a cosmological analysis using a map-based deep learning approach \cite{Fluri:2022rvb, Kacprzak:2022pww}. In this analysis, it was shown that the baryonification model leads to accurate results without biasing cosmological parameter estimates. However, the method relied on a filtering procedure of the map, smoothing out the smallest scales that are the hardest to model correctly. More recently, Ref.~\cite{Zhou:2025tdd} has shown promising agreement between a number of weak lensing higher-order statistics and hydrodynamical simulations from the FLAMINGO suite.

\begin{figure*} 
\centering
\includegraphics[width=0.99\textwidth,trim=12.0cm 0.8cm 1.7cm 0.8cm, clip]{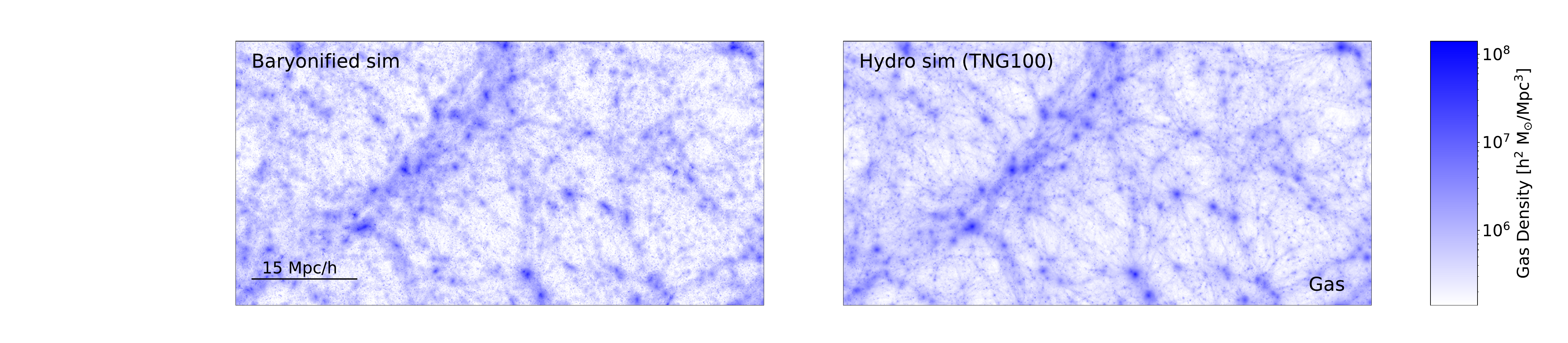}\\
\includegraphics[width=0.99\textwidth,trim=12.0cm 0.8cm 1.7cm 0.8cm, clip]{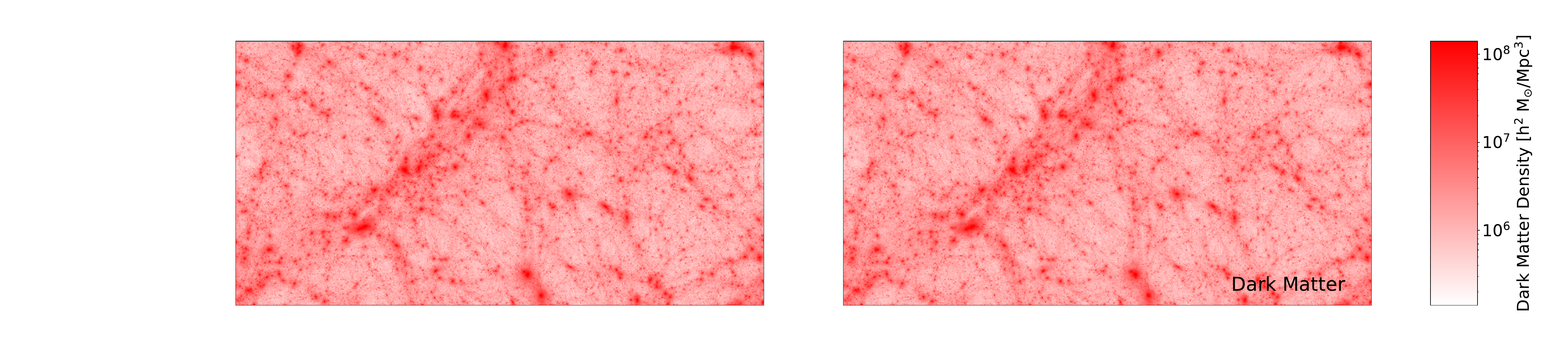}\\
\includegraphics[width=0.99\textwidth,trim=12.0cm 0.8cm 1.7cm 0.8cm, clip]{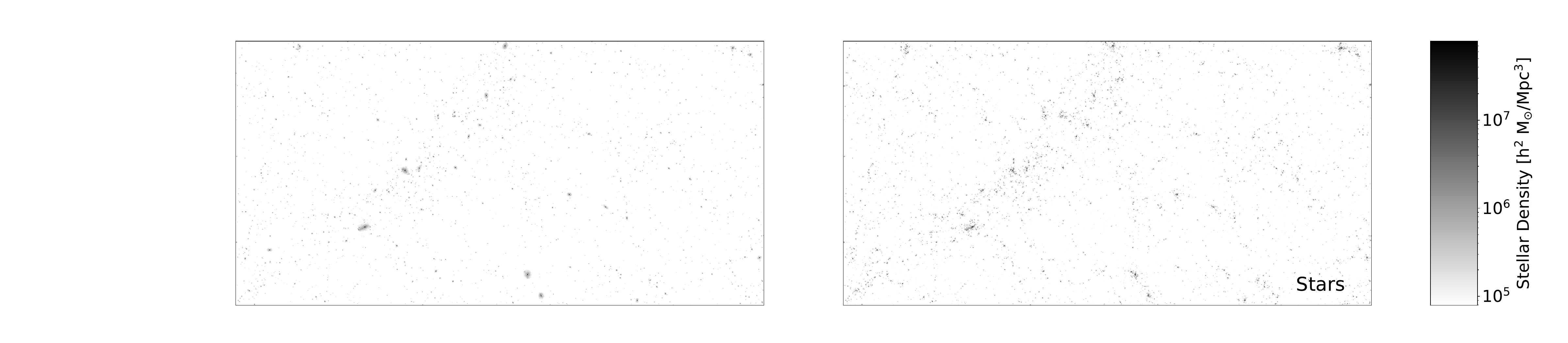}\\
\includegraphics[width=0.99\textwidth,trim=12.0cm 0.8cm 1.7cm 0.8cm, clip]{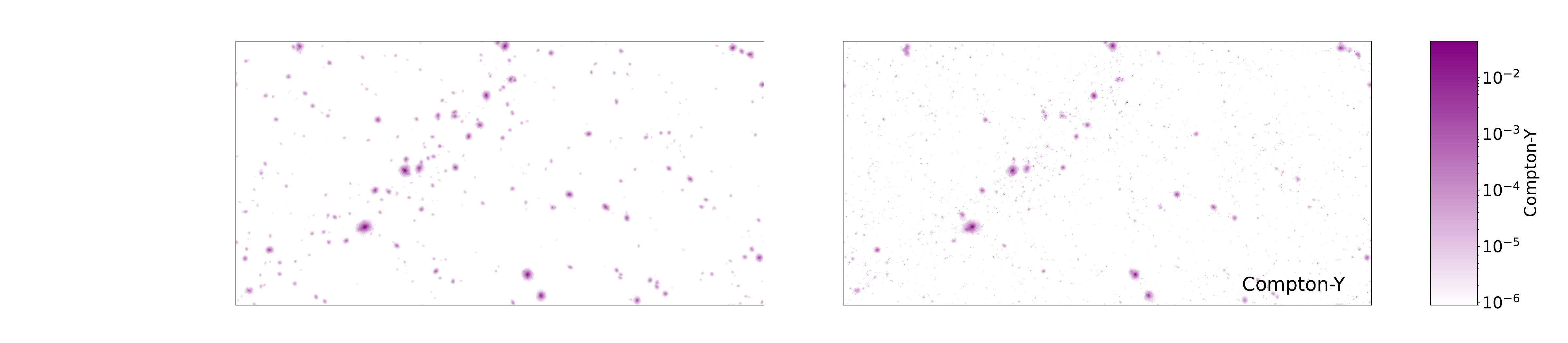}\\
\caption{\label{fig:field} Gas density, dark matter density, stellar density, and Compton-$y$ map (top to bottom) from the baryonified TNG100-DARK simulation (left) and the full hydrodynamical TNG-100 simulation (right).}
\label{fig:profiles}
\end{figure*}

In this section, we take a first look at density and pressure maps from the baryonification model and visually compare them with maps from the TNG-100-2 (mid-resolution) simulation\footnote{We use the TNG-100-2 (mid-resolution) simulation because it has the same feedback prescription and physical resolution as the TNG-300 (high-resolution) simulation while being significantly smaller in size (see Table~\ref{tab:sims}), which simplifies our analysis.}. A more quantitative comparison is left for future work. For the comparison, we baryonify the TNG-100-2-Dark simulation, which is the dark-matter-only counterpart of the TNG-100-2 hydrodynamical simulation. The TNG-100-2-Dark and TNG-100-2 simulations are run with the same code and based on the same realisation of the initial density field, but the former only accounts for gravitational interactions while the latter consists of a full hydrodynamical cosmological simulation with gas cooling and feedback.

In the panels of the top-three rows of Fig.~\ref{fig:field} we show a patch of the density fields of dark matter, gas, and stars of the baryonified TNG-100-2-Dark simulation (left) and the TNG-100-2 hydrodynamical simulation (right) at redshift zero. First, we observe that stars are only visible at the very centres of haloes, while the dark matter and gas distributions extend into the intergalactic space. Compared to the gas, the dark matter distribution is more concentrated around halo centres. This is a direct consequence of the feedback ejection processes affecting the gas component. Comparing the hydrodynamical simulation to the baryonified box, we observe a good general agreement. Regarding the size of structures and overall densities of their peaks, the images show the same qualitative trends. Upon closer inspection, some differences become visible, especially in the stellar and gas maps. There are more small galaxies in the hydrodynamical simulation, especially satellite galaxies. The differences in the gas map are especially visible in the low-density regime, where there are clear filamentary features in the hydrodynamical simulation that are missing or less  prominent in the \bfc{} case. However, it unlikely that such features would affect cosmological observables.

In the bottom panels of Fig.~\ref{fig:field} we show the Compton-$y$ maps of the baryonified box (left) and the hydrodynamical simulation (right). The Compton-$y$ parameter is a measure used to describe CMB distortions caused buy inverse Compton scattering with hot, ionized electrons in clusters. The measure is obtained by integrating the thermal electron pressure ($P_\mathrm{e}$) along the line-of-sight ($l$), i.e., 
\begin{equation}
y(x,y)=\frac{\sigma_\mathrm{T}}{m_\mathrm{e} c^2}\int \mathrm{d}l P_\mathrm{e}(x,y,l),
\end{equation}
where $m_\mathrm{e}$ is the electron mass, $c$ the speed of light, and $\sigma_\mathrm{T}$ denotes the Thomson scattering cross-section.
The electron pressure is obtained from the thermal pressure via 
\begin{equation}
P_{e} = \left(\frac{2+2 X_H}{3+5 X_H}\right)P_{\rm th},
\end{equation}
where $X_H=0.76$ is the Hydrogen mass fraction. In principle, the line-of sight integration should be performed along the light-cone. Here we integrate over the box-length of the simulation for simplicity.

The Compton-$y$ signal is dominated by high-mass haloes. In Fig.~\ref{fig:field} we observe a good general agreement between the \bfc{} results and the TNG-100 simulation. Both the absolute peak values and the extent of the signal look very similar. A closer inspection reveals some differences at the level of individual peaks. This is likely the result of some natural halo-to-halo scatter in the electron pressure that is not included in the \bfc{} model. Such a scatter could easily be implemented, which would further improve the similarity of the \bfc{} and the simulated data at the statistical level. However, adding scatter would not improve the visual resemblance of the two images, given that peaks would be randomly up- and down-scattered and would therefore not necessarily coincide with the ones from the hydrodynamical simulation.

In summary, this first visual comparison between the \bfc{} model and a hydrodynamical simulation shows an encouraging general agreement, with some visible differences at the level of individual haloes. Some differences could be caused by overly simplistic model assumptions. Others could come from the naturally occurring halo-to-halo scatter, which is not included in the \bfc{} modelling. It is also worth noting that a better match could be obtained if the \bfc{} model parameters would be specifically tuned at the map level to a predefined simulation-based discrepancy or loss function.

Future field-level cosmological studies will be performed on realistic maps of observations that include foreground and noise contamination, as well as potential small-scale smoothing. The \bfc{} method will have to be tested against these realistic data sets in order to validate or discard it as a modelling technique for map-level analysis. For a first step in this direction, see also Ref. \cite{Zhou:2025tdd}.

\section{Conclusions}\label{sec:Conclusion}
The baryonification method consists of a semi-analytical technique to displace particles of gravity-only $N$-body simulation outputs in order to account for non-gravitational, baryonic effects on the particle distribution \citep{Schneider:2015wta}. In this paper, we present a new version of baryonification where particles are separated into baryonic and dark matter components before being displaced at different rates. As a result, we obtain individual dark matter, stellar, and gas distributions, mimicking the output of hydrodynamical simulations.

In the first part of the paper, we present the new model and compare it to density profiles from the FLAMINGO m8 and the TNG-300 simulations. We show that simulated gas and stellar profiles can be well reproduced over different mass bins from individual galaxies to galaxy clusters. Regarding the dark matter profile, we present a detailed investigation of back-reaction effects such as adiabatic contraction and relaxation. We suggest a new parametrisation that provides good agreement with the TNG-300 and FLAMINGO m8 simulations, as well as the FLAMINGO m9 simulations with varying feedback.

In the second part of the paper, we validate the baryonification approach by first fitting its parameters to profiles from hydrodynamical simulations before comparing the resulting power spectrum of the baryonified boxes to the same simulations. At redshift zero, we find agreement at the level of two percent or better for all simulations and over all $k$-modes up to $k=10$ $h$/Mpc. As the redshift increases to 0.5, 1.0, and 1.5, the agreement slightly degrades but remains at the 2 percent level up to $k\sim 5$ $h$/Mpc for all simulations. This is very promising, as this wave vector range encompasses most of the scales targeted by Stage-IV surveys such as Euclid and LSST.

The comparison of the power spectra is carried out for both the full 8-parameter \bfc{} model as well as a reduced 2-parameter model, where only the gas parameters $\log M_{\mathrm{c}}$ and $\delta$ are varied. Both models show a similar agreement at the level of profiles and power spectrum. While this may not be very surprising given that we fix the stellar parameters to their ideal values (which differ between simulations), it confirms that two model parameters may be sufficient to describe the baryonic feedback effects at any given individual redshift.

Regarding redshift dependence, we investigate the evolution of individual parameter values from z=0 to z=1.5. We find results in agreement with the general expectation that baryonic feedback effects weaken towards higher redshift. For the reduced 2-parameter model, we see evidence for a linear decrease of $\log M_\mathrm{c}$ while the second parameter $\delta$ remains roughly constant for all simulations. Based on this finding, we propose a 2+1 parameter \bfc{} model that describes baryonic feedback over the redshift regime from $z=0-1.5$ (two parameters regulating the strength of baryonic feedback, one describing the explicit redshift evolution). We show that the model, once fitted to gas and stellar ratio profiles from hydrodynamical simulations, is able to predict the power spectrum at better than 2.5 percent over the full redshift range. We therefore suggest the 2+1 parameter \bfc{} model to be used for cosmological studies.

As a further extension of the original \bfc{} model, we calculate pressure and temperature values for each gas particle. The calculation is based on the hydrostatic equilibrium combined with an empirical, non-thermal pressure correction. We obtain pressure profiles that are in good agreement with both the FLAMINGO m8 and the TNG-300 simulations. The agreement is very encouraging, especially considering the fact that they are obtained using a \bfc{} model parameters that were fitted to the gas and stellar density profiles and not the pressure profiles directly. Our results regarding pressure profiles provide further evidence that relatively simple models based on the hydrostatic equation may be sufficient to model current X-ray and SZ observations.

Finally, we provide a first visual comparison of the \bfc{} model with the (medium resolution) TNG-100 simulation at the map level, focusing on the dark matter, gas, stellar, and Compton-$y$ (SZ) fields. While there are visual differences between the \bfc{} and simulated maps, the general qualitative agreement is encouraging. We leave more quantitative studies of field-level effects to further work.

In summary, we have presented a new baryonification model that provides particle information of stars, gas, and dark matter, emulating outputs of hydrodynamical simulations. We find encouraging agreement with hydrodynamical simulations from the FLAMINGO and TNG suites at scales relevant for cosmology. The model will be used for cosmological studies in further publications.

\section*{Acknowledgments}
AN acknowledges support from the European Research Council (ERC) under the European Union’s Horizon 2020 research and innovation program with Grant agreement No. 101163128. TT acknowledges funding from the Swiss National Science Foundation under the Ambizione project PZ00P2\_193352. SKG acknowledges the support by NWO (grant no OCENW.M.22.307) and Olle Engkvist Stiftelse (grant no. 232-0238).

\appendix

\section{Power spectra from fits to gas and stellar density profiles}\label{app:densityprofiles}

In the main text we have shown both fits to gas/stellar density profiles (see Figs \ref{fig:GASprofiles}, \ref{fig:STELLARprofiles}) and to mass ratio profiles (see Figs.~\ref{fig:ratiofit}, \ref{fig:ratiofit_z0.5}, \ref{fig:ratiofit_z1}, \ref{fig:ratiofit_z1p5}). However, only the latter have so far been used to predict the power suppression caused by baryons. In this Appendix we investigate whether power spectra based on fits to gas and stellar density profiles differ from the ones from mass ratio profiles. We thereby focus on the TNG-300 and the Flamingo m8 simulations at $z=0$, where we already have BFC models fitted to profile data.

\begin{figure*} 
\centering
\includegraphics[width=1.0\textwidth,trim=2.0cm 0.5cm 2.6cm 0.3cm, clip]{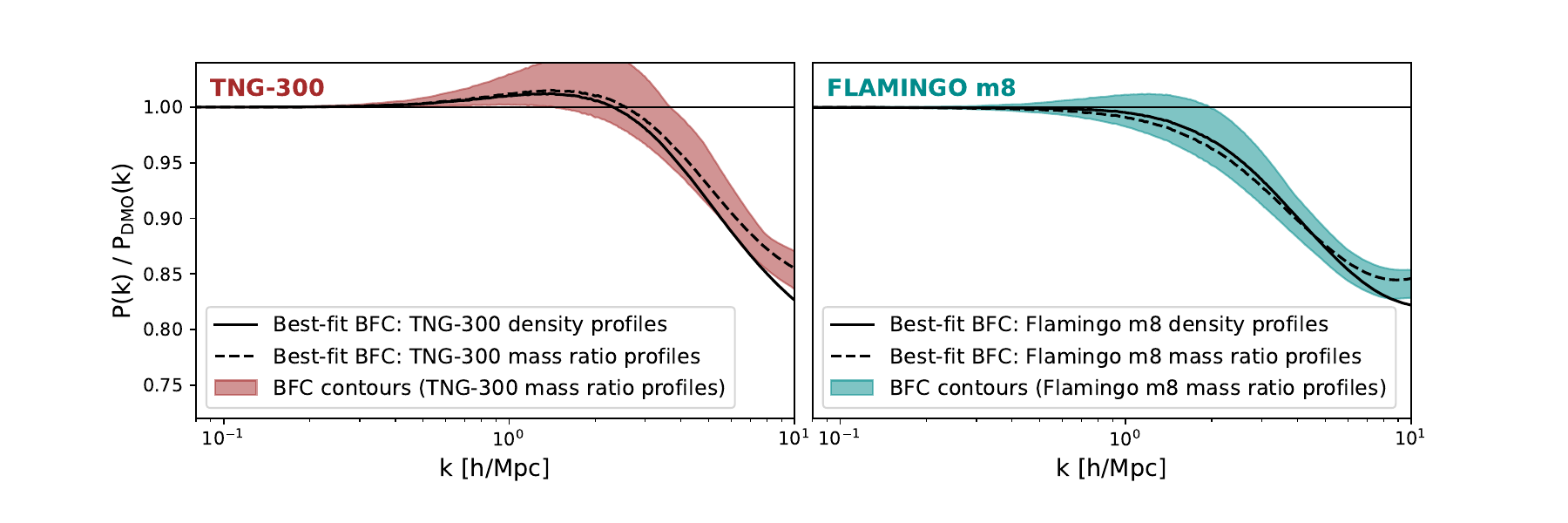}
\caption{\label{fig:PSdensityratio}Predicted baryonic power suppression from BFC models fitted to the TNG-300 (left) and Flamingo m8 (right) simulations at $z=0$. The black solid and dashed lines show the BFC models that best fit the gas and stellar density profiles (see Figs. \ref{fig:GASprofiles}, \ref{fig:STELLARprofiles}) and the mass ratio profiles (see Fig.~\ref{fig:ratiofit}), respectively. The coloured bands correspond to the same contours of the 8-parameter model shown in Fig. \ref{fig:ratiofit}. }
\end{figure*}

The ratios of the power spectra are illustrated in Fig.~\ref{fig:PSdensityratio}. The coloured bands are copied from Fig.~\ref{fig:ratiofit}. They correspond to the 8-parameter model discussed in the main text, resulting from an inference analysis based on gas and stellar mass ratio profiles of the TNG-300 and Flamingo m8 simulations. The highest likelihood model of the same inference analysis is highlighted as a dashed black line. Note that this line is close but not identical to the mean suppression shown in Fig.~\ref{fig:ratiofit}.

The solid black line in Fig.~\ref{fig:PSdensityratio} finally shows the BFC power suppression based on the best fit to the gas and stellar density profiles. The difference with respect to the standard case (based on mass ratio profiles) remains at the level of a few percent. Such an agreement is encouraging given that absolute density profiles contain very different information compared to mass ratio profiles, where the signal from gas and stars is normalised by the total mass distribution.

\section{Back-reaction of Dark Matter}\label{app:back-reaction}
Any non-gravitational effects from the gas and stellar components have an influence on the dark matter distribution. This effect is commonly referred to as dark matter back-reaction or adiabatic contraction and relaxation. In this Appendix, we take a closer look at different analytical models for the back-reaction effect, testing them against the FLAMINGO and the TNG simulations. 

\subsection{Modelling back-reaction}
The simplest model from \cite{Blumenthal:1985qy} is based on angular momentum conservation. It assumes
\begin{equation}\label{ACBlumenthal}
\frac{r_\mathrm{f}}{r_\mathrm{i}}=\frac{M_\mathrm{i}(r_\mathrm{i})}{M_\mathrm{f}(r_\mathrm{f})},
\end{equation}
where $r_\mathrm{i}$ and $r_\mathrm{f}$ are the initial and final radii (before and after the back-reaction effect) and the mass profiles are given by
\begin{eqnarray}
M_\mathrm{i}(r_\mathrm{i})&=&M_{\rm nfw}(r_\mathrm{i}),\\
M_\mathrm{f}(r_\mathrm{f})&=&(f_{\rm cdm}+f_{\rm sga})M_{\rm nfw}(r_\mathrm{i}) + \nonumber\\
&&f_{\rm hga}M_{\rm hga}(r_\mathrm{f}) + f_{\rm iga}M_{\rm iga}(r_\mathrm{f}) + f_{\rm cga}M_{\rm cga}(r_\mathrm{f}),
\end{eqnarray}
standing for the dark-matter-only and the dark matter plus baryon profiles. The former is simply given by the NFW profile. The latter consists of a sum over all mass profiles, whereas the dark matter and satellite galaxy components are described by the NFW profile at $r_\mathrm{i}$ All other components are evaluated at $r_\mathrm{f}$.

In a next step, Eq.~(\ref{ACBlumenthal}) is solved iteratively for $\zeta\equiv r_\mathrm{f}/r_\mathrm{i}$ in order to obtain the final CDM profiles
\begin{equation}\label{ACMprofiles1}
M_{\rm cdm}(r)=f_{\rm cdm}M_{\rm nfw}(r/\zeta),\hspace{0.5cm}\rho_{\rm cdm}=\frac{f_{\rm cdm}}{4\pi r^2}\frac{d}{dr}M_{\rm nfw}(r/\zeta),
\end{equation}
which include the adiabatic contraction and expansion due to the gas and stars. Note that the profiles of the satellite galaxies are assumed to be identical to Eq.~(\ref{ACMprofiles1}) except that $f_{\rm cdm}$ is replaced by $f_{\rm sga}$.

With the first $N$-body simulations, it has become clear that the model of \cite{Blumenthal:1985qy} leads to a overly strong contraction in the centre of haloes. An improved, empirical model of the form
\begin{equation}\label{ACGnedin}
\frac{r_\mathrm{f}}{r_\mathrm{i}}-1=\alpha \left[\left(\frac{M_\mathrm{i}(r_\mathrm{i})}{M_\mathrm{f}(r_\mathrm{f})}\right)^n-1\right]
\end{equation}
was proposed \citep{Gnedin:2004cx,Abadi:2010aaa,Teyssier:2011aaa}. The free parameters $\alpha$ and $n$ allow for a reduction of the overall contraction effect, and they provide more freedom for the original adiabatic equation. \cite{Gnedin:2004cx} used the values $\alpha=0.68$ and $n=1$ (Gnedin04-model), \cite{Abadi:2010aaa} proposed $\alpha=0.3$ and $n=2$ (Abadi10-model). In this paper, we only show results from the Gnedin04-model, which was the default assumption in previous baryonification papers \citep{Schneider:2015wta, Schneider:2018pfw}. Note, however, that the Abadi10-model provides very similar results.

A further extension of the adiabatic correction model has recently been proposed by \cite{Velmani:2022una}. Their functional form
\begin{equation}\label{ACVelmani}
\frac{r_\mathrm{f}}{r_\mathrm{i}}-1=\alpha \left[\frac{M_\mathrm{i}(r_\mathrm{i})}{M_\mathrm{f}(r_\mathrm{f})}-1\right] + q_0,
\end{equation}
produces an upturn at large radii in qualitative agreement with Fig.~\ref{fig:ACM}. However, the upturn appears at radii that are somewhat larger than what the TNG-300 and FLAMINGO simulations suggest. \cite{Velmani:2022una}  also allow the parameter $\alpha$ to become dependent on the radius. Our analysis based on MCMC fits suggest very little radial dependence. The different results can be explained by the fact that \cite{Velmani:2022una} predominantly investigated smaller halo masses.

A better agreement with the simulations is obtained by replacing the constant $q_0$ in Eq.~(\ref{ACVelmani}) with a smooth step function that becomes zero outside of the virial radius. Further improvement can be obtained by replacing the parameter $\alpha$ with two parameters $Q_1$ and $Q_2$ leveraging the effect of contraction and expansion, respectively. This can be achieved by replacing Eq.~(\ref{ACVelmani}) with
\begin{eqnarray}\label{ACModel1}
\frac{r_\mathrm{f}}{r_\mathrm{i}}-1=\left(\frac{M_\mathrm{i}}{M_\mathrm{f}}\right) - \frac{Q_0}{1+(r/r_{\rm step})^n},
\end{eqnarray}
where $r_{\rm step} = \varepsilon(\nu)/\varepsilon_0\times r_{200}$ and $n=3/2$. The mass profiles $M_\mathrm{i}$ and $M_\mathrm{f}$ are given by
\begin{eqnarray}
M_\mathrm{i}&=& M_{\rm nfw}(r_\mathrm{i}),\\
M_\mathrm{f}&=& (f_{\rm cdm}+f_{\rm sga})M_{\rm nfw}(r_\mathrm{i}) +\nonumber\\
&&(1-Q_1)(f_{\rm cga}+f_{\rm iga})M_{\rm nfw}(r_\mathrm{i}) + 
(1-Q_2)f_{\rm hga}M_{\rm nfw}(r_\mathrm{i})+\nonumber\\
&&Q_1 M_{\rm cga}(r_\mathrm{f})+Q_1 M_{\rm iga}(r_\mathrm{f}) + Q_2 M_{\rm hga}(r_\mathrm{f}),
\end{eqnarray}
where $Q_1$ and $Q_2$ are free parameters with values between 0 and 1. They are down-weighting both the contracting effects of stars and cold gas as well as the expanding effects of hot gas.

The function shown in Eq.~(\ref{ACModel1}) is still vaguely motivated by the original idea of adiabatic contraction and expansion \citep[originally introduced by][]{Blumenthal:1985qy}. However, it has been modified substantially, adding three free parameters to account for a generic, step-like contraction around the virial radius and to individually adjust the effects of the inner galaxy and the outer gas distributions.

It turns out that by completely abandoning the functional form introduced by \cite{Blumenthal:1985qy}, it is possible to obtain a simpler empirical function that does not rely on any iterative process, returning nearly identical results than Eq.~(\ref{ACModel1}). It is given by
\begin{eqnarray}\label{ACModel2}
\frac{r_\mathrm{i}}{r_\mathrm{f}}-1=\frac{Q_0}{1+(r_\mathrm{i}/r_{\rm step})^{n_{\rm step}}}
+Q_1 f_{\rm cga}\left[\frac{M_{\rm cga}(r_\mathrm{i})}{M_{\rm nfw}(r_\mathrm{i})}-1\right]+\nonumber\\
Q_1f_{\rm iga}\left[\frac{M_{\rm iga}(r_\mathrm{i})}{M_{\rm nfw}(r_\mathrm{i})}-1\right]+Q_2 f_{\rm hga}\left[\frac{M_{\rm hga}(r_\mathrm{i})}{M_{\rm nfw}(r_\mathrm{i})}-1\right].
\end{eqnarray}
This functional form of the adiabatic correction is used in the main part of the paper. It is purely empirical, but can be interpreted to be shaped by three dominant processes. The first process (captured by the term proportional to $Q_0$) introduces a generic contraction affecting scales around and below the virial radius. This contraction is present in all simulations and is likely caused by the cooling and star-formation processes prior to any significant feedback events. The second process (driven by the terms proportional to $Q_1$) describes the contraction affecting the inner part of the halo. It is caused by the presence of the central galaxy and the inner, cold gas component. The third process (described by the term proportional to $Q_2$) is causing an expansion of the dark matter profile around and outside of the virial radius. This expansion is the response to the AGN feedback effect pushing gas into the intergalactic medium.

The model parameters $Q_1$ and $Q_2$ could, at least in principle, vary with redshift. To account for this by assuming
\begin{equation}
Q_i(z)=q_i(1+z)^{q_{\rm exp,i}},\hspace{1.5cm}i=\lbrace 1,2\rbrace.
\end{equation}
The parameter $Q_0$, on the other hand, describes the contraction happening before AGN feedback becomes important, which means that in the low-redshift universe it should not evolve with redshift. We therefore assume $Q_0=q_0$.

\subsection{Comparing to simulations at redshift zero}
In this section we test different back-reaction models against the TNG-300 simulation as well as the FLAMINGO m8 and m9 simulations at redshift zero. The tests are based on matched haloes between the hydrodynamical simulations and their dark-matter-only counterparts. For each matched halo, we measure the baryonic and dark matter mass ratio profiles ($M_{\rm dm}$[hydro] and $M_{\rm dm}$[dmo]) to determine the mass ratio profiles
\begin{eqnarray}\label{matchedratioprofile1}
&R^{\rm bar}_{\rm matched}(r)&=M_{\rm bar}{\rm [hydro]}/M_{\rm bar}{\rm [dmo]},\\\label{matchedratioprofile2}
&R^{\rm dm}_{\rm matched}(r)&=M_{\rm dm}{\rm [hydro]}/M_{\rm dm}{\rm [dmo]}, 
\end{eqnarray}
where $M_{\rm bar}$[hydro], $M_{\rm bar}$[dmo], $M_{\rm dm}$[hydro], $M_{\rm dm}$[dmo] are the baryonic and dark-matter masses from the hydrodynamical and the dark-matter-only simulations. The dark-matter-only masses are defined as
\begin{equation}
M_{\rm bar}{\rm [dmo]} = f_{\rm b}M_{\rm tot}{\rm [dmo]},\hspace{0.5cm}
M_{\rm dm}{\rm [dmo]} = f_{\rm dm}M_{\rm tot}{\rm [dmo]},
\end{equation}
with $f_{\rm bar}=\Omega_{\rm b}/\Omega_{\rm m}$ and $f_{\rm dm}=\Omega_{\rm dm}/\Omega_{\rm m}$, respectively.

A comparison of back-reaction models with the median matched mass ratio profiles of the simulations reveals that the models of Eq.~(\ref{ACModel1}) and Eq.~(\ref{ACModel2}) provide the most accurate descriptions of the effect. We therefore use Eq.~(\ref{ACModel2}) as the standard model in the paper.

Regarding the model parameters of Eq.~(\ref{ACModel2}), we first fix the $q_0$ parameter to
\begin{equation}
q_0 = 0.075
\end{equation}
for all simulations, regardless of the implemented feedback recipe. This choice is motivated by the fact that $q_0$ describes the contraction effect from the in-fall of gas prior to the onset of AGN feedback processes.

\begin{figure*} 
\centering
\includegraphics[width=0.49\textwidth,trim=0.5cm 0.5cm 0.9cm 0.0cm, clip]{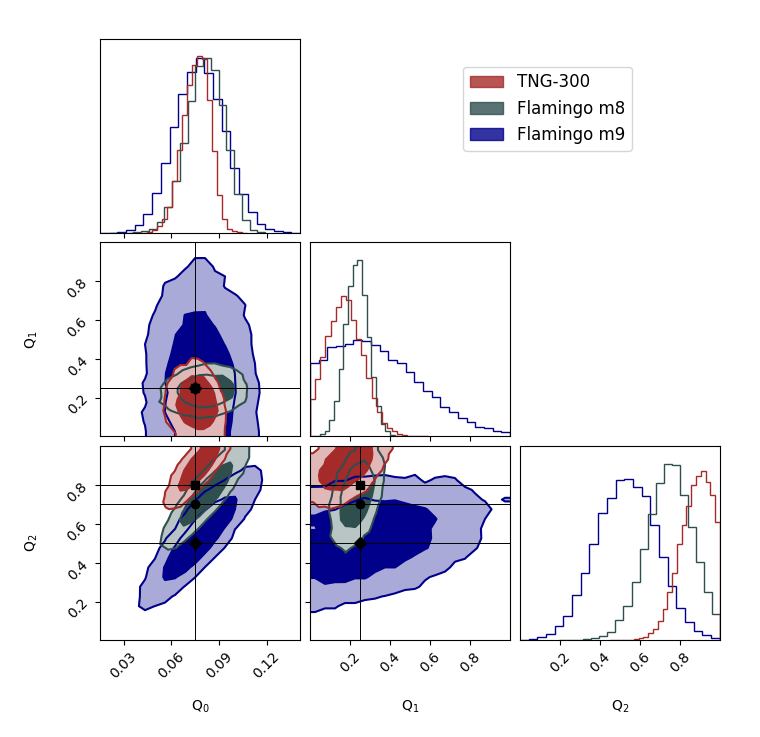}
\includegraphics[width=0.49\textwidth,trim=0.5cm 0.5cm 0.9cm 0.0cm, clip]{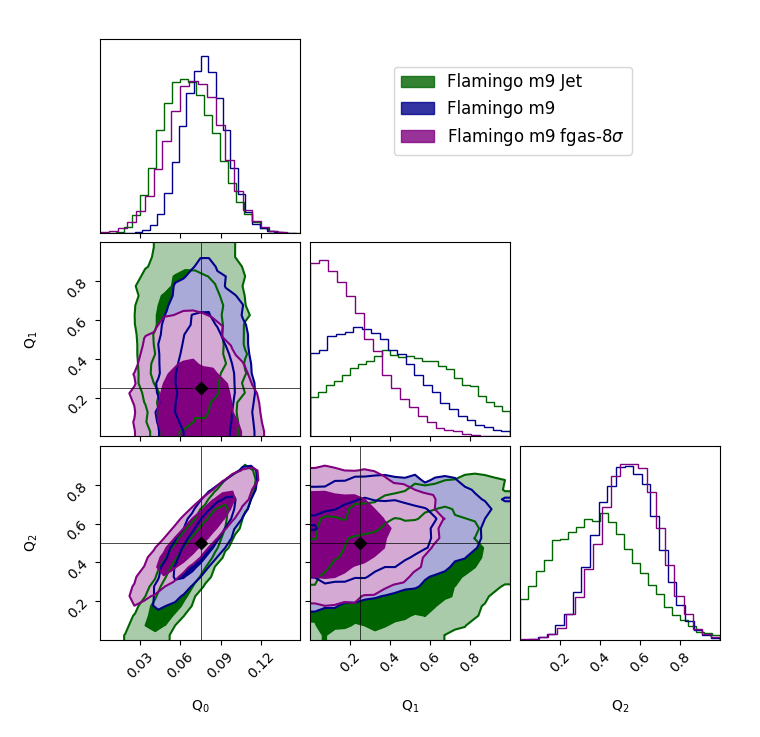}
\caption{\label{fig:AC_params}Posteriors of AC parameters $Q_0$ and $Q_1$ at redshift 0. \emph{Left:} Comparison of TNG-300, FLAMINGO HR, and FLAMINGO fiducial. \emph{Right:} Comparison of three FLAMINGO simulations with different feedback implementations.}
\label{fig:Q0Q1posteriors}
\end{figure*}

The other parameters ($q_1$ and $q_2$) may potentially be different for different simulations and feedback recipes. We determine the values of $q_1$ and $q_2$ with the help of a Bayesian inference analysis. We simultaneously fit the model to the median mass ratios $R_{\rm matched}^{\rm bar}$ and $R_{\rm matched}^{\rm dm}$ for five different halo mass bins in the range $\log(M_{200}h/M_{\odot})=$ [12.5, 13.0, 13.5, 14.0, 14.5, 15.0].

The resulting posterior distributions are plotted in Fig.~\ref{fig:Q0Q1posteriors}. The panels to the left show the posteriors of the TNG-300, as well as the FLAMINGO m8 and m9 simulations. The last two assume the same feedback strength and only differ in their resolution. For the gas parameter, the three simulations give a similar best-fitting value of
\begin{equation}
q_1 = 0.18.
\end{equation}
The stellar parameter $q_2$, on the other hand, differs between the simulations. This is, at least partially, due to the different resolutions between the simulations. The higher the resolution, the higher the value for $q_2$. This can be understood intuitively by the fact that the central galaxies are better resolved in high resolution simulations, thereby also leading to stronger adiabatic contraction. We therefore assume a resolution-dependent value of $q_2$ with 
\begin{equation}
q_2= 0.5,\,0.7,\,0.8,
\end{equation}
for the FLAMINGO m9, FLAMINGO m8, and TNG-300 simulations.

The panels on the right-hand side of Fig.~\ref{fig:Q0Q1posteriors} show the same analysis, this time for the three FLAMINGO m9 simulations that have different feedback but the same numerical resolution. We again observe very similar values for the $q_1$ parameter, while there is a moderate shift in $q_2$, with a higher value for the fgas-8$\sigma$ compared to the Jet simulation. Since the shift is smaller than the resolution effect discussed before, we use the same value ($q_2= 0.5$) for all FLAMINGO m9 simulations.

\begin{figure*} 
\centering
\includegraphics[width=0.99\textwidth,trim=2.5cm 1.95cm 3.5cm 1.0cm, clip]{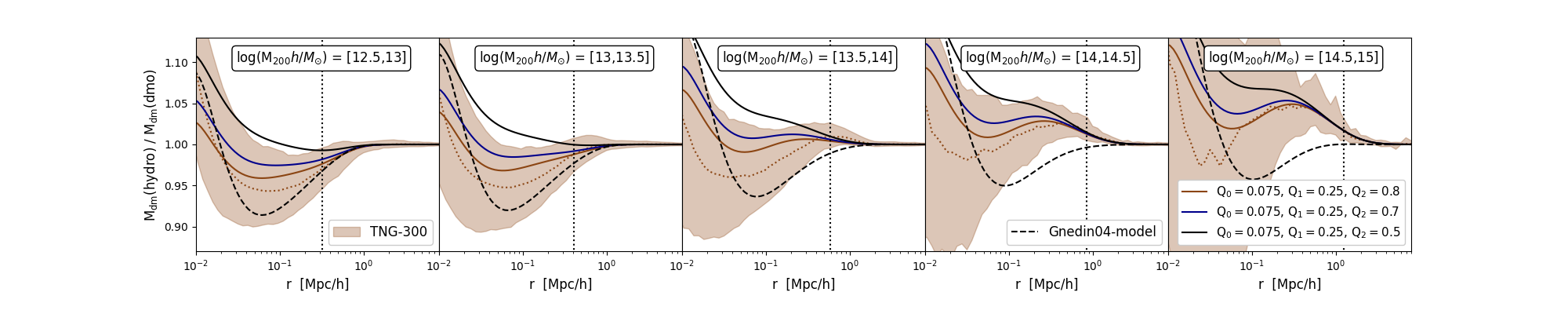}\\
\includegraphics[width=0.99\textwidth,trim=2.5cm 1.95cm 3.5cm 1.2cm, clip]{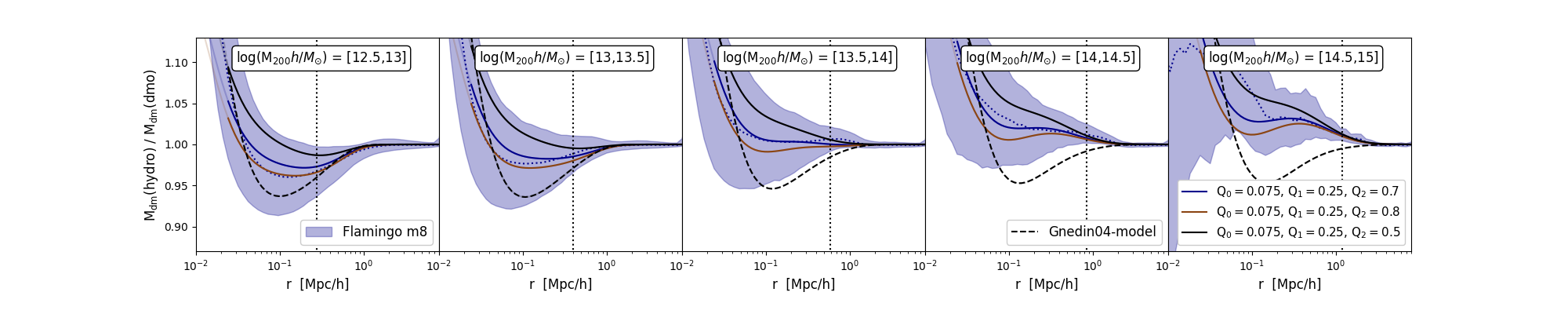}\\
\includegraphics[width=0.99\textwidth,trim=2.5cm 1.95cm 3.5cm 1.2cm, clip]{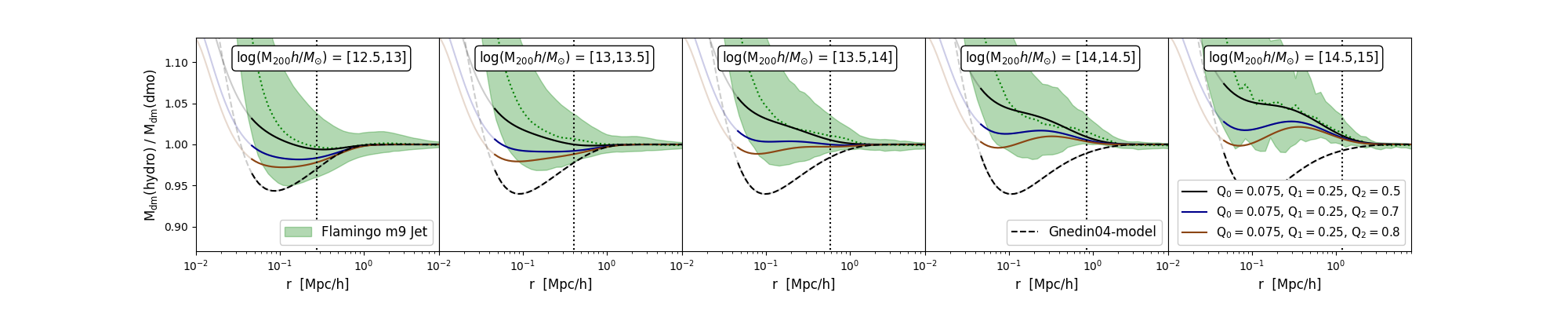}\\
\includegraphics[width=0.99\textwidth,trim=2.5cm 1.95cm 3.5cm 1.2cm, clip]{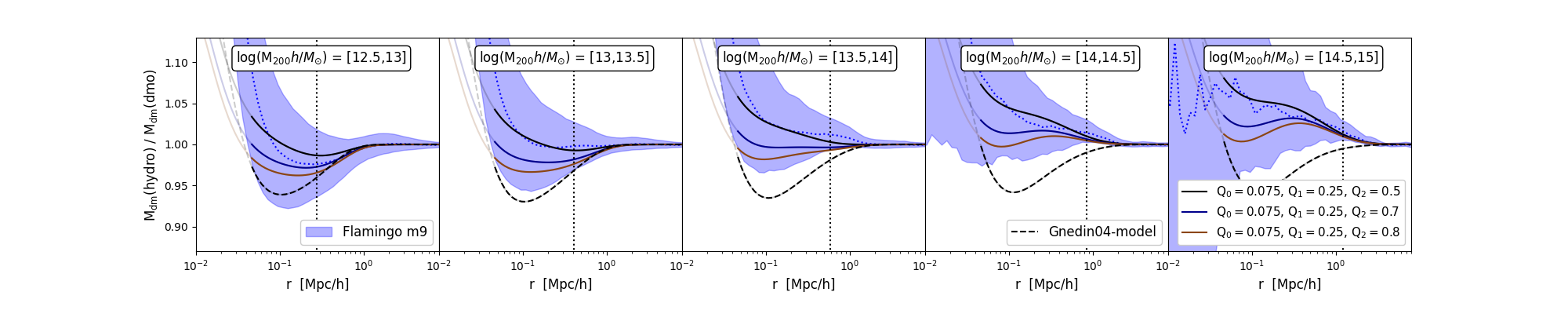}\\
\includegraphics[width=0.99\textwidth,trim=2.5cm 0.0cm 3.5cm 1.2cm, clip]{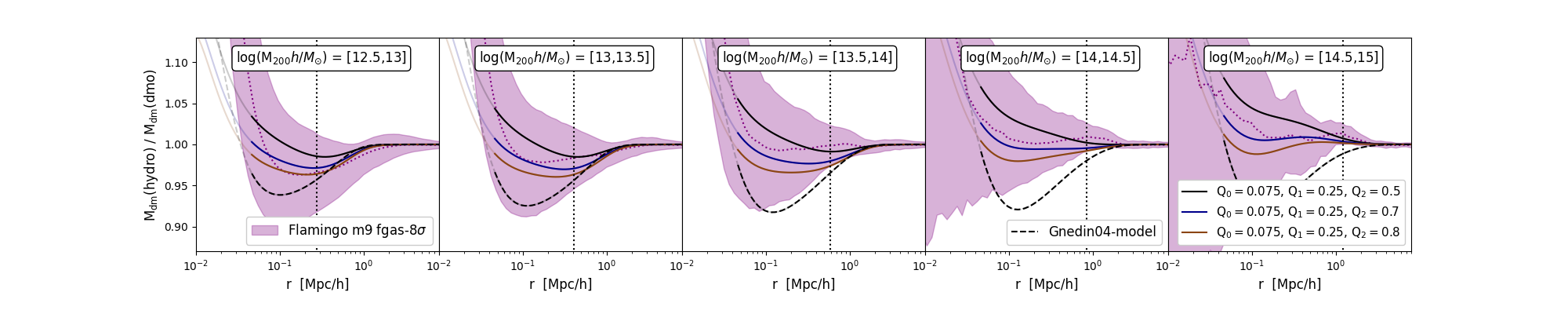}\\
\caption{\label{fig:ACM}Ratio of matched dark matter mass from the hydro and the dmo-run of TNG-300, FLAMINGO HR, FLAMINGO Jets, FLAMINGO fiducial, and FLAMINGO sAGN (from top to bottom). The best fitting \bfc{} model is shown as a solid line. The dashed lines correspond to the fits to the other simulations. The old model from Gnedin04 is shown as dashed black line. All lines are shaded at the resolution limit of the simulations (three times the gravitational softening).}
\label{fig:profiles}
\end{figure*}

In Fig.~\ref{fig:ACM} we plot the resulting mass ratio profiles for the TNG300, the FLAMINGO m8, as well as the FLAMINGO m9 Jet, m9, and m9 fgas-8$\sigma$ simulations (from top to bottom). The coloured dotted lines show the mean of the $R^{\rm dm}_{\rm matched}(r)$ ratio profiles, while the surrounding shaded areas correspond to the 68 percent scatter. In Appendix~\ref{app:FLAMINGOResolution} we show that the three FLAMINGO m9 simulations are only converged down to $r\sim 0.3$ Mpc/$h$. For the FLAMINGO m8 and the TNG300 simulations we do not have any convergence limits, we only know that they should be converged at smaller radii than this.

At the mass-scales of galaxy clusters (panels to the right) all simulations are characterized by a general contraction starting outside of the virial radius and increasing towards small scales. Moving to smaller mass scales, the contraction at large radii is still visible, but now followed by a relaxation below the virial radius. At very small scales, the mass ratio profile increases again due to the presence of the central galaxy. At small masses below the galaxy group scale (panels to the left), the contraction at large scales is mostly gone and replaced by a more and more prominent suppression signal caused by feedback. The small scale contraction due to the central galaxy remains visible at all mass scales.

The solid black blue and brown lines correspond to the results from the back-reaction model of Eq.~(\ref{ACModel2}) with $Q_2 = 0.5,\, 0.6,\, 0.75$, respectively. The $Q_0$ and $Q_1$ parameters are fixed to the values defined above. For the remaining \bfc{} model parameters, we use the best fitting values obtained from the gas and stellar profiles shown in Fig.~\ref{fig:GASprofiles} and \ref{fig:STELLARprofiles} in the main text.

In general, the back-reaction model from Eq.~(\ref{ACModel2}) is able to reproduce the main features of the mass ratio profiles, at least at scales above $r\sim 0.05$ Mpc/$h$. The agreement is not perfect but, for the relevant scales, the model remains within the simulation scatter and no more than a few percent away from the mean. 

As a comparison, we also show the adiabatic correction model of Gnedin04 which was used as the standard model in previous work using baryonification. The Gnedin04 model is plotted as a dashed black line in Fig.~\ref{fig:ACM}. At small halo mass scales it shows a reasonable qualitative behaviour but over-predicts the suppression of the mass ratio at the quantitative level. At larger halo masses, the model fails to reproduce the contraction around the virial radius, showing a suppression instead.

\subsection{Redshift dependence}
Let us look more into the potential redshift evolution of the $Q_1$, $Q_2$ parameters. To do so, we focus on the FLAMINGO m8 simulation where we fit to the matched mass ratio profiles at redshift 0.5, 1.0, and 1.5.

\begin{figure*} 
\centering
\includegraphics[width=0.49\textwidth,trim=0.5cm 0.5cm 0.9cm 0.0cm, clip]{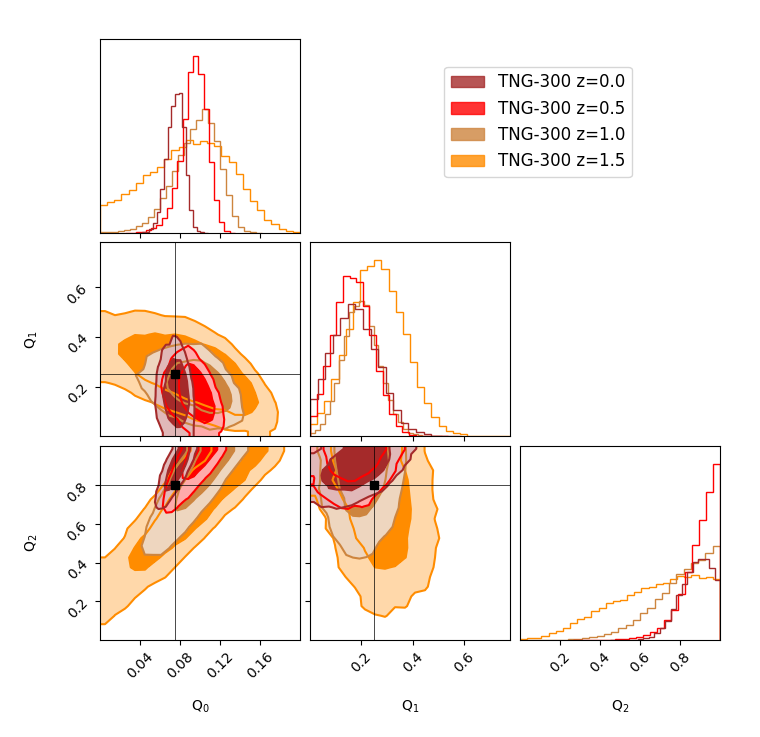}
\includegraphics[width=0.49\textwidth,trim=0.5cm 0.5cm 0.9cm 0.0cm, clip]{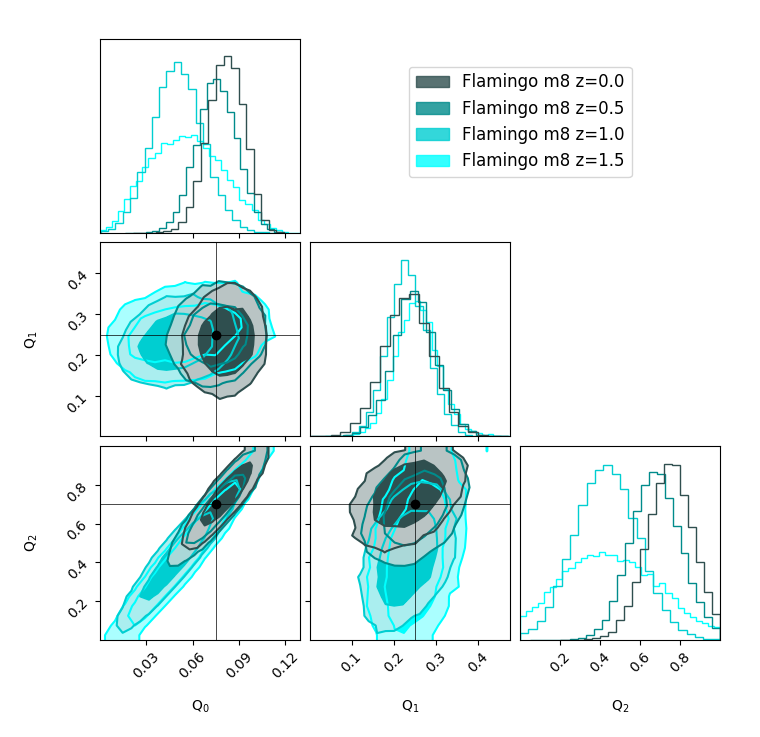}
\caption{\label{fig:AC_params_z}Redshift evolution of the posteriors of AC parameters for the TNG-300 (left) and the FLAMINGO m8 (right) simulations. Nor redshift dependence is assumed for the AC parameters $Q_0$, $Q1$, and $Q_2$.}
\label{fig:Q0Q1redshiftposteriors}
\end{figure*}

The redshift dependence of $Q_1$ and $Q_2$ are shown in Fig.~\ref{fig:Q0Q1redshiftposteriors}. In the panels to the left we plot the case of $q_{\rm 1,exp}=0$ and $q_{\rm 1,exp}=0$ so that the redshift evolution becomes visible in the plot. The posteriors show that $Q_1$ gets pushed to higher values for increasing redshifts while $Q_2$ parameters stays approximately constant.

Based on the results from our redshift study for FLAMINGO m8 we set
\begin{equation}
q_{\rm 1,exp} = 1,\hspace{1cm}q_{\rm 2,exp} = 0,
\end{equation}
meaning that only $Q_1$ evolves with redshift.

As a consistency check, we recalculate the posterior contours using the redshift evolution of $Q_1$ stated above. The results are shown in the panel on the right-hand-side of Fig.~\ref{fig:Q0Q1redshiftposteriors}. As expected, the posterior contours agree once the redshift evolution is considered.

\section{Convergence of FLAMINGO simulations}\label{app:FLAMINGOResolution}
In this Appendix we investigate the convergence of the simulated matched ratio profiles. We restrict ourselves to the Flamingo m8 and m9 runs which are based on the same feedback model but have a factor of eight difference in the mass resolution.

\begin{figure*} 
\centering
\includegraphics[width=0.95\textwidth,trim=2.5cm 1.5cm 3.5cm 1.8cm, clip]{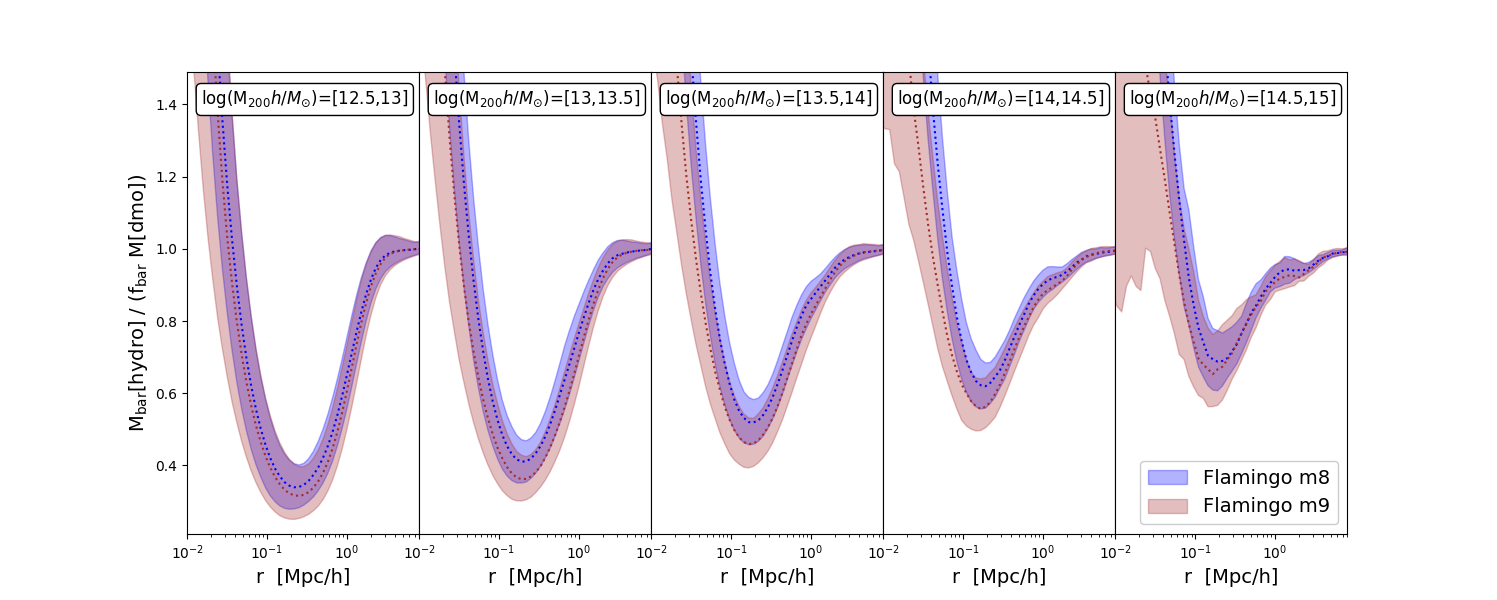}\\
\includegraphics[width=0.95\textwidth,trim=2.5cm 0.0cm 3.5cm 1.8cm, clip]{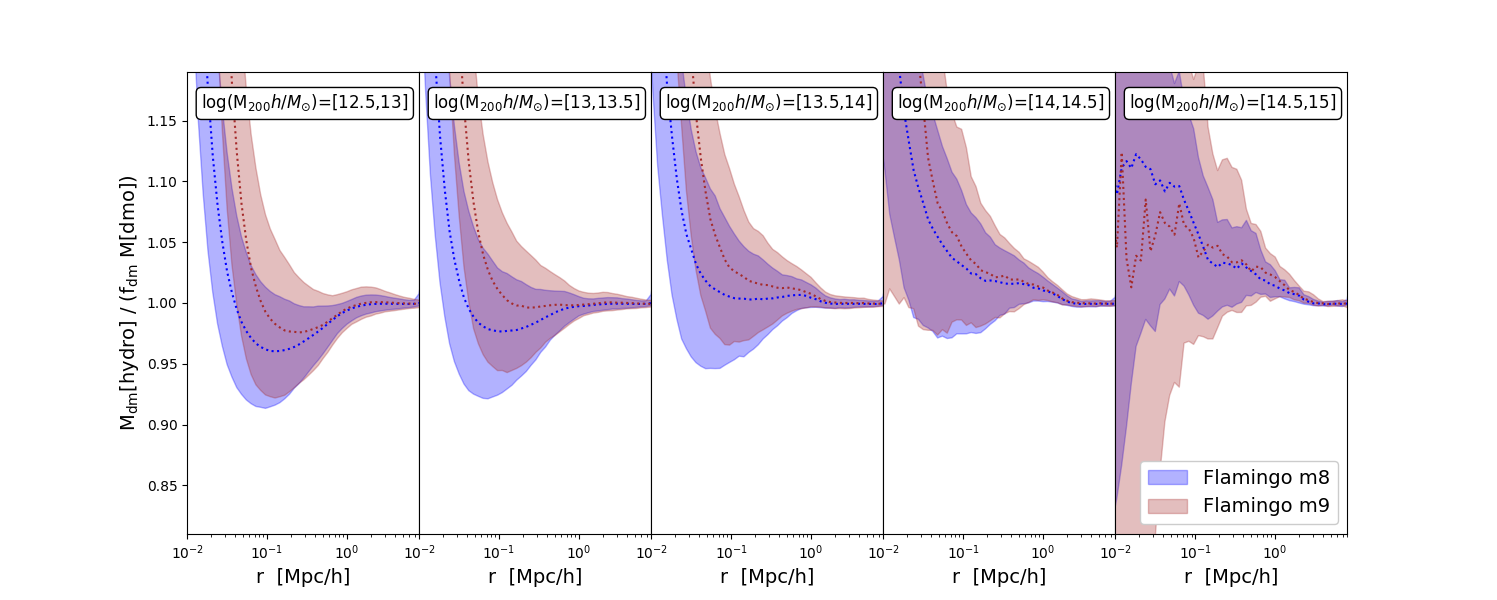}
\caption{\label{fig:FlamingoConvergence}Ratio of matched dark matter mass between hydro and gravity-only (dark-matter-only) runs for FLAMINGO m9 (standard resolution) and FLAMINGO m8 (high resolution). The top panel shows the matched ratio between baryonic mass from the hydro simulation and the gravity-only mass normalised to the baryonic fraction. The bottom panel shows the matched ratio between the dark matter mass from the hydro simulation and the gravity only mass normalised to the dark matter fraction.}
\label{fig:profiles}
\end{figure*}

Fig.~\ref{fig:FlamingoConvergence} illustrates the matched mass ratio profiles between the hydrodynamical and dark-matter-only runs (defined in Eqs.~\ref{matchedratioprofile1}, \ref{matchedratioprofile2}). In the the top-panels we focus on the baryonic mass ratio ($R_{\rm matched}^{\rm bar}$) observing a reasonable general agreement between the higher- and lower-resolution simulations. A more detailed look reveals a slightly shallower baryonic suppression signal for the Flamingo m8 simulation, which means that the gas expulsion becomes slightly less efficient when going to higher resolution. In the bottom-panels, we show the dark-matter mass ratio ($R_{\rm matched}^{\rm dm}$). Intriguingly, in this case the situation is reversed, the Flamingo m8 simulation exhibiting a deeper suppression signal. This means that when going to higher resolution, the back-reaction effect becomes stronger even though slightly less gas is being expelled.

By itself, it is not surprising to see some differences between the Flamingo m8 and m9 simulations as it is well known that hydrodynamical sub-grid feedback effects may depend on resolution. However, the fact that we observe opposite trends for the baryonic and the dark matter distribution shows that the dark-matter back-reaction effect is not converged. This is also reflected in Fig.~\ref{fig:AC_params}, where the Flamingo m8 and m9 simulations show different best-fitting values for the $Q_2$-parameter.

\bibliographystyle{JHEP}
\bibliography{bfc.bib}

@PREAMBLE{
 "\providecommand{\noopsort}[1]{}" 
 # "\providecommand{\singleletter}[1]{#1}%" 
}

@article{Kovac:2025aaa,
    author = "Kovac, Michael and Nicola, Andrina and Bucko,  Jozef and Reischke, Robert and Giri, Sambit K. and Teyssier, Romain",
    title = "{Baryonification II: Constraining feedback with X-ray and kinematic
Sunyaev-Zeldovich observations}",
    eprint = "2025.xxxx",
    archivePrefix = "arXiv",
    primaryClass = "astro-ph.CO",
    doi = "ccc5",
    journal = "\jcap",
    volume = "xxx",
    number = "x",
    pages = "xxxxxx",
    year = "2025"
}

@article{Dolag:2008ar,
    author = "Dolag, K. and Borgani, S. and Murante, G. and Springel, V.",
    title = "{Substructures in hydrodynamical cluster simulations}",
    eprint = "0808.3401",
    archivePrefix = "arXiv",
    primaryClass = "astro-ph",
    doi = "10.1111/j.1365-2966.2009.15034.x",
    journal = "Mon. Not. Roy. Astron. Soc.",
    volume = "399",
    pages = "497",
    year = "2009"
}

@article{Springel:2000qu,
    author = "Springel, Volker and White, Simon D. M. and Tormen, Giuseppe and Kauffmann, Guinevere",
    title = "{Populating a cluster of galaxies. 1. Results at z = 0}",
    eprint = "astro-ph/0012055",
    archivePrefix = "arXiv",
    doi = "10.1046/j.1365-8711.2001.04912.x",
    journal = "Mon. Not. Roy. Astron. Soc.",
    volume = "328",
    pages = "726",
    year = "2001"
}

@article{Elahi:2019wap,
    author = "Elahi, Pascal J. and Ca{\~n}as, Rodrigo and Poulton, Rhys J. J. and Tobar, Rodrigo J. and Willis, James S. and Lagos, Claudia del P. and Power, Chris and Robotham, Aaron S. G.",
    title = "{Hunting for galaxies and halos in simulations with VELOCIraptor}",
    eprint = "1902.01010",
    archivePrefix = "arXiv",
    primaryClass = "astro-ph.CO",
    doi = "10.1017/pasa.2019.12",
    journal = "Publ. Astron. Soc. Austral.",
    volume = "36",
    pages = "e021",
    year = "2019"
}

@article{vanDaalen:2019pst,
    author = "van Daalen, Marcel P. and McCarthy, Ian G. and Schaye, Joop",
    title = "{Exploring the effects of galaxy formation on matter clustering through a library of simulation power spectra}",
    eprint = "1906.00968",
    archivePrefix = "arXiv",
    primaryClass = "astro-ph.CO",
    doi = "10.1093/mnras/stz3199",
    journal = "Mon. Not. Roy. Astron. Soc.",
    volume = "491",
    number = "2",
    pages = "2424--2446",
    year = "2020"
}

@article{Schneider:2019snl,
    author = "Schneider, Aurel and Stoira, Nicola and Refregier, Alexandre and Weiss, Andreas J. and Knabenhans, Mischa and Stadel, Joachim and Teyssier, Romain",
    title = "{Baryonic effects for weak lensing. Part I. Power spectrum and covariance matrix}",
    eprint = "1910.11357",
    archivePrefix = "arXiv",
    primaryClass = "astro-ph.CO",
    doi = "10.1088/1475-7516/2020/04/019",
    journal = "JCAP",
    volume = "04",
    pages = "019",
    year = "2020"
}

@article{Bryan:2012mw,
    author = "Bryan, S. E. and Kay, S. T. and Duffy, A. R. and Schaye, J. and Vecchia, C. Dalla and Booth, C. M.",
    title = "{The impact of baryons on the spins and shapes of dark matter haloes}",
    eprint = "1207.4555",
    archivePrefix = "arXiv",
    primaryClass = "astro-ph.CO",
    doi = "10.1093/mnras/sts587",
    journal = "Mon. Not. Roy. Astron. Soc.",
    volume = "429",
    pages = "3316",
    year = "2013"
}

@article{Chua:2021oqe,
    author = "Chua, Kun Ting Eddie and Vogelsberger, Mark and Pillepich, Annalisa and Hernquist, Lars",
    title = "{The impact of galactic feedback on the shapes of dark matter haloes}",
    eprint = "2109.00012",
    archivePrefix = "arXiv",
    primaryClass = "astro-ph.GA",
    doi = "10.1093/mnras/stac1897",
    journal = "Mon. Not. Roy. Astron. Soc.",
    volume = "515",
    number = "2",
    pages = "2681--2697",
    year = "2022"
}

@article{Ondaro-Mallea:2024lhp,
    author = "Ondaro-Mallea, Lurdes and Angulo, Raul E. and Aric{\`o}, Giovanni and Schaye, Joop and McCarthy, Ian G. and Schaller, Matthieu",
    title = "{FLAMINGO: Galaxy formation and feedback effects on the gas density and velocity fields}",
    eprint = "2412.09526",
    archivePrefix = "arXiv",
    primaryClass = "astro-ph.CO",
    doi = "10.1051/0004-6361/202453480",
    journal = "Astron. Astrophys.",
    volume = "697",
    pages = "A63",
    year = "2025"
}

@article{Schneider:2019xpf,
    author = "Schneider, Aurel and Refregier, Alexandre and Grandis, Sebastian and Eckert, Dominique and Stoira, Nicola and Kacprzak, Tomasz and Knabenhans, Mischa and Stadel, Joachim and Teyssier, Romain",
    title = "{Baryonic effects for weak lensing. Part II. Combination with X-ray data and extended cosmologies}",
    eprint = "1911.08494",
    archivePrefix = "arXiv",
    primaryClass = "astro-ph.CO",
    doi = "10.1088/1475-7516/2020/04/020",
    journal = "JCAP",
    volume = "04",
    pages = "020",
    year = "2020"
}

@ARTICLE{2024arXiv241117682R,
       author = {{Reischke}, Robert and {Kova{\v{c}}}, Michael and {Nicola}, Andrina and {Hagstotz}, Steffen and {Schneider}, Aurel},
        title = "{An analytical model for the dispersion measure of Fast Radio Burst host galaxies}",
      journal = {arXiv e-prints},
         year = 2024,
        month = nov,
          eid = {arXiv:2411.17682},
        pages = {arXiv:2411.17682},
          doi = {10.48550/arXiv.2411.17682},
archivePrefix = {arXiv},
       eprint = {2411.17682},
 primaryClass = {astro-ph.CO},
       adsurl = {https://ui.adsabs.harvard.edu/abs/2024arXiv241117682R}
}

@article{Fedeli:2014gja,
    author = "Fedeli, C. and Semboloni, E. and Velliscig, M. and Van Daalen, M. and Schaye, J. and Hoekstra, H.",
    title = "{The clustering of baryonic matter. II: halo model and hydrodynamic simulations}",
    eprint = "1406.5013",
    archivePrefix = "arXiv",
    primaryClass = "astro-ph.CO",
    doi = "10.1088/1475-7516/2014/08/028",
    journal = "\jcap",
    volume = "08",
    pages = "028",
    year = "2014"
}

@article{ACT:2020lcv,
    author = "Hilton, M. and others",
    collaboration = "ACT, DES",
    title = "{The Atacama Cosmology Telescope: A Catalog of \ensuremath{>}4000 Sunyaev\textendash{}Zel\textquoteright{}dovich Galaxy Clusters}",
    eprint = "2009.11043",
    archivePrefix = "arXiv",
    primaryClass = "astro-ph.CO",
    reportNumber = "DES-2020-0547, FERMILAB-PUB-20-458-AE",
    doi = "10.3847/1538-4365/abd023",
    journal = "Astrophys. J. Suppl.",
    volume = "253",
    number = "1",
    pages = "3",
    year = "2021"
}

@article{vanDaalen:2011xb,
    author = "van Daalen, Marcel P. and Schaye, Joop and Booth, C. M. and Vecchia, Claudio Dalla",
    title = "{The effects of galaxy formation on the matter power spectrum: A challenge for precision cosmology}",
    eprint = "1104.1174",
    archivePrefix = "arXiv",
    primaryClass = "astro-ph.CO",
    doi = "10.1111/j.1365-2966.2011.18981.x",
    journal = "Mon. Not. Roy. Astron. Soc.",
    volume = "415",
    pages = "3649--3665",
    year = "2011"
}

@article{Fluri:2019qtp,
    author = "Fluri, Janis and Kacprzak, Tomasz and Lucchi, Aurelien and Refregier, Alexandre and Amara, Adam and Hofmann, Thomas and Schneider, Aurel",
    title = "{Cosmological constraints with deep learning from KiDS-450 weak lensing maps}",
    eprint = "1906.03156",
    archivePrefix = "arXiv",
    primaryClass = "astro-ph.CO",
    doi = "10.1103/PhysRevD.100.063514",
    journal = "Phys. Rev. D",
    volume = "100",
    number = "6",
    pages = "063514",
    year = "2019"
}

@article{Schaller:2024jiq,
       author = {{Schaller}, Matthieu and {Schaye}, Joop and {Kugel}, Roi and {Broxterman}, Jeger C. and {van Daalen}, Marcel P.},
        title = "{The FLAMINGO project: baryon effects on the matter power spectrum}",
      journal = {\mnras},
         year = 2025,
        month = may,
       volume = {539},
       number = {2},
        pages = {1337-1351},
          doi = {10.1093/mnras/staf569},
archivePrefix = {arXiv},
       eprint = {2410.17109},
 primaryClass = {astro-ph.CO},
       adsurl = {https://ui.adsabs.harvard.edu/abs/2025MNRAS.539.1337S}
}

@article{Semboloni2011aaa,
       author = {{Semboloni}, Elisabetta and {Hoekstra}, Henk and {Schaye}, Joop and {van Daalen}, Marcel P. and {McCarthy}, Ian G.},
        title = "{Quantifying the effect of baryon physics on weak lensing tomography}",
      journal = {\mnras},
         year = 2011,
        month = nov,
       volume = {417},
       number = {3},
        pages = {2020-2035},
          doi = {10.1111/j.1365-2966.2011.19385.x},
archivePrefix = {arXiv},
       eprint = {1105.1075},
 primaryClass = {astro-ph.CO},
       adsurl = {https://ui.adsabs.harvard.edu/abs/2011MNRAS.417.2020S}
}

@article{Zhou:2025tdd,
     author = {{Zhou}, Alan Junzhe and {Gatti}, Marco and {Anbajagane}, Dhayaa and {Dodelson}, Scott and {Schaller}, Matthieu and {Schaye}, Joop},
        title = "{Map-level baryonification: unified treatment of weak lensing two-point and higher-order statistics}",
      journal = {arXiv e-prints},
         year = 2025,
        month = may,
          eid = {arXiv:2505.07949},
        pages = {arXiv:2505.07949},
          doi = {10.48550/arXiv.2505.07949},
archivePrefix = {arXiv},
       eprint = {2505.07949},
 primaryClass = {astro-ph.CO},
       adsurl = {https://ui.adsabs.harvard.edu/abs/2025arXiv250507949Z}
}

@article{Debackere:2019cec,
    author = "Debackere, Stijn N. B. and Schaye, Joop and Hoekstra, Henk",
    title = "{The impact of the observed baryon distribution in haloes on the total matter power spectrum}",
    eprint = "1908.05765",
    archivePrefix = "arXiv",
    primaryClass = "astro-ph.CO",
    doi = "10.1093/mnras/stz3446",
    journal = "Mon. Not. Roy. Astron. Soc.",
    volume = "492",
    number = "2",
    pages = "2285--2307",
    year = "2020"
}

@article{Mead:2020qgo,
    author = {Mead, A. J. and Tr\"oster, T. and Heymans, C. and Van Waerbeke, L. and McCarthy, I. G.},
    title = "{A hydrodynamical halo model for weak-lensing cross correlations}",
    eprint = "2005.00009",
    archivePrefix = "arXiv",
    primaryClass = "astro-ph.CO",
    doi = "10.1051/0004-6361/202038308",
    journal = "Astron. Astrophys.",
    volume = "641",
    pages = "A130",
    year = "2020"
}

@article{Semboloni2013aaa,
       author = {{Semboloni}, Elisabetta and {Hoekstra}, Henk and {Schaye}, Joop},
        title = "{Effect of baryonic feedback on two- and three-point shear statistics: prospects for detection and improved modelling}",
      journal = {\mnras},
         year = 2013,
        month = sep,
       volume = {434},
       number = {1},
        pages = {148-162},
          doi = {10.1093/mnras/stt1013},
archivePrefix = {arXiv},
       eprint = {1210.7303},
 primaryClass = {astro-ph.CO},
       adsurl = {https://ui.adsabs.harvard.edu/abs/2013MNRAS.434..148S}
}

@article{McCarthy:2016mry,
    author = "McCarthy, Ian G. and Schaye, Joop and Bird, Simeon and Le Brun, Amandine M. C.",
    title = "{The BAHAMAS project: Calibrated hydrodynamical simulations for large-scale structure cosmology}",
    eprint = "1603.02702",
    archivePrefix = "arXiv",
    primaryClass = "astro-ph.CO",
    doi = "10.1093/mnras/stw2792",
    journal = "Mon. Not. Roy. Astron. Soc.",
    volume = "465",
    number = "3",
    pages = "2936--2965",
    year = "2017"
}

@article{Schaller:2014uwa,
    author = "Schaller, Matthieu and Frenk, Carlos S. and Bower, Richard G. and Theuns, Tom and Jenkins, Adrian and Schaye, Joop and Crain, Robert A. and Furlong, Michelle and Vecchia, Claudio Dalla and McCarthy, I. G.",
    title = "{Baryon effects on the internal structure of \ensuremath{\Lambda}CDM haloes in the EAGLE simulations}",
    eprint = "1409.8617",
    archivePrefix = "arXiv",
    primaryClass = "astro-ph.CO",
    doi = "10.1093/mnras/stv1067",
    journal = "Mon. Not. Roy. Astron. Soc.",
    volume = "451",
    number = "2",
    pages = "1247--1267",
    year = "2015"
}

@article{ROMAN:2015aaa,
       author = {{Spergel}, D. and {Gehrels}, N. and {Baltay}, C. and {Bennett}, D. and {Breckinridge}, J. and {Donahue}, M. and {Dressler}, A. and {Gaudi}, B.~S. and {Greene}, T. and {Guyon}, O. and {Hirata}, C. and {Kalirai}, J. and {Kasdin}, N.~J. and {Macintosh}, B. and {Moos}, W. and {Perlmutter}, S. and {Postman}, M. and {Rauscher}, B. and {Rhodes}, J. and {Wang}, Y. and {Weinberg}, D. and {Benford}, D. and {Hudson}, M. and {Jeong}, W. -S. and {Mellier}, Y. and {Traub}, W. and {Yamada}, T. and {Capak}, P. and {Colbert}, J. and {Masters}, D. and {Penny}, M. and {Savransky}, D. and {Stern}, D. and {Zimmerman}, N. and {Barry}, R. and {Bartusek}, L. and {Carpenter}, K. and {Cheng}, E. and {Content}, D. and {Dekens}, F. and {Demers}, R. and {Grady}, K. and {Jackson}, C. and {Kuan}, G. and {Kruk}, J. and {Melton}, M. and {Nemati}, B. and {Parvin}, B. and {Poberezhskiy}, I. and {Peddie}, C. and {Ruffa}, J. and {Wallace}, J.~K. and {Whipple}, A. and {Wollack}, E. and {Zhao}, F.},
        title = "{Wide-Field InfrarRed Survey Telescope-Astrophysics Focused Telescope Assets WFIRST-AFTA 2015 Report}",
      journal = {arXiv e-prints},
         year = 2015,
        month = mar,
          eid = {arXiv:1503.03757},
        pages = {arXiv:1503.03757},
          doi = {10.48550/arXiv.1503.03757},
archivePrefix = {arXiv},
       eprint = {1503.03757},
 primaryClass = {astro-ph.IM},
       adsurl = {https://ui.adsabs.harvard.edu/abs/2015arXiv150303757S}
}

@article{Dalal:2023olq,
    author = "Dalal, Roohi and others",
    title = "{Hyper Suprime-Cam Year 3 results: Cosmology from cosmic shear power spectra}",
    eprint = "2304.00701",
    archivePrefix = "arXiv",
    primaryClass = "astro-ph.CO",
    doi = "10.1103/PhysRevD.108.123519",
    journal = "Phys. Rev. D",
    volume = "108",
    number = "12",
    pages = "123519",
    year = "2023"
}

@article{Bucko:2022kss,
    author = "Bucko, Jozef and Giri, Sambit K. and Schneider, Aurel",
    title = "{Constraining dark matter decay with cosmic microwave background and weak-lensing shear observations}",
    eprint = "2211.14334",
    archivePrefix = "arXiv",
    primaryClass = "astro-ph.CO",
    doi = "10.1051/0004-6361/202245562",
    journal = "Astron. Astrophys.",
    volume = "672",
    pages = "A157",
    year = "2023"
}

@article{Bulbul:2024mfj,
    author = "Bulbul, E. and others",
    title = "{The SRG/eROSITA All-Sky Survey - The first catalog of galaxy clusters and groups in the Western Galactic Hemisphere}",
    eprint = "2402.08452",
    archivePrefix = "arXiv",
    primaryClass = "astro-ph.CO",
    doi = "10.1051/0004-6361/202348264",
    journal = "Astron. Astrophys.",
    volume = "685",
    pages = "A106",
    year = "2024"
}

@article{Planck:2015vgm,
    author = "Aghanim, N. and others",
    collaboration = "Planck",
    title = "{Planck 2015 results. XXII. A map of the thermal Sunyaev-Zeldovich effect}",
    eprint = "1502.01596",
    archivePrefix = "arXiv",
    primaryClass = "astro-ph.CO",
    doi = "10.1051/0004-6361/201525826",
    journal = "Astron. Astrophys.",
    volume = "594",
    pages = "A22",
    year = "2016"
}

@article{SPT-SZ:2021gsa,
    author = "Bleem, L. E. and others",
    collaboration = "SPT-SZ",
    title = "{CMB/kSZ and Compton-y Maps from 2500 deg$^{2}$ of SPT-SZ and Planck Survey Data}",
    eprint = "2102.05033",
    archivePrefix = "arXiv",
    primaryClass = "astro-ph.CO",
    reportNumber = "FERMILAB-PUB-21-117-AE",
    doi = "10.3847/1538-4365/ac35e9",
    journal = "Astrophys. J. Supp.",
    volume = "258",
    number = "2",
    pages = "36",
    year = "2022"
}

@article{Wright:2025xka,
    author = "Wright, Angus H. and others",
    title = "{KiDS-Legacy: Cosmological constraints from cosmic shear with the complete Kilo-Degree Survey}",
    eprint = "2503.19441",
    archivePrefix = "arXiv",
    primaryClass = "astro-ph.CO",
    month = "3",
    year = "2025"
}

@article{Kacprzak:2022pww,
    author = "Kacprzak, Tomasz and Fluri, Janis and Schneider, Aurel and Refregier, Alexandre and Stadel, Joachim",
    title = "{CosmoGridV1: a simulated \ensuremath{\mathsf{w}}CDM theory prediction for map-level cosmological inference}",
    eprint = "2209.04662",
    archivePrefix = "arXiv",
    primaryClass = "astro-ph.CO",
    doi = "10.1088/1475-7516/2023/02/050",
    journal = "\jcap",
    volume = "02",
    pages = "050",
    year = "2023"
}

@article{DES:2024iny,
    author = "Bigwood, L. and others",
    collaboration = "DES",
    title = "{Weak lensing combined with the kinetic Sunyaev\textendash{}Zel\textquoteright{}dovich effect: a study of baryonic feedback}",
    eprint = "2404.06098",
    archivePrefix = "arXiv",
    primaryClass = "astro-ph.CO",
    reportNumber = "DES-2024-0827, FERMILAB-PUB-24-0130-PPD",
    doi = "10.1093/mnras/stae2100",
    journal = "Mon. Not. Roy. Astron. Soc.",
    volume = "534",
    number = "1",
    pages = "655--682",
    year = "2024"
}

@article{Grandis:2023qwx,
    author = "Grandis, Sebastian and Arico', Giovanni and Schneider, Aurel and Linke, Laila",
    title = "{Determining the baryon impact on the matter power spectrum with galaxy clusters}",
    eprint = "2309.02920",
    archivePrefix = "arXiv",
    primaryClass = "astro-ph.CO",
    doi = "10.1093/mnras/stae259",
    journal = "Mon. Not. Roy. Astron. Soc.",
    volume = "528",
    number = "3",
    pages = "4379--4392",
    year = "2024"
}

@article{Moster:2012aaa,
      author = {{Moster}, Benjamin P. and {Naab}, Thorsten and {White}, Simon D.~M.},
        title = "{Galactic star formation and accretion histories from matching galaxies to dark matter haloes}",
      journal = {\mnras},
         year = 2013,
        month = feb,
       volume = {428},
       number = {4},
        pages = {3121-3138},
          doi = {10.1093/mnras/sts261},
archivePrefix = {arXiv},
       eprint = {1205.5807},
 primaryClass = {astro-ph.CO},
       adsurl = {https://ui.adsabs.harvard.edu/abs/2013MNRAS.428.3121M}
}

@ARTICLE{Rudd:2008,
       author = {{Rudd}, Douglas H. and {Zentner}, Andrew R. and {Kravtsov}, Andrey V.},
        title = "{Effects of Baryons and Dissipation on the Matter Power Spectrum}",
      journal = {\apj},
         year = 2008,
        month = jan,
       volume = {672},
       number = {1},
        pages = {19-32},
          doi = {10.1086/523836},
archivePrefix = {arXiv},
       eprint = {astro-ph/0703741},
 primaryClass = {astro-ph},
       adsurl = {https://ui.adsabs.harvard.edu/abs/2008ApJ...672...19R}
}

@article{Gnedin:2004cx,
    author = "Gnedin, Oleg Y. and Kravtsov, Andrey V. and Klypin, Anatoly A. and Nagai, Daisuke",
    title = "{Response of dark matter halos to condensation of baryons: Cosmological simulations and improved adiabatic contraction model}",
    eprint = "astro-ph/0406247",
    archivePrefix = "arXiv",
    doi = "10.1086/424914",
    journal = "Astrophys. J.",
    volume = "616",
    pages = "16--26",
    year = "2004"
}

@article{Arico:2019ykw,
    author = "Aric\`o, Giovanni and Angulo, Raul E. and Hern\'andez-Monteagudo, Carlos and Contreras, Sergio and Zennaro, Matteo and Pellejero-Iba\~nez, Marcos and Rosas-Guevara, Yetli",
    title = "{Modelling the large-scale mass density field of the universe as a function of cosmology and baryonic physics}",
    eprint = "1911.08471",
    archivePrefix = "arXiv",
    primaryClass = "astro-ph.CO",
    doi = "10.1093/mnras/staa1478",
    journal = "Mon. Not. Roy. Astron. Soc.",
    volume = "495",
    number = "4",
    pages = "4800--4819",
    year = "2020"
}

@article{Osato:2022znr,
    author = "Osato, Ken and Nagai, Daisuke",
    title = "{Baryon pasting algorithm: halo-based and particle-based pasting methods}",
    eprint = "2201.02632",
    archivePrefix = "arXiv",
    primaryClass = "astro-ph.CO",
    reportNumber = "YITP-21-156",
    doi = "10.1093/mnras/stac3669",
    journal = "Mon. Not. Roy. Astron. Soc.",
    volume = "519",
    number = "2",
    pages = "2069--2082",
    year = "2023"
}

@article{Fluri:2022rvb,
    author = "Fluri, Janis and Kacprzak, Tomasz and Lucchi, Aurelien and Schneider, Aurel and Refregier, Alexandre and Hofmann, Thomas",
    title = "{Full wCDM analysis of KiDS-1000 weak lensing maps using deep learning}",
    eprint = "2201.07771",
    archivePrefix = "arXiv",
    primaryClass = "astro-ph.CO",
    doi = "10.1103/PhysRevD.105.083518",
    journal = "Phys. Rev. D",
    volume = "105",
    number = "8",
    pages = "083518",
    year = "2022"
}

@article{Anbajagane:2024nzx,
       author = {{Anbajagane}, Dhayaa and {Pandey}, Shivam and {Chang}, Chihway},
        title = "{Map-level baryonification: Efficient modelling of higher-order correlations in the weak lensing and thermal Sunyaev-Zeldovich fields}",
      journal = {The Open Journal of Astrophysics},
         year = 2024,
        month = dec,
       volume = {7},
          eid = {108},
        pages = {108},
          doi = {10.33232/001c.126788},
archivePrefix = {arXiv},
       eprint = {2409.03822},
 primaryClass = {astro-ph.CO},
       adsurl = {https://ui.adsabs.harvard.edu/abs/2024OJAp....7E.108A}
}

@article{Chisari:2019tus,
    author = "Chisari, Nora Elisa and others",
    title = "{Modelling baryonic feedback for survey cosmology}",
    eprint = "1905.06082",
    archivePrefix = "arXiv",
    primaryClass = "astro-ph.CO",
    doi = "10.21105/astro.1905.06082",
    journal = "Open J. Astrophys.",
    volume = "2",
    number = "1",
    pages = "4",
    year = "2019"
}

@article{DES:2022eua,
    author = "Chen, A. and others",
    collaboration = "DES",
    title = "{Constraining the baryonic feedback with cosmic shear using the DES Year-3 small-scale measurements}",
    eprint = "2206.08591",
    archivePrefix = "arXiv",
    primaryClass = "astro-ph.CO",
    reportNumber = "FERMILAB-PUB-22-540-AD-PPD-SCD-V",
    doi = "10.1093/mnras/stac3213",
    journal = "Mon. Not. Roy. Astron. Soc.",
    volume = "518",
    number = "4",
    pages = "5340--5355",
    year = "2023"
}

@article{Arico:2020yyf,
    author = "Aric\`o, Giovanni and Angulo, Raul E. and Hern\'andez-Monteagudo, Carlos and Contreras, Sergio and Zennaro, Matteo",
    title = "{Simultaneous modelling of matter power spectrum and bispectrum in the presence of baryons}",
    eprint = "2009.14225",
    archivePrefix = "arXiv",
    primaryClass = "astro-ph.CO",
    doi = "10.1093/mnras/stab699",
    journal = "Mon. Not. Roy. Astron. Soc.",
    volume = "503",
    number = "3",
    pages = "3596--3609",
    year = "2021"
}

@article{Arico:2023ocu,
    author = "Aric\`o, Giovanni and Angulo, Raul E. and Zennaro, Matteo and Contreras, Sergio and Chen, Angela and Hern\'andez-Monteagudo, Carlos",
    title = "{DES Y3 cosmic shear down to small scales: Constraints on cosmology and baryons}",
    eprint = "2303.05537",
    archivePrefix = "arXiv",
    primaryClass = "astro-ph.CO",
    doi = "10.1051/0004-6361/202346539",
    journal = "Astron. Astrophys.",
    volume = "678",
    pages = "A109",
    year = "2023"
}

@article{Weiss:2019jfx,
    author = "Weiss, Andreas J. and Schneider, Aurel and Sgier, Raphael and Kacprzak, Tomasz and Amara, Adam and Refregier, Alexandre",
    title = "{Effects of baryons on weak lensing peak statistics}",
    eprint = "1905.11636",
    archivePrefix = "arXiv",
    primaryClass = "astro-ph.CO",
    doi = "10.1088/1475-7516/2019/10/011",
    journal = "\jcap",
    volume = "10",
    pages = "011",
    year = "2019"
}

@article{Williams:2022zma,
    author = "Williams, Ian M. and Khan, Adnan and McQuinn, Matthew",
    title = "{Baryonic post-processing of N-body simulations, with application to fast radio bursts}",
    eprint = "2207.05233",
    archivePrefix = "arXiv",
    primaryClass = "astro-ph.CO",
    doi = "10.1093/mnras/stad293",
    journal = "Mon. Not. Roy. Astron. Soc.",
    volume = "520",
    number = "3",
    pages = "3626--3640",
    year = "2023"
}

@article{Abadi:2010aaa,
       author = {{Abadi}, Mario G. and {Navarro}, Julio F. and {Fardal}, Mark and {Babul}, Arif and {Steinmetz}, Matthias},
        title = "{Galaxy-induced transformation of dark matter haloes}",
      journal = {\mnras},
         year = 2010,
        month = sep,
       volume = {407},
       number = {1},
        pages = {435-446},
          doi = {10.1111/j.1365-2966.2010.16912.x},
archivePrefix = {arXiv},
       eprint = {0902.2477},
 primaryClass = {astro-ph.GA},
       adsurl = {https://ui.adsabs.harvard.edu/abs/2010MNRAS.407..435A}
}

@ARTICLE{Potter_pkdgrav3_2017,
       author = {{Potter}, Douglas and {Stadel}, Joachim and {Teyssier}, Romain},
        title = "{PKDGRAV3: beyond trillion particle cosmological simulations for the next era of galaxy surveys}",
      journal = {Computational Astrophysics and Cosmology},
         year = 2017,
        month = may,
       volume = {4},
       number = {1},
          eid = {2},
        pages = {2},
          doi = {10.1186/s40668-017-0021-1},
archivePrefix = {arXiv},
       eprint = {1609.08621},
 primaryClass = {astro-ph.IM},
       adsurl = {https://ui.adsabs.harvard.edu/abs/2017ComAC...4....2P}
}

@ARTICLE{Knollmann:2009,
       author = {{Knollmann}, Steffen R. and {Knebe}, Alexander},
        title = "{AHF: Amiga's Halo Finder}",
      journal = {\apjs},
         year = 2009,
        month = jun,
       volume = {182},
       number = {2},
        pages = {608-624},
          doi = {10.1088/0067-0049/182/2/608},
archivePrefix = {arXiv},
       eprint = {0904.3662},
 primaryClass = {astro-ph.CO},
       adsurl = {https://ui.adsabs.harvard.edu/abs/2009ApJS..182..608K}
}

@article{Teyssier:2011aaa,
       author = {{Teyssier}, Romain and {Moore}, Ben and {Martizzi}, Davide and {Dubois}, Yohan and {Mayer}, Lucio},
        title = "{Mass distribution in galaxy clusters: the role of Active Galactic Nuclei feedback}",
      journal = {\mnras},
         year = 2011,
        month = jun,
       volume = {414},
       number = {1},
        pages = {195-208},
          doi = {10.1111/j.1365-2966.2011.18399.x},
archivePrefix = {arXiv},
       eprint = {1003.4744},
 primaryClass = {astro-ph.CO},
       adsurl = {https://ui.adsabs.harvard.edu/abs/2011MNRAS.414..195T}
}

@article{Schneider:2021wds,
    author = "Schneider, Aurel and Giri, Sambit K. and Amodeo, Stefania and Refregier, Alexandre",
    title = "{Constraining baryonic feedback and cosmology with weak-lensing, X-ray, and kinematic Sunyaev\textendash{}Zeldovich observations}",
    eprint = "2110.02228",
    archivePrefix = "arXiv",
    primaryClass = "astro-ph.CO",
    doi = "10.1093/mnras/stac1493",
    journal = "Mon. Not. Roy. Astron. Soc.",
    volume = "514",
    number = "3",
    pages = "3802--3814",
    year = "2022"
}

@article{Schaye:2023jqv,
    author = "Schaye, Joop and others",
    title = "{The FLAMINGO project: cosmological hydrodynamical simulations for large-scale structure and galaxy cluster surveys}",
    eprint = "2306.04024",
    archivePrefix = "arXiv",
    primaryClass = "astro-ph.CO",
    doi = "10.1093/mnras/stad2419",
    journal = "Mon. Not. Roy. Astron. Soc.",
    volume = "526",
    number = "4",
    pages = "4978--5020",
    year = "2023"
}

@article{Kugel:2023wte,
    author = "Kugel, Roi and others",
    title = "{FLAMINGO: calibrating large cosmological hydrodynamical simulations with machine learning}",
    eprint = "2306.05492",
    archivePrefix = "arXiv",
    primaryClass = "astro-ph.CO",
    doi = "10.1093/mnras/stad2540",
    journal = "Mon. Not. Roy. Astron. Soc.",
    volume = "526",
    number = "4",
    pages = "6103--6127",
    year = "2023"
}

@article{Springel:2017tpz,
    author = "Springel, Volker and others",
    title = "{First results from the IllustrisTNG simulations: matter and galaxy clustering}",
    eprint = "1707.03397",
    archivePrefix = "arXiv",
    primaryClass = "astro-ph.GA",
    doi = "10.1093/mnras/stx3304",
    journal = "Mon. Not. Roy. Astron. Soc.",
    volume = "475",
    number = "1",
    pages = "676--698",
    year = "2018"
}

@article{Dutton:2014xda,
    author = "Dutton, Aaron A. and Macci\`o, Andrea V.",
    title = "{Cold dark matter haloes in the Planck era: evolution of structural parameters for Einasto and NFW profiles}",
    eprint = "1402.7073",
    archivePrefix = "arXiv",
    primaryClass = "astro-ph.CO",
    doi = "10.1093/mnras/stu742",
    journal = "Mon. Not. Roy. Astron. Soc.",
    volume = "441",
    number = "4",
    pages = "3359--3374",
    year = "2014"
}

@article{Pillepich:2017fcc,
    author = "Pillepich, Annalisa and others",
    title = "{First results from the IllustrisTNG simulations: the stellar mass content of groups and clusters of galaxies}",
    eprint = "1707.03406",
    archivePrefix = "arXiv",
    primaryClass = "astro-ph.GA",
    doi = "10.1093/mnras/stx3112",
    journal = "Mon. Not. Roy. Astron. Soc.",
    volume = "475",
    number = "1",
    pages = "648--675",
    year = "2018"
}

@article{Baltz:2007vq,
    author = "Baltz, Edward A. and Marshall, Phil and Oguri, Masamune",
    title = "{Analytic models of plausible gravitational lens potentials}",
    eprint = "0705.0682",
    archivePrefix = "arXiv",
    primaryClass = "astro-ph",
    reportNumber = "SLAC-PUB-12497",
    doi = "10.1088/1475-7516/2009/01/015",
    journal = "\jcap",
    volume = "01",
    pages = "015",
    year = "2009"
}

@article{Nelson:2017cxy,
    author = "Nelson, Dylan and others",
    title = "{First results from the IllustrisTNG simulations: the galaxy colour bimodality}",
    eprint = "1707.03395",
    archivePrefix = "arXiv",
    primaryClass = "astro-ph.GA",
    doi = "10.1093/mnras/stx3040",
    journal = "Mon. Not. Roy. Astron. Soc.",
    volume = "475",
    number = "1",
    pages = "624--647",
    year = "2018"
}

@article{Marinacci:2017wew,
    author = "Marinacci, Federico and others",
    title = "{First results from the IllustrisTNG simulations: radio haloes and magnetic fields}",
    eprint = "1707.03396",
    archivePrefix = "arXiv",
    primaryClass = "astro-ph.CO",
    doi = "10.1093/mnras/sty2206",
    journal = "Mon. Not. Roy. Astron. Soc.",
    volume = "480",
    number = "4",
    pages = "5113--5139",
    year = "2018"
}

@article{Naiman:2017aaa,
       author = {{Naiman}, Jill P. and {Pillepich}, Annalisa and {Springel}, Volker and {Ramirez-Ruiz}, Enrico and {Torrey}, Paul and {Vogelsberger}, Mark and {Pakmor}, R{\"u}diger and {Nelson}, Dylan and {Marinacci}, Federico and {Hernquist}, Lars and {Weinberger}, Rainer and {Genel}, Shy},
        title = "{First results from the IllustrisTNG simulations: a tale of two elements - chemical evolution of magnesium and europium}",
      journal = {\mnras},
         year = 2018,
        month = jun,
       volume = {477},
       number = {1},
        pages = {1206-1224},
          doi = {10.1093/mnras/sty618},
archivePrefix = {arXiv},
       eprint = {1707.03401},
 primaryClass = {astro-ph.GA},
       adsurl = {https://ui.adsabs.harvard.edu/abs/2018MNRAS.477.1206N}
}

@article{Schneider:2015wta,
    author = "Schneider, Aurel and Teyssier, Romain",
    title = "{A new method to quantify the effects of baryons on the matter power spectrum}",
    eprint = "1510.06034",
    archivePrefix = "arXiv",
    primaryClass = "astro-ph.CO",
    doi = "10.1088/1475-7516/2015/12/049",
    journal = "\jcap",
    volume = "12",
    pages = "049",
    year = "2015"
}

@article{Velmani:2022una,
    author = "Velmani, Premvijay and Paranjape, Aseem",
    title = "{The quasi-adiabatic relaxation of haloes in the IllustrisTNG and EAGLE cosmological simulations}",
    eprint = "2206.07733",
    archivePrefix = "arXiv",
    primaryClass = "astro-ph.GA",
    doi = "10.1093/mnras/stad297",
    journal = "Mon. Not. Roy. Astron. Soc.",
    volume = "520",
    number = "2",
    pages = "2867--2886",
    year = "2023"
}

@article{Blumenthal:1985qy,
    author = "Blumenthal, George R. and Faber, S. M. and Flores, Ricardo and Primack, Joel R.",
    title = "{Contraction of Dark Matter Galactic Halos Due to Baryonic Infall}",
    reportNumber = "SCIPP 85/42",
    doi = "10.1086/163867",
    journal = "Astrophys. J.",
    volume = "301",
    pages = "27",
    year = "1986"
}

@article{Hayashi:2007uk,
    author = "Hayashi, E. and White, S. D. M.",
    title = "{Understanding the shape of the halo-mass and galaxy-mass cross-correlation functions}",
    eprint = "0709.3933",
    archivePrefix = "arXiv",
    primaryClass = "astro-ph",
    doi = "10.1111/j.1365-2966.2008.13371.x",
    journal = "Mon. Not. Roy. Astron. Soc.",
    volume = "388",
    pages = "2",
    year = "2008"
}

@article{Diemer:2014xya,
    author = "Diemer, Benedikt and Kravtsov, Andrey V.",
    title = "{Dependence of the outer density profiles of halos on their mass accretion rate}",
    eprint = "1401.1216",
    archivePrefix = "arXiv",
    primaryClass = "astro-ph.CO",
    doi = "10.1088/0004-637X/789/1/1",
    journal = "Astrophys. J.",
    volume = "789",
    pages = "1",
    year = "2014"
}

@article{Oguri:2011vj,
    author = "Oguri, Masamune and Hamana, Takashi",
    title = "{Detailed cluster lensing profiles at large radii and the impact on cluster weak lensing studies}",
    eprint = "1101.0650",
    archivePrefix = "arXiv",
    primaryClass = "astro-ph.CO",
    doi = "10.1111/j.1365-2966.2011.18481.x",
    journal = "Mon. Not. Roy. Astron. Soc.",
    volume = "414",
    pages = "1851--1861",
    year = "2011"
}

@article{Schneider:2018pfw,
    author = "Schneider, Aurel and Teyssier, Romain and Stadel, Joachim and Chisari, Nora Elisa and Le Brun, Amandine M. C. and Amara, Adam and Refregier, Alexandre",
    title = "{Quantifying baryon effects on the matter power spectrum and the weak lensing shear correlation}",
    eprint = "1810.08629",
    archivePrefix = "arXiv",
    primaryClass = "astro-ph.CO",
    doi = "10.1088/1475-7516/2019/03/020",
    journal = "\jcap",
    volume = "03",
    pages = "020",
    year = "2019"
}

@article{Troster:2021gsz,
    author = {Tr{\"o}ster, Tilman and others},
    title = "{Joint constraints on cosmology and the impact of baryon feedback: Combining KiDS-1000 lensing with the thermal Sunyaev{\textendash}Zeldovich effect from Planck and ACT}",
    eprint = "2109.04458",
    archivePrefix = "arXiv",
    primaryClass = "astro-ph.CO",
    doi = "10.1051/0004-6361/202142197",
    journal = "Astron. Astrophys.",
    volume = "660",
    pages = "A27",
    year = "2022"
}

@article{Planck:2015koh,
    author = "Ade, P. A. R. and others",
    collaboration = "Planck",
    title = "{Planck 2015 results. XXVII. The Second Planck Catalogue of Sunyaev-Zeldovich Sources}",
    eprint = "1502.01598",
    archivePrefix = "arXiv",
    primaryClass = "astro-ph.CO",
    doi = "10.1051/0004-6361/201525823",
    journal = "Astron. Astrophys.",
    volume = "594",
    pages = "A27",
    year = "2016"
}

@article{Zennaro:2024dyy,
    author = "Zennaro, Matteo and Aric{\`o}, Giovanni and Garc{\'\i}a-Garc{\'\i}a, Carlos and Angulo, Ra{\'u}l E. and Ondaro-Mallea, Lurdes and Contreras, Sergio and Nicola, Andrina and Schaller, Matthieu and Schaye, Joop",
    title = "{A 1{\%} accurate method to include baryonic effects in galaxy-galaxy lensing models}",
    eprint = "2412.08623",
    archivePrefix = "arXiv",
    primaryClass = "astro-ph.CO",
    month = "12",
    year = "2024"
}

@article{Burger:2025meh,
    author = "Burger, P. A. and others",
    title = "{Euclid: An emulator for baryonic effects on the matter bispectrum}",
    eprint = "2506.18974",
    archivePrefix = "arXiv",
    primaryClass = "astro-ph.CO",
    month = "6",
    year = "2025"
}

@article{ACT:2025llb,
    author = "Pandey, S. and others",
    collaboration = "ACT, DES",
    title = "{Constraints on cosmology and baryonic feedback with joint analysis of Dark Energy Survey Year 3 lensing data and ACT DR6 thermal Sunyaev-Zel'dovich effect observations}",
    eprint = "2506.07432",
    archivePrefix = "arXiv",
    primaryClass = "astro-ph.CO",
    reportNumber = "FERMILAB-PUB-25-0394-PPD",
    month = "6",
    year = "2025"
}

@article{Lee:2022jyg,
    author = "Lee, Max E. and Lu, Tianhuan and Haiman, Zolt\'an and Liu, Jia and Osato, Ken",
    title = "{Comparing weak lensing peak counts in baryonic correction models to hydrodynamical simulations}",
    eprint = "2201.08320",
    archivePrefix = "arXiv",
    primaryClass = "astro-ph.CO",
    doi = "10.1093/mnras/stac3592",
    journal = "Mon. Not. Roy. Astron. Soc.",
    volume = "519",
    number = "1",
    pages = "573--584",
    year = "2023"
}

@article{Euclid:2024yrr,
       author = {{Euclid Collaboration} and {Mellier}, Y. and {Abdurro'uf} and {Acevedo Barroso}, J.~A. and {Ach{\'u}carro}, A. and {Adamek}, J. and {Adam}, R. and {Addison}, G.~E. and {Aghanim}, N. and {Aguena}, M. and {Ajani}, V. and {Akrami}, Y. and {Al-Bahlawan}, A. and {Alavi}, A. and {Albuquerque}, I.~S. and {Alestas}, G. and {Alguero}, G. and {Allaoui}, A. and {Allen}, S.~W. and {Allevato}, V. and {Alonso-Tetilla}, A.~V. and {Altieri}, B. and {Alvarez-Candal}, A. and {Alvi}, S. and {Amara}, A. and {Amendola}, L. and {Amiaux}, J. and {Andika}, I.~T. and {Andreon}, S. and {Andrews}, A. and {Angora}, G. and {Angulo}, R.~E. and {Annibali}, F. and {Anselmi}, A. and {Anselmi}, S. and {Arcari}, S. and {Archidiacono}, M. and {Aric{\`o}}, G. and {Arnaud}, M. and {Arnouts}, S. and {Asgari}, M. and {Asorey}, J. and {Atayde}, L. and {Atek}, H. and {Atrio-Barandela}, F. and {Aubert}, M. and {Aubourg}, E. and {Auphan}, T. and {Auricchio}, N. and {Aussel}, B. and {Aussel}, H. and {Avelino}, P.~P. and {Avgoustidis}, A. and {Avila}, S. and {Awan}, S. and {Azzollini}, R. and {Baccigalupi}, C. and {Bachelet}, E. and {Bacon}, D. and {Baes}, M. and {Bagley}, M.~B. and {Bahr-Kalus}, B. and {Balaguera-Antolinez}, A. and {Balbinot}, E. and {Balcells}, M. and {Baldi}, M. and {Baldry}, I. and {Balestra}, A. and {Ballardini}, M. and {Ballester}, O. and {Balogh}, M. and {Ba{\~n}ados}, E. and {Barbier}, R. and {Bardelli}, S. and {Baron}, M. and {Barreiro}, T. and {Barrena}, R. and {Barriere}, J. -C. and {Barros}, B.~J. and {Barthelemy}, A. and {Bartolo}, N. and {Basset}, A. and {Battaglia}, P. and {Battisti}, A.~J. and {Baugh}, C.~M. and {Baumont}, L. and {Bazzanini}, L. and {Beaulieu}, J. -P. and {Beckmann}, V. and {Belikov}, A.~N. and {Bel}, J. and {Bellagamba}, F. and {Bella}, M. and {Bellini}, E. and {Benabed}, K. and {Bender}, R. and {Benevento}, G. and {Bennett}, C.~L. and {Benson}, K. and {Bergamini}, P. and {Bermejo-Climent}, J.~R. and {Bernardeau}, F. and {Bertacca}, D. and {Berthe}, M. and {Berthier}, J. and {Bethermin}, M. and {Beutler}, F. and {Bevillon}, C. and {Bhargava}, S. and {Bhatawdekar}, R. and {Bianchi}, D. and {Bisigello}, L. and {Biviano}, A. and {Blake}, R.~P. and {Blanchard}, A. and {Blazek}, J. and {Blot}, L. and {Bosco}, A. and {Bodendorf}, C. and {Boenke}, T. and {B{\"o}hringer}, H. and {Boldrini}, P. and {Bolzonella}, M. and {Bonchi}, A. and {Bonici}, M. and {Bonino}, D. and {Bonino}, L. and {Bonvin}, C. and {Bon}, W. and {Booth}, J.~T. and {Borgani}, S. and {Borlaff}, A.~S. and {Borsato}, E. and {Bose}, B. and {Botticella}, M.~T. and {Boucaud}, A. and {Bouche}, F. and {Boucher}, J.~S. and {Boutigny}, D. and {Bouvard}, T. and {Bouwens}, R. and {Bouy}, H. and {Bowler}, R.~A.~A. and {Bozza}, V. and {Bozzo}, E. and {Branchini}, E. and {Brando}, G. and {Brau-Nogue}, S. and {Brekke}, P. and {Bremer}, M.~N. and {Brescia}, M. and {Breton}, M. -A. and {Brinchmann}, J. and {Brinckmann}, T. and {Brockley-Blatt}, C. and {Brodwin}, M. and {Brouard}, L. and {Brown}, M.~L. and {Bruton}, S. and {Bucko}, J. and {Buddelmeijer}, H. and {Buenadicha}, G. and {Buitrago}, F. and {Burger}, P. and {Burigana}, C. and {Busillo}, V. and {Busonero}, D. and {Cabanac}, R. and {Cabayol-Garcia}, L. and {Cagliari}, M.~S. and {Caillat}, A. and {Caillat}, L. and {Calabrese}, M. and {Calabro}, A. and {Calderone}, G. and {Calura}, F. and {Camacho Quevedo}, B. and {Camera}, S. and {Campos}, L. and {Ca{\~n}as-Herrera}, G. and {Candini}, G.~P. and {Cantiello}, M. and {Capobianco}, V. and {Cappellaro}, E. and {Cappelluti}, N. and {Cappi}, A. and {Caputi}, K.~I. and {Cara}, C. and {Carbone}, C. and {Cardone}, V.~F. and {Carella}, E. and {Carlberg}, R.~G. and {Carle}, M. and {Carminati}, L. and {Caro}, F. and {Carrasco}, J.~M. and {Carretero}, J. and {Carrilho}, P. and {Carron Duque}, J. and {Carry}, B.},
        title = "{Euclid: I. Overview of the Euclid mission}",
      journal = {\aap},
         year = 2025,
        month = may,
       volume = {697},
          eid = {A1},
        pages = {A1},
          doi = {10.1051/0004-6361/202450810},
archivePrefix = {arXiv},
       eprint = {2405.13491},
 primaryClass = {astro-ph.CO},
       adsurl = {https://ui.adsabs.harvard.edu/abs/2025A&A...697A...1E}
}

@article{Arico:2024pvt,
    author = "Aric{\`o}, Giovanni and Angulo, Raul E.",
    title = "{Baryonification extended to thermal Sunyaev Zel{\textquoteright}dovich}",
    eprint = "2406.01672",
    archivePrefix = "arXiv",
    primaryClass = "astro-ph.CO",
    doi = "10.1051/0004-6361/202451055",
    journal = "Astron. Astrophys.",
    volume = "690",
    pages = "A188",
    year = "2024"
}

@article{DES:2020aks,
    author = "Sevilla-Noarbe, I. and others",
    collaboration = "DES",
    title = "{Dark Energy Survey Year 3 Results: Photometric Data Set for Cosmology}",
    eprint = "2011.03407",
    archivePrefix = "arXiv",
    primaryClass = "astro-ph.CO",
    reportNumber = "FERMILAB-PUB-20-569-AE, DES-2019-0451",
    doi = "10.3847/1538-4365/abeb66",
    journal = "Astrophys. J. Suppl.",
    volume = "254",
    number = "2",
    pages = "24",
    year = "2021"
}

@article{SPT:2014wbo,
    author = "Bleem, L. E. and others",
    collaboration = "SPT",
    title = "{Galaxy Clusters Discovered via the Sunyaev-Zel'dovich Effect in the 2500-square-degree SPT-SZ survey}",
    eprint = "1409.0850",
    archivePrefix = "arXiv",
    primaryClass = "astro-ph.CO",
    reportNumber = "FERMILAB-PUB-14-548-AE",
    doi = "10.1088/0067-0049/216/2/27",
    journal = "Astrophys. J. Suppl.",
    volume = "216",
    number = "2",
    pages = "27",
    year = "2015"
}

@article{AtacamaCosmologyTelescope:2020wtv,
    author = "Schaan, Emmanuel and others",
    collaboration = "Atacama Cosmology Telescope",
    title = "{Atacama Cosmology Telescope: Combined kinematic and thermal Sunyaev-Zel\textquoteright{}dovich measurements from BOSS CMASS and LOWZ halos}",
    eprint = "2009.05557",
    archivePrefix = "arXiv",
    primaryClass = "astro-ph.CO",
    doi = "10.1103/PhysRevD.103.063513",
    journal = "Phys. Rev. D",
    volume = "103",
    number = "6",
    pages = "063513",
    year = "2021"
}

@article{LSST:2008ijt,
    author = "Ivezi\'c, \v{Z}eljko and others",
    collaboration = "LSST",
    title = "{LSST: from Science Drivers to Reference Design and Anticipated Data Products}",
    eprint = "0805.2366",
    archivePrefix = "arXiv",
    primaryClass = "astro-ph",
    reportNumber = "SLAC-PUB-16076",
    doi = "10.3847/1538-4357/ab042c",
    journal = "Astrophys. J.",
    volume = "873",
    number = "2",
    pages = "111",
    year = "2019"
}

@article{LSSTDarkEnergyScience:2018jkl,
       author = {{The LSST Dark Energy Science Collaboration} and {Mandelbaum}, Rachel and {Eifler}, Tim and {Hlo{\v{z}}ek}, Ren{\'e}e and {Collett}, Thomas and {Gawiser}, Eric and {Scolnic}, Daniel and {Alonso}, David and {Awan}, Humna and {Biswas}, Rahul and {Blazek}, Jonathan and {Burchat}, Patricia and {Chisari}, Nora Elisa and {Dell'Antonio}, Ian and {Digel}, Seth and {Frieman}, Josh and {Goldstein}, Daniel A. and {Hook}, Isobel and {Ivezi{\'c}}, {\v{Z}}eljko and {Kahn}, Steven M. and {Kamath}, Sowmya and {Kirkby}, David and {Kitching}, Thomas and {Krause}, Elisabeth and {Leget}, Pierre-Fran{\c{c}}ois and {Marshall}, Philip J. and {Meyers}, Joshua and {Miyatake}, Hironao and {Newman}, Jeffrey A. and {Nichol}, Robert and {Rykoff}, Eli and {Sanchez}, F. Javier and {Slosar}, An{\v{z}}e and {Sullivan}, Mark and {Troxel}, M.~A.},
        title = "{The LSST Dark Energy Science Collaboration (DESC) Science Requirements Document}",
      journal = {arXiv e-prints},
         year = 2018,
        month = sep,
          eid = {arXiv:1809.01669},
        pages = {arXiv:1809.01669},
          doi = {10.48550/arXiv.1809.01669},
archivePrefix = {arXiv},
       eprint = {1809.01669},
 primaryClass = {astro-ph.CO},
       adsurl = {https://ui.adsabs.harvard.edu/abs/2018arXiv180901669T}
}

@article{Euclid:2019clj,
    author = "Blanchard, A. and others",
    collaboration = "Euclid",
    title = "{Euclid preparation. VII. Forecast validation for Euclid cosmological probes}",
    eprint = "1910.09273",
    archivePrefix = "arXiv",
    primaryClass = "astro-ph.CO",
    doi = "10.1051/0004-6361/202038071",
    journal = "Astron. Astrophys.",
    volume = "642",
    pages = "A191",
    year = "2020"
}

@article{Sheth:1999mn,
    author = "Sheth, Ravi K. and Tormen, Giuseppe",
    title = "{Large scale bias and the peak background split}",
    eprint = "astro-ph/9901122",
    archivePrefix = "arXiv",
    doi = "10.1046/j.1365-8711.1999.02692.x",
    journal = "Mon. Not. Roy. Astron. Soc.",
    volume = "308",
    pages = "119",
    year = "1999"
}

@article{giri2021emulation,
       author = {{Giri}, Sambit K. and {Schneider}, Aurel},
        title = "{Emulation of baryonic effects on the matter power spectrum and constraints from galaxy cluster data}",
      journal = {\jcap},
         year = 2021,
        month = dec,
       volume = {2021},
       number = {12},
          eid = {046},
        pages = {046},
          doi = {10.1088/1475-7516/2021/12/046},
archivePrefix = {arXiv},
       eprint = {2108.08863},
 primaryClass = {astro-ph.CO},
       adsurl = {https://ui.adsabs.harvard.edu/abs/2021JCAP...12..046G}
}

@article{Shaw2010,
       author = {{Shaw}, Laurie D. and {Nagai}, Daisuke and {Bhattacharya}, Suman and {Lau}, Erwin T.},
        title = "{Impact of Cluster Physics on the Sunyaev-Zel'dovich Power Spectrum}",
      journal = {\apj},
         year = 2010,
        month = dec,
       volume = {725},
       number = {2},
        pages = {1452-1465},
          doi = {10.1088/0004-637X/725/2/1452},
archivePrefix = {arXiv},
       eprint = {1006.1945},
 primaryClass = {astro-ph.CO},
       adsurl = {https://ui.adsabs.harvard.edu/abs/2010ApJ...725.1452S}
}

\end{document}